\documentclass[reprint,aps,prl,groupedaddress,10pt]{revtex4-2}
\usepackage{packages}
\usepackage{notations}

\usepackage{stackengine}
\begin{document}
\setlength{\abovedisplayskip}{4pt}
\setlength{\belowdisplayskip}{4pt}
\title{Integrability and exact large deviations of the weakly-asymmetric exclusion process}
\author{Alexandre Krajenbrink}
\email{alexandre.krajenbrink@quantinuum.com}
\affiliation{Quantinuum, Partnership House, Carlisle Place, London SW1P 1BX, United Kingdom}
\affiliation{Le Lab Quantique, 58 rue d'Hauteville, 75010, Paris, France}
\author{Pierre Le Doussal}
\affiliation{Laboratoire de Physique de l'\'Ecole Normale Sup\'erieure, CNRS, ENS $\&$ PSL University, Sorbonne Universit\'e, Universit\'e de Paris, 75005 Paris, France}
\date{\today}
\begin{abstract}
The weakly asymmetric exclusion process (WASEP) in one dimension is a paradigmatic
system of interacting particles described by the macroscopic fluctuation theory (MFT)
in the presence of driving. We consider an initial condition with densities $\rho_1,\rho_2$
on either side of the origin, so that for $\rho_1=\rho_2$ the gas is stationary.
Starting from the microscopic description, we obtain exact formulae
for the cumulant generating functions,
and
large deviation rate functions of the time-integrated current
and the position
of a tracer. As the asymmetry/driving is increased, these describe the crossover between the
symmetric exclusion process (SSEP) and the weak noise regime of the Kardar-Parisi-Zhang (KPZ) equation:
we recover the two limits and describe the crossover from
the WASEP cubic tail to the $5/2$ and $3/2$ KPZ tail exponents.
Finally, we show that the MFT of the
WASEP is classically integrable, by exhibiting the explicit Lax pairs,
which are obtained through a novel mapping between the MFT of the WASEP
and a complex extension of the classical anisotropic Landau-Lifshitz spin chain.
This shows integrability of all MFTs of asymmetric models with quadratic mobility
as well as their dual versions.
\end{abstract}
\maketitle
{\bf Introduction}.
Interacting particle systems in one dimension are a playground
to test the developing tools of non-equilibrium physics \cite{spohn2012large,chou2011non,derrida2007Review}. It has practical applications, e.g., to single-file diffusion which describes the transport of non-crossing particles in narrow channels
\cite{lea1963permeation,hahn1996single,wei2000single,kollmann2003single,lin2005random,chowdhury2000statistical,lipowsky2001random}.
Solvable microscopic models, and fluctuating hydrodynamic theories play an important role.
When the particle hopping is symmetric, as for the symmetric exclusion process (SSEP),
the fluctuations of the
density and current obey diffusive scaling at large scale.
The macroscopic fluctuation theory (MFT) \cite{bertini2015macroscopic,footnoteHydroSSEP} then provides a powerful framework in which all microscopic details are replaced by two model-dependent transport coefficients $D(\rho)$ and $\sigma(\rho)$
\cite{arita2017variational},
and the particle density $\rho$ obeys a conservative stochastic equation with weak noise.
While the typical fluctuations are Gaussian, these systems exhibit nontrivial
cumulants and large deviation tails well described by MFT \cite{derrida2007Review}.
For driven systems, such as the asymmetric exclusion process (ASEP)
\cite{ferrari1994current,mallick2015exclusion}
these fluctuations
fall instead in the realm of the
1D Kardar-Parisi-Zhang (KPZ) universality class \cite{tracy2009asymptotics,sasamoto2010crossover},
and are beyond the reach of the MFT. Nevertheless,
a MFT/weak noise approach is possible when the asymmetry/driving
is small and scaled appropriately: this is the {\it weakly
asymmetric exclusion process} (WASEP)
\cite{bodineau2005current,prolhac2009cumulants,vanicat2021mapping,zhao2024moderate}.
Note that the KPZ equation itself, in its short time regime, can also
be studied using a weak noise theory (WNT) \cite{kolokolov2009explicit,kolokolov2007optimal,meerson2016large,kamenev2016short}.
Another avenue of applications of MFT was opened recently by considering quantum extensions
of the SSEP \cite{bauer2017stochastic,bernard2021solution,Bernard_2021,costa2025emergence}.

Recently there has been notable progress in obtaining exact results for particle systems described
by MFT or weak noise in one dimension. The fluctuations of the current and density
in the steady state have been much studied for an open
interval with two reservoirs \cite{bodineau2005current,derrida2004current},
often using
exact solutions for microscopic models (such as WASEP and ASEP) \cite{derrida2002exact,Enaud2004,de2011large,gorissen2012exact,LazarescuThesis},
as well as for particles on a ring
\cite{derrida1998exact,lee1999large,derrida1999universal,appert2008universal,prolhac2009cumulants,dagallier2023fluctuations}.
The problem on the infinite line, with an initial condition which may or may not be stationary,
is another challenge.
Some observables of interest are the time-integrated current,
i.e., the total particle flux through the origin up to time $t$,
or
the position of a tracer particle at time $t$. The aim is
to obtain the cumulants of their fluctuations, encoded
in the cumulant generating
function (CGF), and the corresponding
large deviation rate functions.
Whenever the noise is weak, one can use saddle point methods (akin to a "classical" limit)
and reduce the problem to systems of non linear PDE's with mixed boundary conditions.
Perturbative methods then allow to compute a few lowest order cumulants (quickly
intractable with increasing order) \cite{krapivsky2014large,grabsch2024tracer,berlioz2024tracer},
as well as some asymptotics of the
rate functions \cite{meerson2016large,kamenev2016short}.
To go beyond and obtain the complete exact solution there are presently two main routes. When the microscopic model is
integrable, e.g., from Bethe ansatz methods, moment or Fredholm determinant
formula are available for some observables, and it may be
possible to extract their weak noise/"classical" asymptotics.
This was applied to obtain the integrated current and tracer
CGF at large time for the SSEP with step or stationary initial condition
\cite{derrida2009currentExact,derrida2009currentMFT,imamura2017large,Imamura2021,dandekar2022macroscopic}.
It was also done for the KPZ equation at short time,
with various initial conditions
\cite{le2016exact,krajenbrink2017exact,smith2018exact}.
The second route, opened by us in the case of the WNT of the KPZ equation
\cite{UsWNTDroplet2021,UsWNTFlat2021,tsai2023integrability}, and subsequently applied to the MFT of the SSEP
\cite{SasamotoExactSSEP,mallick2024exact}, and of the KMP model
\cite{NaftaliKMP1,UsWNTCrossover,NaftaliKMP2},
is to directly identify the classical Lax integrability structure of the saddle point equations
and to use inverse scattering methods to obtain large deviation quantities.
This was also achieved in a parallel series of works
\cite{grabsch2022exact,poncet2021generalized,grabsch2024joint},
and using duality
\cite{Rizkallah_2023}, for review see \cite{grabsch2023exact}.
Until now however the WASEP has resisted these approaches,
a notable exception.

The aim of this Letter is to obtain the exact cumulant generating
functions, and large deviations rate functions for the integrated current and the tracer position
in the WASEP. We will start with the first method, taking
a limit from the ASEP. Then we will indicate how
the second method can be applied, by
exhibiting an explicit Lax pair.

Let us first recall that in the ASEP
each particle attempts a simple
random walk on $\Z$, with continuous time denoted ${\sf t} \in [0,{\sf T}]$,
jumping left at rate $L$ and right at rate $R$, with the caveat that jumps
are suppressed if the destination site is already occupied.
The WASEP is defined as the large-time/weak-asymmetry limit, where the asymmetry
is scaled as $1/\sqrt{{\sf T}}$ where ${\sf T} \gg 1$ is the observation time.
To this aim, we introduce a small parameter $\varepsilon$,
choose $R=1+\varepsilon \nu$ and $L=1-\varepsilon \nu$,
and rescale time and space
as $t=\varepsilon^2 {\sf t}$, $x= \varepsilon m$, where $m \in \Z$ \cite{footnotescaling}.
so that $x \in \mathbb{R}$ in the limit $\varepsilon \to 0$. The rescaled final time is thus
$T = {\sf T} \varepsilon^2$, typically chosen $T = \mathcal{O}(1)$.

In the limit $\varepsilon \ll 1$ it has been shown \cite{derrida1998exact,derrida2007Review,bodineau2004current,bodineau2005current,mallick2015exclusion,bertini2023concurrent,kipnis1989hydrodynamics}
that the (coarse grained) density field $\rho(x,t)$ of the
WASEP is described by fluctuating hydrodynamics, i.e., by a stochastic equation in terms of the current $j(x,t)$
\begin{subequations}
\label{eq:StochCurrent}
\begin{align}
&j = - D_0 \partial_x \rho + \nu \sigma(\rho) + \sqrt{\varepsilon \sigma(\rho)}  \eta \\
&\partial_t \rho + \partial_x j = 0 , \\
&\sigma(\rho) = 2 D_0 \rho(1-\rho)
\:
\end{align}

\end{subequations}
with $D_0=(L+R)/2=1$, and where $\eta(x,t)$ is a Gaussian space-time white noise. Since there is
a factor $\varepsilon$ in front of the noise, the system is in a weak noise regime.
Setting $\nu=0$ recovers the MFT of the SSEP, and (see below), letting $\nu \to +\infty$ leads to
the weak noise regime of the KPZ equation.

It is useful to define a height function $h(x,t)=\int_0^{+\infty} dy  \rho(y,0) -\int_x^{+\infty} dy  \rho(y,t)
= h(0,t) + \int_0^{x} dy  \rho(y,t) $ and the integrated current $J(x,t)=-h(x,t)$. Here $h(0,t)=-Q_t$, where
$Q_t= \int_0^{+\infty} dx (\rho(x,t)-\rho(x,0))$ is the number of particles
which have crossed the origin from left to right minus right to left during time $t$,
and is the integrated current $J(0,T)=Q_t= \int_0^t dt' j(0,t')$
\cite{footnotecurrentQ}.

We consider the ASEP with a two-sided Bernoulli initial condition with
mean density $\rho_1$ for $x<0$ and $\rho_2$ for $x>0$,
with $\rho_2 < \rho_1$. In the WASEP limit it corresponds to choosing
the initial density field $\rho(x,0)$ with probability measure
\be \label{initialP}
{\cal P}[\rho(\cdot,0)] \sim e^{- \frac{1}{\varepsilon} \int_\R \rmd x  \int_{\bar{\varrho}(x)}^{\varrho(x,0)}\rmd z \frac{\varrho(x,0)-r}{r(1-r)}}
\ee
with $\bar \rho(x)=\rho_1 \Theta(-x) + \rho_2 \Theta(x)$. The case $\rho_1=\rho_2=\rho$
corresponds to a stationary initial condition and the case $\rho_1=1$, $\rho_2=0$ is
the step initial condition.

{\bf Main result}.
Our first main result is an exact expression for an appropriately defined generating function (GF)
for the distribution of the observable
$z(X,T) = e^{2\nu h(X,T)}= e^{-2\nu J(X,T)}$ at the observation time $T$ and position $X$.
This GF involves an additional auxiliary real random variable $\omega \geq 1$ of probability density function (PDF)
$P(\omega)= e^{- \frac{1}{2\nu \varepsilon} F(\omega) } $,
where the function
\begin{equation} \label{defF}
F(\omega)=\mathrm{Li}_2\left(1/\omega\right)-\log \omega \log \alpha +\mathrm{Li}_2(\alpha)-\mathrm{Li}_2(1)
\end{equation}
has a unique minimum at the typical value $\omega_{\rm typ}=1/(1-\alpha)$
with $F(\omega_{\rm typ})=0$, $0 \leq \alpha=\frac{\rho_2(1-\rho_1)}{\rho_1(1-\rho_2)} \leq 1$,
and ${\rm Li}_2$ is the dilogarithm function.
Our result is then expressed as
\be   \label{observableB}
\begin{split}
& \int_1^{+\infty} \!\!\!\!\!\! d\omega \,  \big\langle e^{\frac{1}{2\nu \varepsilon} [{\rm Li}_2(- u \omega z(X,T)  ) - F(\omega)]} \big\rangle \underset{\varepsilon \to 0}{\sim} e^{- \frac{1}{ \varepsilon} \Psi(u) }
\end{split}
\ee
where the l.h.s is a double expectation value, one over
the noise $\eta$ in \eqref{eq:StochCurrent} and the initial
density \eqref{initialP}, denoted here and below
as $\langle \cdot \rangle$, and the second over $\omega$.
The rate function on the r.h.s. is given by
\bea
\label{eq:rate-function-wasep-general}
&&   \Psi(u)  = - \frac{1}{2\nu} \int_{\I \mathbb{R} + \delta}
\frac{\rmd y}{2\I \pi y(1-y)} \\
&& \times \mathrm{Li}_2\left(-u\frac{\rho _1 \left(1-\rho _2\right) (1-y) y}{\left(\rho_1-y\right)
\left(y-\rho _2\right)} e^{- 4\nu^2 y(1-y)  T + 2\nu y X } \right) \nn
\eea
where $\rho_2 <\delta < \rho_1$. Here $u$ a generalized Laplace parameter, the above formula hold for $u>0$
but can be continued for $u<0$ as we discuss below. Note that $\Psi(u)$ implicitly depends on $X,T$ and that
each term in the expansion
$\Psi(u)= \sum_{n \geq 1} \frac{u^n}{n!}  \Psi_n(X,T)$ satisfies a heat equation
\be
\partial_T \Psi_n = \frac{1}{n} \partial_X^2 \Psi_n - 2 \nu \partial_X \Psi_n
\ee
In the stationary limit $\varrho_1\to \varrho_2$, $\Psi(u)$ admits an expansion in powers of $\sqrt{u}$, see \cite{SM}.
Finally, the result for the step initial condition is obtained setting $\rho_1=1$ and $\rho_2=0$, i.e.,
$\alpha=0$, where $\omega=1$ becomes deterministic.

The above generating
function \eqref{observableB} has an unusual form, as compared to the
SSEP or to the weak noise KPZ equation.
This happens
because the Fredholm determinant formulae in the ASEP, from which we
obtain our main result, see Appendix~\ref{app:derivation},
involve $q$-deformed
Laplace transforms. In some models such as the $q$-TASEP
it can neatly be traced to underlying Poissonian statistics \cite{us_qtasep-preparation},
but the interpretation here is less clear \cite{YanQSeries}. Despite this
complication, we can extract the cumulants and the large deviation rate
functions of the
integrated current and of the tracer position.
Below one may set $T=1$ \cite{rescalingnu}, in which case
$\frac{1}{\varepsilon} = \sqrt{{\sf T}} \gg 1$.

{\bf Results for the integrated current}.
The PDF $\mathcal{P}(J)$ of the integrated current $J=-h$ (leaving implicit the dependence in $X$)
takes the large deviation form
\be
\mathcal{P}(J) \underset{\varepsilon \to 0}{\sim}  e^{- \frac{1}{\varepsilon} \Phi(J) }
\ee
From the knowledge of $\Psi(u)$ one can obtain $\Phi(J)$ as well
as the cumulants, which scale as
\be \label{cumscale}
\langle J^n \rangle^c
=  \varepsilon^{n-1}  \kappa_n
\ee
The coefficients $ \kappa_n=\mathcal{O}(1)$
can be extracted from their generating function $\phi(P)$, which
takes the form
\be
\label{defphi}
\big\langle e^{\frac{1}{  \varepsilon} P J } \big\rangle \sim e^{ \frac{1}{ \varepsilon}  \phi(P) }
\quad , \quad   \phi(P) = \sum_{n\geq 0} \frac{  \kappa_n}{n!} P^n
\ee
From
the above one shows \cite{SM} that $\phi(P)$ and $\Phi(J)$
are obtained parametrically from $\Psi(u)$ (by eliminating $u$) through the system
\cite{economic,footnote2}
\be
P = \Phi'(J)  = 2\nu u \Psi'(u)  = \log(1+u\omega_{u z} e^{-2\nu J})
\label{eqslegendre1}
\ee
\bea
&& \!\!\!\!\!  \phi'(P) = J \label{eqslegendre2}  \\
&&
\!\!\!\!\! = \frac{-1}{2\nu}  \log \left( \frac{(1- e^{-  2\nu u \Psi'(u)}) (e^{  2\nu u \Psi'(u)}-\alpha)}{u} \right)  \label{eqslegendre3}
\eea
where
$2u \omega_{u}=  \alpha-1+u  +\sqrt{\left(\alpha +u -1\right)^2+4 u }$ and  $z=e^{-2 \nu J}$.

The $n$-th cumulant is thus given by
 $\kappa_n
 =  \left(\frac{1}{2\nu \p_u( u \Psi'(u)) }\p_u\right)^{n-1} J|_{u=0}$ with $J$ given
 by \eqref{eqslegendre3} and $\Psi(u)$ explicit from
 \eqref{eq:rate-function-wasep-general}. Defining the ratios $r_m = \Psi^{(m)}(0)/\Psi'(0)^m$, the first
 two
 cumulants of the integrated
 current \eqref{cumscale} read
\bea
&&    \kappa_1 = -  \frac{1}{2 \nu}   \log \left(2\nu (1-\alpha ) \Psi '(0)\right) \\
&&
\kappa_2 = \frac{1+\alpha}{4 \nu (\alpha-1)} - \frac{r_2}{4 \nu^2}
\eea
This gives the cumulants of $J=J(X,T)$. One finds
\bea
\kappa_1 &=& - \frac{\log( \frac{1}{2} e^{-2 \nu j_{\rho_1} } {\rm Erfc}(\frac{z_{\rho_1}}{ \sqrt{T}})
+ \frac{1}{2} e^{-2 \nu j_{\rho_2} } {\rm Erfc}(-\frac{z_{\rho_2}}{ \sqrt{T}}))}{2 \nu} \nn \\
& \underset{\rho_i=\rho}{=} &
j_\rho:= 2 \nu  \rho(1-\rho) T - \rho X
\eea
with $z_\rho=  \nu (2 \rho-1) T + X/2$,
and for $\rho_2=\rho_1=\rho$
\be  \kappa_2 =
2 \rho (1-\rho ) \sqrt{T}  \, {\cal G}(\frac{z_\rho}{\sqrt{T}})  \label{cum2text}
\ee
where ${\cal G}(y)= y {\rm Erf}(y) + \frac{1}{\sqrt{\pi}} e^{-y^2}$
\cite{footnotesound}.
We give a formula for the higher cumulants in
Appendix~\ref{app:statcum} and further explicit results
in \cite{SM}, which
we checked reproduce in special cases those which appeared in the literature,
namely in the stationary case $\rho_1=\rho_2=\rho$ our $ \kappa_n$ reproduce
the $\hat \kappa_{n}$ in \cite[Eqs.~(10--11)]{berlioz2024tracer} up to $n=3$, and up to $n=4$ in
the special case $\nu=0$, in \cite[Eq.~(S61)]{grabsch2024tracer}.
Surprisingly, the higher cumulants
exhibit non-monotonic behavior as function of the driving $\nu$.

{\bf Results for the tracer}. Let us denote $M_{\sf t}$ the position of a tagged particle in the ASEP, i.e.,
the tracer. Its position in the WASEP is $Y_t = \varepsilon M_{{\sf t}=t/\varepsilon^2}$.
It
is defined by the conservation of particle number to the right of the tracer
\be
\int_{Y_t}^{\infty} dx \rho(x,t) = \int_{Y_0=0}^{+\infty} \rmd x \rho(x,0) \quad \Leftrightarrow h(Y_t,t)= 0
\ee
equivalent to $J(Y_t,t)=0$ (we
choose initial condition $Y_0=0$).
The constraint of following the tracer can thus be implemented
by setting $J=0$ in the above equations. Let us set $T=1$ in this tracer section,
and study the fluctuations of the scaled position $Y_1=M_{\sf T}/\sqrt{{\sf T}}$ for ${\sf T} \gg 1$.
Making explicit the dependence in $X$
in the rate functions, we can obtain the full statistics of the tracer's position
from the $X$ dependence of the rate function $\Phi(J) \equiv \Phi(J,X)$.
We find that its PDF takes the large deviation form
\be
\mathcal{P}( Y_1 = X) = \mathcal{P}(J(X,1)=0 ) \sim e^{ - \frac{1}{\varepsilon}  \Phi(0;X) }
\ee
Next we introduce the CGF
$C(\lambda)$ defined by
\be
\langle e^{  \lambda M_{\sf T}  }  \rangle =
\langle e^{ \frac{1}{\varepsilon} \lambda Y_1 } \rangle \sim e^{  \frac{1}{\varepsilon} C(\lambda)} \\
\ee
It can be obtained from the relation $C(\lambda) = \max_{Y \in \mathbb{R}} [\lambda Y  - \Phi(0;Y)   ]$,
and its Taylor expansion coefficients $c_n$ then give the cumulants of the tracer position
\be \label{tracercumdef}
C(\lambda) = \sum_{n \geq 1} \frac{c_n}{n!}  \lambda^n
\quad , \quad \langle M_{\sf T}^n \rangle^c \simeq c_n \sqrt{{\sf T}}
\ee
i.e., $\langle Y_1^n \rangle^c \simeq \varepsilon^{n-1} c_n$. There is
a simple combinatorics to compute the $c_n$ from the cumulants $\kappa_n=\kappa_n(X)$
(making explicit their $X$ dependence)
obtained above (setting $T=1$). The typical tracer position $c_1$ is the root of
$\kappa_1(c_1) = 0$, which simplifies as $c_1=2 \nu(1-\rho)$ for $\rho_1=\rho_2$.
The variance is given by
\be
c_2 = \frac{\kappa_2\left(c_1\right)}{\kappa
_1'\left(c_1\right){}^2} \underset{\rho_i=\rho}{=} \frac{2 (1-\rho)}{\rho}
{\cal G}(\nu \rho)
\ee
in agreement with the recent perturbative calculation \cite[Eq.~(S135)]{berlioz2024tracer}.
Higher cumulants are displayed in \cite{SM}, and exhibit
a Poissonian ballistic limit for large $\nu$ \cite{Ferrari_Fontes_1996}.

{\bf Limit to the SSEP}. As $\nu \to 0$ our results match the known results for the SSEP \cite{derrida2009currentExact,derrida2009currentMFT,imamura2017large,Imamura2021,mallick2024exact,mallick2024exact}.
Since $z=e^{-2\nu J}$, this limit is quite delicate.
One first shows that $\Psi(u)$ has the following expansion
\be \label{smallnu-2}
\Psi(u) = \frac{1}{2 \nu}  \Psi_{-1}(u) + \Psi_0(u) +\mathcal{O}(\nu)
\ee
with $u \Psi'_{-1}(u) = \log(1+u \omega_{u})$.
Next, evaluating the factor in the l.h.s of \eqref{observableB}
at the saddle point $\omega = \omega_{uz}$ and expanding $z(X,T)=z=1 - 2 \nu J$
the factor $e^{-\frac{1}{2 \nu \varepsilon}  \Psi_{-1}(u)}$ cancels
on both sides, and one obtains
\be
\langle  e^{\frac{\log(1+u \omega_{u})  J}{\varepsilon}} \rangle  \sim e^{-\frac{\Psi_0(u)}{\varepsilon}}
\ee
Hence comparing with \eqref{defphi} we see that in the limit we can identify $P=\log(1+u \omega_{u})$ and
obtain $\phi(P) = - \Psi_0(u)|_{u= \left(1-e^{-P}\right) \left(e^P-\alpha \right)}$.
In the case $X=0$ and $T=1$ we obtain \cite{SM} $\Psi_0(u)= \frac{1}{\sqrt{\pi}} \sum_{n \geq 1} \frac{(- \Omega)^n}{n^{3/2} }$
where $\Omega = u \rho_1 (1-\rho_2)$. One can
then check that the CGF $\phi(P)$ is identical to $F(\omega=\Omega)$ in
\cite[Eqs.~(1-3)]{derrida2009currentExact}, with $P=\lambda$
and $\rho_{a,b}=\rho_{1,2}$.

More generally, the function $\Psi_0(u)$ is computed for any $X$ in \cite{SM},
and it is checked there that the result for $\phi(P)$
(setting $T=1$) coincides with the one displayed in \cite[Eq.~(6.35)]{mallick2024exact}.
This implies that we also recover correctly all the tracer
quantities in the SSEP limit.

{\bf Limit to the KPZ equation}. In the limit of large driving $\nu \to +\infty$, upon proper rescaling one
obtains the KPZ equation. Indeed, let us expand the density field around the maximum of
$\nu \sigma(\varrho)$ by writing $\rho = \frac{1}{2} + \frac{\tilde \rho}{2\nu}$.
Inserting in \eqref{eq:StochCurrent} and
expanding at large $\nu$, we find that the dynamics of $\tilde \rho$ is governed by the
stochastic Burgers equation
\be
\partial_t \tilde \rho =   \partial^2_x \tilde \rho  + \partial_x \tilde \rho^2 + 2 \nu \partial_x \sqrt{\frac{\varepsilon}{2}   }  \eta
\ee
In terms of the height field defined above $\rho=\partial_x h$, upon rescaling
$h = -J = \frac{H}{2 \nu}$, $z=e^H$,
it leads to the KPZ equation for $H$
\be \label{KPZeq}
\partial_t H =   \partial^2_x H  + (\partial_x H)^2 +  \sqrt{2 \varepsilon_{\rm KPZ}} \eta
\ee
with $\varepsilon_{\rm KPZ}= \nu^2 \varepsilon$. Hence
here $\varepsilon_{\rm KPZ} \ll 1$, which
corresponds to the weak noise regime of the KPZ equation,
equivalently to
its short-time regime \cite{rescaling}. We thus expect that the cumulants and rate functions
will converge to those obtained for the
WNT of the KPZ equation. The way it works however
is not so trivial.
First one should rescale the initial condition (IC), i.e.,
one writes $\rho_{1,2}= \frac{1}{2} \pm \frac{\tilde{w}}{2\nu}$,
and the convergence then should be to the KPZ equation with (i) the two-sided
Brownian IC for generic $\tilde w$ (ii) the stationary IC for $\tilde w=0$
(iii) the droplet IC for $\tilde w \to +\infty$. Let us rescale
the Laplace parameter $u$ and the partition function $z$ as
\begin{equation}
u=\frac{4 \tilde{u} e^{\nu ^2 T -\nu  X}}{\nu ^2}, \; z=Ze^{-\nu ^2 T+\nu  X}
\end{equation}
and set $T=1$. Defining then the KPZ time as $T_{\rm KPZ}= \nu^4 \varepsilon^2  \ll 1$ one shows
\cite{SM} that in the limit $\nu\to \infty$ Eq.~\eqref{observableB} becomes
\begin{equation}
\int_0^{+\infty} d\tilde \omega \langle e^{ -\frac{ \tilde{u}\tilde{w}Z +\tilde{F}(\tilde{\omega})}{\sqrt{T_{\rm KPZ}}} } \rangle   \sim e^{- \frac{\tilde{\Psi}(\tilde{u})}{ \sqrt{T_{\rm KPZ}}}  }
\end{equation}
where the auxiliary function $\tilde{F}(\tilde{\omega})$
matches exactly the one in \cite[Eq.~(71)]{krajenbrink2017exact}
(with $\tilde{\omega} = e^{\chi'}$), and the rate function $\tilde{\Psi}(\tilde{u})$
reproduces the one for the double-sided Brownian initial condition.
We performed additional checks of convergence, up to the fourth order cumulant, see \cite{SM}.

{\bf Tails}. It is known \cite{derrida2009currentExact,meerson2014extreme} for
the SSEP that the distribution of the integrated current ${\cal P}(J) \sim e^{- \frac{1}{\varepsilon} \Phi(J)}$
exhibits a cubic tail at large $|J|$, i.e.,  $\Phi(J) \simeq \frac{\pi^2}{12} |J|^3$ (setting $T=1$ here and below).
We show from our result that the WASEP
exhibits the same cubic tail $\Phi(J) \simeq \frac{\pi^2}{12} J^3$ for $J \to +\infty$.
The effect of the driving is thus negligible in the
tail unless it is taken large $\nu \sim J$. In that regime
one finds a nontrivial crossover to the KPZ lower tail.
Defining $\tilde J= \frac{J}{2 \nu}-\frac{1}{4}$
we obtain the crossover scaling form
$\Phi(J) \simeq (2 \nu)^3 \Phi_+(\tilde J)$, which interpolates between the WASEP
$\Phi_+(\tilde{J}) \simeq \frac{\pi^2}{12}\tilde{J}^3$ for $\tilde J \gg 1$,
and the known KPZ tail $5/2$ exponent \cite{kolokolov2007optimal}
and prefactor \cite{le2016exact,krajenbrink2017exact,JanasDynamical,meerson2016large},
$\Phi_+(\tilde{J}) \simeq \frac{16}{15\pi}\tilde{J}^{5/2}$ for $\tilde J \ll 1$.
It admits the parametric representation \cite{footnotesimple}
\bea
&& \Phi'_+(\tilde J)  =
\frac{\left(4 k^2+1\right)
\arctan(2k)-2 k}{2 \pi } \\
&& \tilde J = \frac{k}{\pi } + k^2 - (k^2 + \frac{1}{4}) \frac{2}{\pi} \arctan(2 k)
\eea
where $k \in [0,+\infty[$ should be eliminated.

As for the KPZ equation and other models \cite{le2016exact,krajenbrink2017exact,UsWNTDroplet2021,UsWNTFlat2021,NaftaliKMP2,UsWNTCrossover,krajenbrink2023weak}, to obtain the other tail requires continuing
 \eqref{eq:rate-function-wasep-general}, which gives only
 the {\it main branch} of $\Psi(u)=\Psi_0(u)$, valid
 for $u>u_c=- (\sqrt{\alpha}-1)^2 e^{\nu ^2 T}$ \cite{footnotesimple}
 and which describes $J>J_c$ (where $J_c< \langle J \rangle$).
For $J<J_c$, all formula still apply, but with $\Psi(u)=\Psi_0(u)+ \Delta(u)$,
$u>u_c$,
where $\Delta(u)$ is the contribution of "solitons".
As for stationary KPZ there can be multiple solitonic branches,
and even a phase transition (breaking the symmetry $x \to - x$) \cite{JanasDynamical,meerson-landau,UsWNTFlat2021}
as also found in open driven diffusive systems \cite{Lecomte_phasetransition}.
The details are involved and analyzed elsewhere \cite{usinprep}. We
restrict here to the simpler case of the step initial condition,
$\rho_1=1$, $\rho_2=0$, where there is a "wall" at $J=0$,
since $J=J(0,1)$ is
the number of particles to the right of zero and is positive. In the regime $\nu \gg 1$
we find that $\Phi(J)$ takes the following scaling form
for $0<J<J_c$,
\be \label{Phiupper}
\Phi(J) \simeq 4 J_c (2 \kappa + (4 \kappa ^2-1) \mathrm{arctanh}(2 \kappa )), ~
\ee
where $2 \kappa= \sqrt{1 - \frac{J}{J_c}}$ and $J_c \simeq \langle J \rangle \simeq \frac{\nu}{2}$,
with $\Phi(J) \simeq 4 J_c + 2 J \log(J/4 J_c) + \mathcal{O}(J)$ and
$\Phi(J) \simeq \frac{8}{3} J_c
   (1-\frac{J}{J_c})^{3/2}$ for $J/J_c$ near $1$.
The $3/2$ exponent coincides with the one of the upper-tail of
the KPZ equation. Indeed and remarkably, we find
that the function in the r.h.s. of \eqref{Phiupper} coincides
with the recent result in \cite[Eq.~(1.4)]{Das_2022}
for the upper tail of the ASEP in the a priori very
different regime of fixed asymmetry $R-L=\mathcal{O}(1)$.
This coincidence indicates that no intermediate regime exists
in the upper tail when the asymmetry is scaled to small values.

{\bf Integrability of the MFT of the WASEP}.
The stochastic hydrodynamics equations \eqref{eq:StochCurrent} of the WASEP can also be described using the MSR dynamical path integral formalism \cite{SM}. In the limit of weak asymmetry $\varepsilon \to 0$, the MSR action is concentrated at its saddle point yielding the non-linear system which underpin the macroscopic fluctuation theory of the WASEP
\begin{subequations}
\label{eq:MFTbulkWASEP}
\begin{align}
\partial_t q
&= \partial_x \left[
D_0 \partial_x q - \sigma(q) \partial_x p - \nu \sigma(q)
\right]
\:,
\\
- \partial_t p
&=  D_0 \partial_x^2 p + \frac{1}{2} \sigma'(q) (\partial_x p)^2
+ \nu \sigma'(q) \partial_x p
\:,
\end{align}
\end{subequations}
where $p$ is the response field associated to $q$ which is the optimal density. Here $D_0=1$ and $\sigma(q)=2q(1-q)$. In the case of the Bernoulli initial condition
\eqref{initialP}, these equations are completed with initial and terminal conditions for $p(x,0)$ and $p(x,T)$  while for the step initial condition, the initial condition is replaced by one for $q(x,0)$, see \cite{SM}. Solving \eqref{eq:MFTbulkWASEP} perturbatively to order $n$ in the amplitude of the response field,
one can obtain the cumulant $\kappa_{n+1}$ of $J$, in a calculation which becomes extremely
tedious as $n$ increases \cite{berlioz2024tracer}.
It would thus be of great interest if an integrable structure exist, allowing to go beyond perturbative methods.

We now unveil the integrable structure of the WASEP.
We start by transforming the MFT equations \eqref{eq:MFTbulkWASEP} using the following change of functions
\begin{equation} \label{qpRZ}
\begin{split}
q(x,t)&= \frac{1}{2} + \frac{1}{2\nu} \p_x \log Z(x,t),\\
\p_x p(x,t)&=-\frac{2\nu R(x,t) Z(x,t)}{1+R(x,t)Z(x,t)}
\end{split}
\end{equation}
where the first transformation has the Cole-Hopf form, and the second
defines a response field. The new functions $Z$ and $R$ obey the following system
\begin{equation}
\label{eq:main-stereo-ll-system}
\begin{split}
\p_t Z &= \p_x^2 Z - \frac{  2   R  }{1+  R Z}\left((\p_x Z)^2- (\nu  Z)^2\right)\\
-\p_t R &=  \partial_x^2 R-\frac{   2   Z }{1+  RZ}\left((\p_x R)^2- (\nu  R)^2\right)
\end{split}
\end{equation}
It turns out that this system describes
the dynamics
of a complex extension of the classical anisotropic Landau-Lifshitz spin chain
of Hamiltonian  \cite{SM}
\begin{equation}
\mathcal{H}=\frac{1}{2}\int_\R \rmd x\, (\p_x S_+ \p_x S_-+(\p_x S_z)^2-\nu^2(S_z-1)^2)
\end{equation}
through the following stereographic representation \cite{LandauLifschitzGilbert,LakshmananFascinating}
of the spin $\mathcal{S}$
in terms of the fields
$R$ and $Z$
\begin{equation}
\begin{split}
\mathcal{S}&=\begin{pmatrix}
S_z & S_+\\
S_- & - S_z \\
\end{pmatrix}=\frac{1}{{1+RZ}}\begin{pmatrix}
1-RZ & 2 R \\
2 Z & -(1-RZ) \\
\end{pmatrix}
\end{split}
\end{equation}
The properties of the spin are as follows $\mathcal{S}^2=I_d$, $\det \mathcal{S}=-1$ and $S_- S_+ + S_z^2 =1$
and note that $S_\pm$ are real.
Such mapping have been considered before but only in the case $\nu=0$ (SSEP)
\cite{Tailleur_2008,derrida2009currentMFT} (see \cite{Essler_1996,ALCARAZ1994250}
for related references).
The anisotropic LL model is integrable \cite{Nakamura_1982}.
As a consequence, the system \eqref{eq:main-stereo-ll-system} is also integrable
in the Lax sense, allowing for its exact solvability. The Lax representation of the problem is given by the linear system $\p_x \vec{v} = L\vec{v}$, $\p_t \vec{v}=M\vec{v}$, with $\vec{v}$ a two-component vector, which compatibility gives the zero curvature condition $\p_t L -\p_x M+[L,M]=0$. In the present case, the Lax matrices read
\begin{equation}
\label{eq:main-text-lax-matrix-L}
\begin{split}
&L= -\frac{\I k}{2}\mathcal{S}+\mu [\sigma_3,\mathcal{S}]\\
&= \frac{1}{1+RZ}
\begin{pmatrix}
-\frac{\I k}{2}  \left(1-RZ \right) & (\nu -\I k) R \\
-(\nu+\I k ) Z & \frac{\I k}{2}  \left(1-RZ \right) \\
\end{pmatrix}
\end{split}
\end{equation}
and
\begin{equation}
\begin{split}
M =& \frac{k^2}{2}\mathcal{S}+\frac{\I k}{2}\mathcal{S} \p_x \mathcal{S}+\I \mu k [\sigma_3,\mathcal{S}]\\
&-\mu [\sigma_3,\mathcal{S} \p_x \mathcal{S}]+4\mu^2 \{\sigma_3,\mathcal{S} \}\sigma_3-\frac{\nu^2}{2} \sigma_3
\end{split}
\end{equation}
with $\mu=\pm \frac{\nu}{4}$, $\sigma_3=\rm{Diag}(1,-1)$ and $\{\cdot, \cdot \}$ (resp. $[\cdot,\cdot]$) is the standard anti-commutator (resp. commutator). The zero curvature is verified if the pair $(R,Z)$ verifies \eqref{eq:main-stereo-ll-system}. We have chosen $\mu = \frac{\nu}{4}$ in the second line of \eqref{eq:main-text-lax-matrix-L}.

{\bf Discussion and outlook}. We have obtained the exact large deviation rate functions
and cumulants of the time-integrated current and of the
tracer position for the WASEP.
In addition we constructed the explicit Lax pair for the WASEP,
by unveiling a mapping to the anisotropic Landau-Lifshitz chain.
This opens the way for using inverse scattering
in future work, and investigate a broader
family of initial conditions and observables.
As a byproduct we also obtain, see Appendix \ref{app:quadratic},
rate functions, cumulants and integrability
for several other MFT's. This
includes the weakly asymmetric inclusion process (WASIP),
defined in \cite{GiardinaSIP,Opoku_2015,reuveni2011asymmetric,MeersonInclusion},
the weakly asymmetric KMP model,
as well as the three dual models, including
the weakly asymmetric zero range process
with geometric stationary measure (upon exchanging by duality
integrated current and tracer position
\cite{SM,Rizkallah_2023,berlioz2024tracer}).
It would finally be interesting to consider quantum
extensions of the WASEP and of the
WASIP along the lines of \cite{krajQWASEP,bernard2025large}.
\begin{acknowledgments}
\paragraph{Acknowledgments.}  We thank Aurelien Grabsch for very useful discussions on the WASEP.
We also thank G.~Barraquand, A.~Borodin, B.~Derrida  and V.~Pasquier for discussions on related topics.
PLD acknowledges support from the ANR grant ANR-23-CE30-0020-01 EDIPS,
and hospitality from LPTMS-Orsay. We thank the Harvard University Center of Mathematical Sciences
and Applications for hospitality and partial support.
\end{acknowledgments}
\clearpage
\appendix
\setcounter{equation}{0}
\setcounter{figure}{0}
\renewcommand{\thetable}{S\arabic{table}}
\renewcommand{\theequation}{S\thesection.\arabic{equation}}
\renewcommand{\thefigure}{S\thesection.\arabic{figure}}
\renewcommand{\thetable}{S\thesection.\arabic{table}}
\setcounter{secnumdepth}{2}
\begin{center}
{\Large End matter \\
\vspace{0.22cm}
}
\end{center}
\section{Derivation of the rate function $\Psi(u)$}
\label{app:derivation}
To obtain the rate function $\Psi(u)$ of the WASEP we start from an exact
formula obtained for the ASEP in \cite[Thm.~4.8]{Aggarwal6v},
see also \cite{borodin2014duality,AggarwalBorodin6v}. Upon some
changes of notations, let us define $\tau=L/R<1$,
$\theta_i= \rho_i/(1-\rho_i)$, and $\alpha=\theta_2/\theta_1$. It is assumed that $\rho_2<\rho_1$.
The above theorems characterize the distribution of the
discrete current ${\sf J}_{\sf t}(m)$
of the ASEP, defined by \cite[Eq.~(1.1)]{Aggarwal6v},
in terms of a Fredholm determinant
\be \label{FD1}
\E \left[ \big\langle \frac{1}{(- u \tau^{{\sf J}_{\sf t}(m)-\sf{j}} , \tau)_\infty} \big\rangle \right] = \Det(I+K_u)_{L^2(C_0)}
\ee

    where  $\sf{j}$ is an auxiliary random variable following a $\tau$-deformed geometric distribution with PDF
\begin{equation}
\label{eq:asep-two-bernoulli-q-geometric-proba}
\begin{split}
\mathbf{p}_{\alpha}(\sf{j})&=\alpha^{\sf{j} }\frac{(\alpha ;\tau)_{\infty}}{(\tau;\tau)_{\sf{j}}}\mathds{1}\{0 \leq {\sf{j}} \}
\end{split}
\end{equation}
with $\alpha=\theta_2/\theta_1$ and we recall the notation $(a,q)_{\infty} =\prod_{\ell=0}^{\infty}(1-q^\ell a)$ and $ (a;q)_{n} = \frac{(a;q)_{\infty}}{(a q^n;q)_{\infty}}$. The kernel in \eqref{FD1} has several equivalent expressions, and we use the one in
 \cite[Thm.~4.11]{AggarwalBorodin6v}
\begin{equation} \label{KernelASEP}
K_u(v,v') = \int_{D} \frac{\rmd s}{2 \I \pi} \frac{\pi}{\sin(\pi s)}   \frac{g(v)}{g(\tau^s v)} \frac{u^s}{\tau^s v -v'}
\end{equation}
with the function $g(v)$ being defined as
\be
\label{eq:g-func-ASEP-two-bernoulli}
g(v) = e^{(R-L) {\sf t} \frac{\tau}{\tau + v}} \left(\frac{\tau}{\tau + v}\right)^{-m}  \frac{((\tau \theta_2)/v;\tau)_{\infty}}{(v/(\tau \theta_1);\tau)_{\infty}}
\ee
we refer to \cite{Aggarwal6v,AggarwalBorodin6v} for the precise definition of the contours $D$ and $C_0$.
In order to study the WASEP limit, we choose, with $\nu>0$
\begin{equation} \label{wasep:rescaling}
\begin{split}
&R=1+\varepsilon \nu,  L=1-\varepsilon \nu, \tau \simeq e^{-2\nu \varepsilon}\\
&{\sf{T}}=\frac{T} { \varepsilon^2}, m=\frac{X}{ \varepsilon}, z(X,T)=\tau^{{\sf J}_{\sf t}(m)}
\end{split}
\end{equation}

Using the asymptotics of $q$-Pochhammer functions (see e.g., Section~\ref{sec:supp-mat-asymptotic-pochhammer})
\be
\label{app:eq:asymptotic-q-deformedEM}
\log (x , q)_{\infty}  \underset{q \to 1}{\to}   -\frac{1}{1-q} {\rm Li}_2(x)+\frac{1}{2}\log(1-x)+\mathcal{O}(1-q)
\ee
we obtain that in the WASEP limit $\varepsilon \ll 1$  the auxiliary random variable $\omega=\tau^{-{\sf j}}$ has a pdf proportional to $\mathbf{p}_{\alpha}({\sf{j}}) \sim e^{-\frac{1}{2\nu\varepsilon} F(\omega) } $
so that the double expectation value in the l.h.s of \eqref{FD1} becomes
\begin{equation}
\int_1^{+\infty} \!\! d\omega \, \big\langle  e^{\frac{1}{2\nu \varepsilon} [{\rm Li}_2(- u \omega z(X,T)  ) -F(\omega)]}\big\rangle
\end{equation}

Next we extend to the present case the first cumulant method developed in \cite{KrajLedou2018,ProlhacKrajenbrink,krajenbrink2019beyond}
to obtain the asymptotics of such Fredholm determinants. This requires the following asymptotics
\be
\label{eq:limit_gv-asep-phi}
\begin{split}
& \lim_{\varepsilon\to 0}2\nu\varepsilon\log g(v) = \varphi(v) \\
& = \frac{4\nu^2 T}{(1+v)}+{2\nu X} \log(1+v)+\mathrm{Li}_2\left(\frac{v}{\theta_1}\right)-\mathrm{Li}_2\left(\frac{\theta_2}{v}\right)      \end{split}
\ee
obtained from \eqref{eq:g-func-ASEP-two-bernoulli} and \eqref{app:eq:asymptotic-q-deformedEM}. The details
are given in \cite{SM} and the result is
\begin{equation} \label{sa10}
\Det(I+K_u) \underset{\varepsilon \to 0}{\sim} \exp\left(-\frac{\Psi(u)}{  \varepsilon}\right)
\end{equation}
with
\be \label{sa11}
\Psi(u)=- \frac{1}{2\nu}\int_{C} \frac{\rmd v}{2\I \pi v}\mathrm{Li}_2(-u e^{ v\varphi'(v)})
\ee
where the contour $C$ encloses $0$, remains to the right of $-1$
and crosses the positive real axis between $\theta_2$ and $\theta_1$.
Upon the change of variable $y=v/(1+v)$ and the choice
of contour $y= \delta+ \I \mathbb{R}$, with $\rho_2<\delta<\rho_1$ one obtains
our result \eqref{eq:rate-function-wasep-general}. Note that for the step initial condition $\rho_1=1$, $\rho_2=0$,
we can alternatively use \cite[Thm.~5.3]{borodin2014duality}.

The above derivation is restricted to $\nu>0$. Indeed
since here $\rho_1>\rho_2$, if one considers instead the ASEP
with $L>R$, a shock appears. Here, in the WASEP limit however this
shock is rounded and one can check
that the cumulants are indeed well behaved at $\nu=0$, so their
expressions straightforwardly extend to $\nu<0$.
It is only in the large WASEP time limit $T\gg 1$ (or equivalently for large $\nu$)
that the shock develops \cite{SM}.

\section{Stationary cumulants} \label{app:statcum}

In the case $\rho_1=\rho_2=\rho$ the rate function $\Psi(u)=\psi(v)$
where $v=\sqrt{u}$. The cumulants $\kappa_n$ for $n \geq 2$
can be obtained as polynomials in the ratios $R_n = \frac{\psi^{(n)}(0)}{\psi'(0)^n}$
upon expanding
\bea  \label{kappangen2}
&& \kappa_n = -   \frac{B_{n-1}}{\nu( n-1)} \delta_{n \geq 3,{\rm odd}} \\
&&  - \frac{1}{\nu^n} \left[\left(\frac{1}{\p_v( v \psi'(v)) }\p_v\right)^{n-1} \log(\nu\psi'(v)) \right]_{v=0}
\nonumber
\eea
where $B_n$ are Bernoulli numbers,
giving $\kappa_2=-\frac{R_2}{\nu^2}$,
$\kappa_3 =  - \frac{1}{12 \nu} + \frac{1}{\nu^3} (3 R_2^2 - R_3)$
and so on. We found explicit formula for the $R_n$
(setting $T=1$)
\be  \label{psievenmain}
\begin{split}
R_{2q} =& - \frac{(\nu^2 \rho (1- \rho))^q}{2 q \nu}
\partial_p^{2q-1} \bigg(
((\rho + p) (1- \rho - p))^{q-1}  \\
& \times
e^{4 q \nu  p (\nu p - y) }
{\rm Erf}(\sqrt{q} (2 p \nu  - y) ) \bigg)
|_{p=0}
\end{split}
\ee
and
\be
R_{2 q+1}  = \frac{\nu^{2q}}{(2 q+1)  } \p_{\tilde{z}}^{2 q} \left(
\frac{\partial_{\tilde{z}}p}{(\rho+ p)(1-\rho-p)} \right)|_{\tilde z=0}
\ee
with $y = \nu (1-2 \rho)  - X/2$ and where $p=p(\tilde z)$
is obtained by inverting the series $\tilde z=\tilde z(p)$
with
\be
\tilde z
=  \frac{p \, e^{-2 \nu p (\nu p - y) }}{\sqrt{\rho (\rho-1) (\rho+p )  (\rho+p -1)}}
\ee
These formulae allow iterative calculation of the cumulants to an arbitrary order,
for more details see \cite{SM}.

\section{Extension to other quadratic MFT models} \label{app:quadratic}

Extending to the driven case the mapping introduced in \cite[Eq.~(33)]{derrida2009currentMFT}
between the MFT's where $\sigma(\rho)$ is
quadratic in $\rho$ and $D(\rho)=1$, we obtain \cite{SM} the cumulants of the integrated current
for two additional MFT models starting from their (two-sided)
stationary initial conditions (with densities $\rho_1,\rho_2$ on each
side of the origin). For the WASIP, i.e.,
for $\sigma(\rho)=2 \rho (1+\rho)$, we find
\be
\kappa_n^{\rm WASIP}(\rho_1,\rho_2,\nu)= - \kappa_n^{\rm WASEP}(- \rho_1 , -\rho_2 , \nu)
\ee
understood as an analytical continuation in the parameters.
For the weakly asymmetric KMP model, i.e., for $\sigma(\rho)=2 \rho^2$,
we find
\bea
&& \kappa_n^{\rm WAKMP}(\rho_1,\rho_2,\nu) \\
&& = (-1)^{n+1} \lim_{B \to 0} B^n
\kappa_n^{\rm WASEP}(\frac{\rho_1}{B} , \frac{\rho_2}{B} , - B \nu) \nn
\eea
Since the map \eqref{qpRZ} extends to these models, see \cite{SM},
it also shows their integrability. Finally these results extend to
the three dual models, see \cite{SM}.

\section{Second Lax matrix for the MFT of the WASEP}
We give here the explicit form of the second Lax matrix $M$ in terms
of the fields $R$ and $Z$ defined in \eqref{qpRZ}.
\begin{widetext}
\begin{equation}
\begin{split}
&M = \frac{1}{(1+RZ)^2}\\
&\begin{pmatrix}
\frac{k^2}{2}- \nu ^2   ZR-\I k (Z \p_x R-R\p_x Z)-R^2 Z^2 \left(\frac{k^2}{2}+\nu ^2\right) & (k+\I\nu ) \left(R \left(k+R \left(k Z+\I\p_x Z\right)\right)+\I\p_x R\right) \\
(k-\I\nu ) \left(Z \left(k+Z \left(k R-\I\p_x R\right)\right)-\I\p_x Z\right) &-\frac{k^2}{2}+ \nu ^2   ZR+\I k (Z \p_x R-R\p_x Z)+R^2 Z^2 \left(\frac{k^2}{2}+\nu ^2\right) \\
\end{pmatrix}
\end{split}
\end{equation}
\end{widetext}

\clearpage

\bibliography{discrete-wnt-modified-4}

\begin{thebibliography}{118}
\bibitem [{foo({\natexlab{b}})}]{footnotecurrentQ}%
  \BibitemOpen
  \href@noop {} {} ({\natexlab{b}}),\ \bibinfo {note} {note that the number of particles which have crossed $x$ from left to right minus right to left during time $t$ is $J(x,t)-J(0,t)$.}\BibitemShut {Stop}%
\bibitem [{eco()}]{economic}%
  \BibitemOpen
  \href@noop {} {}\bibinfo {note} {To economize notations we denote by the same letters the random variables $z=z(X,T)$, $J=J(X,T)$, e.g. in \eqref{observableB}, and their values at the saddle point, e.g. in \eqref{eqslegendre1}, which enter as arguments of rate functions, and one should not confuse them}\BibitemShut {NoStop}%
\bibitem [{foo({\natexlab{c}})}]{footnote2}%
  \BibitemOpen
  \href@noop {} {} ({\natexlab{c}}),\ \bibinfo {note} {although it appears to involve one additional equation, the system is compatible}\BibitemShut {NoStop}%
\end{thebibliography}

\newpage

\onecolumngrid
\begin{center}
{\Large Supplementary Material for\\  {\it Macroscopic fluctuations in the weakly-asymmetric exclusion process  } }
\end{center}
{\hypersetup{linkcolor=black}
\setcounter{tocdepth}{1}
\tableofcontents
}
\section{First cumulant approximation for the Fredholm determinant of the ASEP}
\label{sec:first-cum-approximation}
{\bf Important remark}. Everywhere in this section the parameter $\tau$ defined in the text is
denoted $q$.
\subsection{Fredholm determinant factorization}
We first rewrite the Fredholm determinant $\Det(I+K_u)_{L^2(C_0)}$ in \eqref{FD1} in a more convenient form
for the asymptotic analysis. The ASEP kernel
\eqref{KernelASEP} is part a of a larger family of kernels for the various $q$-deformed models
which have the following form.
\be
K_u(v,v') = \int_{D}\frac{\rmd s}{2 \I \pi} \frac{\pi}{\sin(\pi s)}  u^s \frac{g(v)}{g(q^s v)} \frac{1}{q^s v-v'}
\ee
These kernels differ mainly with the choice of function $g(v)$ as well as some details about the choice of the integration contour $D$ on $s$, and the contour $C_0$ on which $v$, $v'$ belong. \\

Let us perform some manipulations on the kernel valid for any $q$. One first sets $\zeta = q^s v$ and gets
\be
K_u(v,v') = \frac{1}{\log q} \int_{D'} \frac{\rmd \zeta}{2 \I \pi \zeta } \frac{\pi}{\sin(\pi s)}  u^s \frac{g(v)}{g(\zeta)} \frac{1}{\zeta -v'}
\ee
with the identification $s \log q = \log \zeta - \log v$.
Introducing the identity
\begin{equation}
\frac{\pi}{\sin(\pi s)}u^s = |\log q| \int_\R \rmd r \, \frac{u}{u+q^{-r}}e^{-sr \log q}, \quad 0<\Re(s)<1
\end{equation}

The kernel further reads (since $q<1$)
\be
K_u(v,v') = - \int_\R \rmd r \,  \int_{D'} \frac{\rmd \zeta}{2 \I \pi \zeta} \frac{u}{u+q^{-r}} \frac{v^r}{\zeta^r}
\frac{g(v)}{g(\zeta)} \frac{1}{\zeta -v'}
\ee
One now factorises the kernel into
\be
K_u(v,v') = - \int_\R \rmd r \,  \int_{D'} \frac{\rmd \zeta}{2 \I \pi \zeta}
A_1(v,r) \sigma(r) A_2(r,\zeta) A_3(\zeta,v')
\ee
where we have defined the kernels
\begin{equation}
A_1(v,r)=v^r g(v), \quad , \quad A_2(r,\zeta) = \frac{1}{\zeta^r g(\zeta)} \quad , \quad A_3(\zeta,v') = \frac{1}{\zeta-v'}
\quad , \quad \sigma_u(r)= \frac{1}{1+q^{-r}/u}
\end{equation}
Inserting this factorization into the Fredholm determinant one obtains
\begin{equation}
\begin{split}
\Det(I+K_u)_{L^2(C_0)} &=\Det(I-A_1\sigma_u A_2 A_3)_{L^2(C_0)} \\&=\Det(I-\sigma_u A_2 A_3 A_1)_{L^2(\R)} \\
&= \mathbb{E}\left[\prod_{\ell=1}^{+\infty}(1-\sigma_u(a_\ell))\right]
\end{split}
\end{equation}
where from the first to the second line we have used Sylvester's identity. In the third line we
have formally interpreted the Fredholm determinant of the second line as an expectation value over some determinantal point process $\{ a_\ell \}_{\ell \geq 1}$ which correlations are controlled by the kernel $\mathcal{A}=A_2 A_3 A_1$.
\begin{equation}
\begin{split}
\label{eq:dpp-q-deformed}\Det(I+K_u)_{L^2(C_0)}&=\mathbb{E}\left[\prod_{\ell=1}^{+\infty}(1-\sigma_u(a_\ell))\right] =\mathbb{E}\left[\prod_{\ell=1}^{+\infty}e^{-\varsigma(a_\ell)}\right]
\end{split}
\end{equation}

where we have introduced the function $\varsigma$ defined as
$e^{-\varsigma}=1-\sigma_u= \frac{1}{1+ u q^r}$, which allows to interpret Eq.~\eqref{eq:dpp-q-deformed} as a linear statistics of the point process $\{ a_\ell \}_{\ell \geq 1}$ over the observable $\varsigma(r)=\log(1+ u q^r)=-\textrm{Li}_1(- u q^r) $.

\subsection{First cumulant approximation in the limit $q\to 1$}

We now apply the first cumulant method \cite{KrajLedou2018,ProlhacKrajenbrink,krajenbrink2019beyond} which asserts that as some control parameter (here $\log q$) goes to 0, the linear statistics over the point process self-averages, i.e., to the leading order one has
\begin{equation}
\begin{split}
\Det(I+K_u)_{L^2(C_0)}=\Det(I-\sigma_u A_2 A_3 A_1)_{L^2(\R)}=\mathbb{E}\left[\prod_{\ell=1}^{+\infty}e^{-\varsigma(a_\ell)}\right]
\sim e^{-\Tr(\varsigma A_2 A_3 A_1)}
\end{split}
\end{equation}
This means that to leading order, the quantity of interest only involves the diagonal part of the kernel $\mathcal{A}= A_2 A_3 A_1$, i.e., its density $\varrho(a)$, via $\Tr(\varsigma \mathcal{A})=\int \rmd a  \varrho(a)\varsigma(a)$
One has
\be
\rho(r)= (A_2 A_3 A_1)(r,r) = \int_{C_0} \frac{\rmd v}{2 \I \pi } \int_{D'} \frac{\rmd \zeta}{2 \I \pi \zeta}
\frac{v^r g(v)}{\zeta^r g(\zeta)}  \frac{1}{\zeta-v}
\ee

The first cumulant reads
\begin{equation}
\begin{split}
\Tr(\varsigma A_2 A_3 A_1)& = \int_\R \rmd r \rho(r) \log(1+ u q^r) \\
&=- \int_\R \rmd r \int_{C_0} \frac{\rmd v}{2 \I \pi } \int_{D'} \frac{\rmd \zeta}{2 \I \pi \zeta}  \mathrm{Li}_1(-u q^r)
\frac{v^r g(v)}{\zeta^r g(\zeta)}  \frac{1}{\zeta-v} \\
&= \frac{1}{\log q}\int_\R \rmd r \int_{C_0} \frac{\rmd v}{2 \I \pi } \int_{D'} \frac{\rmd \zeta}{2 \I \pi \zeta}  \mathrm{Li}_2(-u q^r)
\frac{v^r g(v)}{\zeta^r g(\zeta)}  \frac{\log{\zeta}-\log v}{\zeta-v} \\
\end{split}
\end{equation}

We have proceeded to an integration by part on the variable $r$ from the second to the third line, note that there is no boundary term. In the regime where we want to apply the first cumulant approximation, we rescale the variables as
\begin{equation}
\{ q=e^{-\eta}, r=\frac{\tilde{r}}{\eta}\}
\end{equation}
and we will study the limit $\eta \to 0$. Note that to apply this method to the main text we will use $\eta = 2 \nu \varepsilon$.
This will allow us to further evaluate the integrals at their saddle point. Further dropping the tilde, we find that at leading order
\begin{equation}
\begin{split}
\Tr(\varsigma A_2 A_3 A_1)
&= - \frac{1}{ \eta^2}\int_\R \rmd r \int_{C_0} \frac{\rmd v}{2 \I \pi }  \int_{D'} \frac{\rmd \zeta}{2 \I \pi \zeta}  \mathrm{Li}_2(-u e^{-r}) \frac{e^{\frac{r}{\eta}\log v +\log g(v)}}{e^{\frac{r}{\eta}\log \zeta + \log g (\zeta)}} \frac{\log{\zeta}-\log v}{\zeta-v} \\
&= - \frac{1}{ \eta^2}\int_\R \rmd r \int_{C_0} \frac{\rmd v}{2 \I \pi }  \int_{D'} \frac{\rmd \zeta}{2 \I \pi \zeta}  \mathrm{Li}_2(-u e^{-r}) \frac{e^{\frac{1}{\eta} \tilde{\Phi}(v)}}{e^{\frac{1}{\eta} \tilde{\Phi}(\zeta)}} \frac{\log{\zeta}-\log v}{\zeta-v}
\end{split}
\end{equation}

where we have defined $\tilde{\Phi}(v)= r \log v + \eta\log g(v)$.
One obtains
\begin{equation}
\Tr(\varsigma \mathcal{A})=\frac{1}{\eta^2} \int_\R \rmd r \mathrm{Li}_2(-u e^r)I(r)
\end{equation}
where the integrand is
\begin{equation}
\label{eq:first-cumulant-saddle-IR}
I(r)=-\int_{C_0} \frac{\rmd v}{2 \I \pi }  \int_{D'} \frac{\rmd \zeta}{2 \I \pi \zeta}   \frac{e^{\frac{1}{\eta} \tilde{\Phi}(v)}}{e^{\frac{1}{\eta} \tilde{\Phi}(\zeta)}} \frac{\log{\zeta}-\log v}{\zeta-v}
\end{equation}
where the rate function at the lowest order reads
\begin{equation}
\begin{split}
\tilde{\Phi}(v)&  r \log v + \varphi(v)\\
\end{split}
\end{equation}
since the function $g(v)$ is generally $q$-dependent, we only need to study its first order
\begin{equation}
\varphi(v)= \lim_{\eta\to 0}\eta\log g(v)|_{q=e^{-\eta}}
\end{equation}
The function depends on $g(v)$, hence $\varphi(v)$ depends on the model.
For the present model (ASEP with Bernoulli initial conditions) the function $\varphi(v)$ was obtained in  \eqref{eq:limit_gv-asep-phi}. The saddle point of the integral in \eqref{eq:first-cumulant-saddle-IR} is thus the solution of
\begin{equation}
\tilde{\Phi}'(v)=0 \Rightarrow \varphi'(v)=-\frac{r}{v}
\end{equation}
Note that this leads to
\begin{equation}
\begin{cases}
\rmd r = -(\varphi'(v)+v\varphi''(v))\rmd v, \\
\tilde{\Phi}''(v)=\varphi''(v)+ \frac{\varphi'(v)}{v}
\end{cases}
\end{equation}
This implies the following Jacobian between $v$ and $r$
\begin{equation}
\rmd r = - v \tilde{\Phi}''(v)\rmd v
\end{equation}
The saddle point on $v$ and $\zeta$ is the same by contour deformation and the factor $\frac{\log{\zeta}-\log v}{\zeta-v} $ simply becomes the derivative $\frac{1}{v}$. Using the saddle point formula we obtain
\begin{equation}
\begin{split}
\Tr(\varsigma \mathcal{A})
&\simeq  \frac{1}{\eta}\int_\R \frac{\rmd r}{2\I \pi }\frac{1 }{v(r)^2}\mathrm{Li}_2(-u e^{-r})\frac{1}{\tilde{\Phi}''(v(r))}\\
&\simeq -\frac{1}{ \eta}\int_{C} \frac{\rmd v}{2\I \pi }\frac{  1 }{v}\mathrm{Li}_2(-u e^{ v\varphi'(v)})\\
\end{split}
\end{equation}
where $C$ is the limit of $C_0$ when $\eta \to 0$ (i.e., $q \to 1$). Hence, we obtain
\begin{equation}
\Det(I+K_u)_{L^2(C_0)} \sim \exp \left(\frac{1}{ \eta}\int_{C} \frac{\rmd v}{2\I \pi }\frac{  1 }{v}\mathrm{Li}_2(-u e^{ v\varphi'(v)}) \right)
\end{equation}

Setting $\eta=2 \nu \varepsilon$ this leads to \eqref{sa10} and \eqref{sa11} in the text.
In the case of the ASEP the contours are as follows \cite{Aggarwal6v,AggarwalBorodin6v}
$D$ is a positively oriented contour in the complex plane containing $0$ and $\tau \theta_2$, but leaving outside $-\tau$ and $\tau \theta_1$. $C_0$ is a positively oriented contour in the complex plane contained in $\tau^{-1}D$, containing $-\tau$, $\tau \theta_2$, 0 and the contour $D$,  but leaving outside $-1$ and $\tau \theta_1$. In that
case the contour $C$ encloses $0$, remains to the right of $-1$ (it contains $-1$ as an asymptotic point), and crosses the positive real axis between $\theta_2$ and $\theta_1$.

\section{$q$-deformed functions and useful asymptotic formulae} \label{sec:supp-mat-asymptotic-pochhammer}

We define the infinite $q$-Pochhammer symbol, its finite $n$ version and the the $q$-factorial as
\begin{equation}
\label{eq:def-q-infinite-pochhammer}
(a,q)_{\infty} =\prod_{\ell=0}^{\infty}(1-q^\ell a), \quad (a;q)_{n} = \frac{(a;q)_{\infty}}{(a q^n;q)_{\infty}}= \prod_{\ell=0}^{n-1}(1-aq^\ell), \quad [n]_q!=\frac{(q;q)_n}{(1-q)^n}
\end{equation}
leading to a definition of the $q$-deformed binomial factor as
\begin{equation}
\label{eq:def-q-binomial}
\frac{(q;q)_m}{(q;q)_{m-j}(q;q)_j}=\binom{m}{j}_q
\end{equation}

We now recall the following two useful asymptotics identities of $q$-Pochhammer functions which are used in this work
\be
\label{app:eq:asymptotic-q-deformed}
\log (x , q)_{\infty}  \underset{q \to 1}{\to}   -\frac{1}{1-q} {\rm Li}_2(x)+\frac{1}{2}\log(1-x)+\mathcal{O}(1-q)
\ee
implying the following result for the finite $n$ $q$-Pochhammer symbol
\begin{equation}
\label{eq:limit-pochhammer-dilogarithm}
\log (a;q)_n \underset{\substack{q\to 1,\\ n\to \infty, \\q^n=b \, \mathrm{finite}}}{\to} -\frac{1}{1-q}\mathrm{Li}_2(a)+\frac{1}{1-q}\mathrm{Li}_2(a b)+\frac{1}{2}\log \left(\frac{1-a}{1-a b} \right)+\mathcal{O}(1-q)
\end{equation}
The asymptotic formula \eqref{eq:limit-pochhammer-dilogarithm} was also used in the context of high energy physics in \cite{faddeev1994quantum}. A more complete series can be found in Ref.~\cite{DilogarithmStackExchange}. Finally, we need the following asymptotic to obtain estimates on the tails of the current distribution
\be
\label{eq:asymptotic-polylog}
\mathrm{Li}_s(-e^{\mu})\underset{\mu \to +\infty}{\simeq} - \frac{\mu^s}{\Gamma(s+1)}
\ee

\section{
Definitions and observables} \label{app:definitions}

We study in this work the WASEP from different perspectives:
\begin{itemize}
\item a Eulerian perspective where we focus on a single position in space and investigate the number of particles that have crossed this position. For instance we define $Q_t= \int_0^{+\infty} dy (\rho(y,t)-\rho(y,0))$ which is the total flux through the origin, i.e., the number of particles which have crossed the origin from left to right minus right to left during time $t$. We can also define an integrated current $J(x,t)=\int_x^{+\infty} dy  \rho(y,t)-\int_{0}^{+\infty} dy  \rho(y,0) $ so that $Q_T = J(0,T)$. More generally the number of particles
$Q_t(x)$ which have crossed the point $x$ from left to right minus right to left during time $t$
is then $Q_t(x)= \int_x^{+\infty} dy (\rho(y,t)-\rho(y,0))=
J(x,t)-J(x,0)$. This definition of $J(x,t)$ is the analog in the continuum of
the current (called ${\sf J}_{\sf t}(m)$ in our notations) defined in
in \cite[Eq.~(1.1)]{Aggarwal6v} for the discrete model. More precisely upon the WASEP rescaling
\eqref{wasep:rescaling}, one has $\varepsilon {\sf J}_{\sf t}(m) \to J(x,t)$
in the limit $\varepsilon \to 0$.
\item
a Lagrangian perspective where we focus on a single tagged particle (the tracer) and investigate the properties of its displacement over time
If such a tracer starts at initial time at a position $X_0=0$, then by defining the height field
${h}(x,t)=\int_{0}^{+\infty} dy  \rho(y,0) -\int_x^{+\infty} dy  \rho(y,t) $, we have by the conservation of the number of particles to the right of the tracer, that its position at time $t$, $X_t$ can be obtained as
\begin{equation}
{h}(X_t,t)= 0
\end{equation}
\end{itemize}
These two points of view provide dual and equivalent representations of the initial problem. Additionally, we will define $z(x,t)$ as the exponential of the height or current which we will identify as a partition function. Here and  below, we focus on the fields at the "observation point" $x=X$ and $t=T$, and we often
make implicit the dependence in $X,T$, i.e.,
\begin{equation}
z(X,T)= z = e^{2\nu h} = e^{-2 \nu J}=e^H
\end{equation}
In order to study the convergence of the WASEP to the KPZ equation, we additionally define a KPZ height field
\be
\label{eq:def-hkpz-compendium}
H_{\rm KPZ}= - 2 \nu J  + \nu^2 T - \nu X \, ,
\ee
Note the minus sign between the definitions of the height and the integrated current $J$, i.e., $h=-J$, so that events with lower than
average $J$ (few particles have moved from left to right)
correspond to the upper tail of the height field, and vice-versa.
\begin{remark}
In most of the sections below we find useful to keep the explicit dependence in the WASEP
time $T$. In fact since the MFT stochastic equation \eqref{eq:StochCurrent} is statistically invariant under
the change from $(\nu,T,x,t)$ to $ (\nu \sqrt{T},1,x/\sqrt{T},t/T)$ it is not
strictly necessary. Dimensionless quantities, such as $\rho$ or $2 \nu J$, are invariant
under this change, and so is the (random) initial condition studied here,
implying e.g., the statistical equivalence (i.e., in law)
\be
J(\nu,X,T) \equiv \sqrt{T} J(\nu \sqrt{T},X/\sqrt{T},1)
\ee

As a consequence the coefficients $\kappa_n$ defined in the text have the scaling form
\be
\kappa_n(\nu,X,T)= \sqrt{T} \kappa_n(\nu \sqrt{T}, X/\sqrt{T},1)
\ee
Finally, note that large WASEP time $T$ is thus equivalent to large $\nu$.
\end{remark}

\section{Derivation of the Legendre transforms Eqs.~\eqref{eqslegendre1}--\eqref{eqslegendre2}}
\label{sec:derivation}

Here we give a derivation of the various Legendre transforms given in the main text in Eqs.~\eqref{eqslegendre1}--\eqref{eqslegendre2}.  From the unusual generating function in Eq.~\eqref{observableB} and since the PDF of $J$ takes the large deviation form
$P(J) \sim e^{- \frac{1}{\varepsilon} \Phi(J) }$, the rate function $\Psi(u)$ of the WASEP with a two-sided Bernoulli initial condition is determined by the following variational problem
\be
\Psi(u) = \min_{J \in \mathbb{R} } [ \Phi(J) +  \frac{1}{2 \nu} \min_{\omega \in [1,+\infty[} [ F(\omega) - {\rm Li}_2(- u \omega e^{-2 \nu J} ) ]  ]   \label{PsiMin}
\ee
where we recall that $F(\omega)=-\log \omega \log \alpha +\mathrm{Li}_2(\alpha)-\mathrm{Li}_2(1)+\mathrm{Li}_2(\frac{1}{\omega})$
with $\alpha$ being defined from the densities $\varrho_1$ and $\varrho_2$ as
\begin{equation}
\label{eq:def_alpha_densities}
\alpha=\frac{\left(1-\rho _1\right) \rho _2}{\rho _1 \left(1-\rho _2\right)} \quad , \quad
0 < \alpha \leq 1
\end{equation}

Let us denote $\omega_{u,z}$ the value of $\omega$ which realizes $\min_\omega$ in \eqref{PsiMin} for a given $z=e^{-2 \nu J}$.
One finds that it must obey $(\omega-1)  (1 + u \omega z) = \alpha \omega$. Hence it is only a function of the product $u z$, and
we denote $\omega_{u,z}= \omega_{uz}$. The correct root of this equation, i.e., the
branch which vanishes for $u=0$, reads
\begin{equation} \label{omegauz}
\omega= \omega_{u,z} = \omega_{u z} \quad , \quad \omega_u = \frac{\alpha-1+u  +\sqrt{\left(\alpha +u -1\right)^2+4 u }}{2 u
} = \frac{2}{1-\alpha -u +\sqrt{\left(\alpha +u -1\right)^2+4 u }}
\end{equation}
\begin{remark}
Note that $\omega_{u}$ is well defined as long as $u  \geq -(\sqrt{\alpha}-1)^2$,
with $\omega_{u} \geq 1$.
\end{remark}
On the other hand taking a derivative of \eqref{PsiMin} w.r.t. $u$ it is sufficient to take only the explicit derivative
w.r.t. $u$, which leads to
\be  \label{uPsi}
\Psi'(u)= \frac{1}{2 \nu} \frac{\log (1+u \omega_{uz} z)}{u}   \quad , \quad  z=e^{-2 \nu J}
\ee

which shows one of the equalities in \eqref{eqslegendre1}. Next we can
write the condition of minimization of \eqref{PsiMin} w.r.t. $J$. This gives, since again
one can take only the explicit derivative w.r.t. $J$
\be
\Phi'(J) = \log (1+u \omega_{u z} z)    \quad , \quad  z=e^{-2 \nu J}
\ee
showing another equality in \eqref{eqslegendre1}. Next the equation \eqref{defphi} implies that
\be \label{Lagrangephi}
\phi(P) = \max_{J} [ P J -  \Phi(J) ]
\ee
The value of $J$ which realizes the maximum is thus such that
\be \label{Lagrangephi2}
P =  \Phi'(J) \quad , \quad  \phi'(P) = J
\ee

Finally let us return to \eqref{uPsi}. One has
\be
\log(1 + u \omega_{u z} z) = 2 \nu u \Psi'(u)  = P
\ee
It turns out that, inserting \eqref{omegauz}, the equation $1 + u \omega_{u z} z  = e^P$ is inverted simply as
\be \label{inversionuz}
u z = (1 - e^{-P}) (e^P - \alpha)
\ee
which, using $z=e^{-2 \nu J}$ and inserting $P= 2 \nu \Psi'(u)$
gives the last equation in \eqref{eqslegendre2}.
\begin{remark}
The step initial condition $\rho_1=1$, $\rho_2=0$ implies $\alpha=0$ and amounts to set
$\omega_{u z}=1$ with $F(1)=0$ (which means that the variable $\omega$ becomes deterministic).
All the above relations, as well as \eqref{eqslegendre1}--\eqref{eqslegendre2},
hold with this substitution as a special case.
\end{remark}
\section{Parametric representations}
\subsection{Parametric representations of $\Phi(J)$}
\label{sec:parametricphi}
Knowing $\Psi(u)$, the rate function $\Phi(J)$ of the PDF of the integrated current, and its derivative $\Phi'(J)$,
have the parametric representation (obtained by varying $u$)
\bea \label{ParamPhi}
&& \Phi(J)  = \Psi(u) - \frac{1}{2 \nu} [ F(\omega_{\zeta(u)}) - {\rm Li}_2(-  \zeta(u) \omega_{\zeta(u)}    ) ] \\
&& \Phi'(J) = \log (1+ \zeta(u) \omega_{\zeta(u)}  ) =  2\nu u\Psi'(u) \\
&&  J  = - \frac{1}{2\nu}  \log \frac{\zeta(u)}{u} \quad ,\quad \zeta(u) := (1- e^{-  2\nu u \Psi'(u)}) (e^{  2\nu u \Psi'(u)}-\alpha)
\eea

where the function $u \mapsto \omega_u$ is given in \eqref{omegauz} and
$\alpha=\frac{\left(1-\rho _1\right) \rho _2}{\rho _1 \left(1-\rho _2\right)}<1$.
To obtain this result we have performed the inverse Legendre transform of \eqref{PsiMin}, and used the relations in Section~\ref{sec:derivation},
denoting $u z=\zeta(u)$. The case of the step initial condition, $\alpha=0$,
is treated separately in Remark~\ref{remarkstep}.
\begin{remark}
This is valid for $J>J_c$ which corresponds to $u>u_c$ and the main branch of $\Psi(u)$
given by \eqref{eq:rate-function-wasep-general}. The other branches are discussed
in Section~\ref{sec:domain}, and $\Phi(J)$ is
given for $J>J_c$ in Section~\ref{subsec:paramsmallJ}
for $X=0$, $\rho_1+\rho_2=1$ and in some parameter range defined in \eqref{domain}.
\end{remark}
\subsection{Parametric representations of $\phi(P)$}
It is also possible to obtain a parametric representation of $\phi(P)$. We integrate the relation
\bea
\phi'(P) = J
= -\frac{1}{2\nu} \log \left(  \frac{(1- e^{-P}) (e^{P}-\alpha)}{u}  \right)
\eea

to obtain
\be
\label{eq:param-repn-phi-P}
\begin{split}
\phi(P) &= - \frac{1}{2\nu} \int_0^P \rmd P \log( \alpha  (e^{-P}-1)+e^{P}-1))    + \int_0^{u(P)} du (\partial_u (u \Psi'(u) ) \log u \\
&=  - \Psi(u(P)) + \frac{P}{2\nu} \log u(P)  - \frac{P^2}{4\nu} - \frac{1}{2 \nu} \left( \text{Li}_2\left(\alpha e^{-P}\right) - \text{Li}_2\left(\alpha\right) \right)  - \frac{1}{2 \nu}
\left( \text{Li}_2\left(e^{-P}\right)  - \text{Li}_2\left(1\right) \right)
\end{split}
\ee
which provides the parametric representation of $\phi(P)$, where $u(P)$ is defined by inverting the
equation $P  =2 \nu u \Psi'(u)  $.

\section{Explicit formula for the derivatives of the rate function $\Psi(u)$}
\label{sec:derPsi}

Let us recall the expression for the rate function $\Psi(u)$ (which is an implicit function of the observation space-time point $X,T$)
\be
\Psi(u) = - \frac{1}{2\nu} \int_{\I \mathbb{R} + \delta} \frac{\rmd y}{2\I \pi y(1-y)}\mathrm{Li}_2\left(-u\frac{\rho _1 \left(1-\rho _2\right) (1-y) y}{\left(y-\rho _1\right)
\left(\rho_2-y\right)} e^{- 4\nu^2 y(1-y)  T + 2\nu yX } \right)
\ee
where the integration contour on $y$ is taken along $\I \R +\delta$ with $\rho_2 <\delta < \rho_1$.
To calculate the cumulants of the current it is useful to use the Taylor series of $\Psi$ as
\be
\begin{split}
\Psi(u) \label{expansionPsi}
&=  \sum_{n=1}^{\infty} \frac{u^n}{n!} \Psi^{(n)}(0)      \quad , \quad \Psi^{(n)}(0)  := \Psi_n(X,T)
\end{split}
\ee
and to first obtain an explicit formula for the multiple derivatives $\Psi^{(n)}(0)$ for arbitrary $(X,T)$.
Using the series definition of the dilogarithm ${\rm Li}_2(z)=\sum_{n \geq 1} \frac{z^n}{n^2}$ we obtain
\begin{equation}
\begin{split}
\Psi^{(n)}(0)
&=(-1)^{n-1} (n-1)! \frac{(\rho _1 \left(1-\rho _2\right))^n}{2\nu n }\int_{\I \mathbb{R} + \delta} \frac{\rmd y}{2\I \pi }\frac{ (y(1-y))^{n-1} }{\left(y-\rho _1\right)^n
\left(\rho _2-y\right)^n}  e^{- 4\nu^2 n y(1-y)  T + 2\nu n yX }
\end{split}
\end{equation}

To compute this integral it is useful to note that each term in the expansion \eqref{expansionPsi} verifies the following heat equation
\be \label{heat2}
\partial_T \Psi_n = \frac{1}{n} \partial_X^2 \Psi_n - 2 \nu \partial_X \Psi_n
\ee
where the diffusion coefficient depends on the order of the expansion $n$. Hence we can first compute the integral for $T=0$, which can be done using residues, and then propagate it
using the heat equation. We obtain the initial condition at $T=0$ as follows.
Let us recall the formula for the residue of a multiple pole. One has, on a positively oriented closed contour ${\cal C}$
around $z$, if $g(y)$ is analytic within the region enclosed by the contour
\be
\int_{\cal C} \frac{dy}{2 \I \pi} \frac{g(y)}{(y-z)^p} = \frac{1}{(p-1)!} g^{(p-1)}(z)
\ee
Hence we find for $X<0$, closing the contour at infinity on the side $\Re \, y >0$
\be
\Psi^{(n)}(0)\underset{T=0,X<0}{=} (-1)^n
\frac{\rho _1^n \left(1-\rho _2\right)^n}{2\nu n }\frac{d^{n-1}}{dy^{n-1}} \left( \frac{e^{ 2\nu n y X}(y (1-y))^{n-1}}{  (\rho_2-y)^n } \right)|_{y=\rho_1}
\ee
And for $X>0$, closing the contour at infinity on the side $\Re  \, y <0$
\be
\Psi^{(n)}(0)\underset{T=0,X>0}{=}
- \frac{\rho _1^n \left(1-\rho _2\right)^n}{2\nu n }\frac{d^{n-1}}{dy^{n-1}} \left( \frac{e^{ 2\nu n y X}(y (1-y))^{n-1}}{  (y-\rho_1)^n } \right)|_{y=\rho_2}
\ee
\begin{remark}
We can check that these expressions are continuous at $X=0$, however the derivative w.r.t $X$ has a jump at $X=0$.
\end{remark}
We then need to propagate this initial condition through the heat equation \eqref{heat2}, using its associated kernel
$G_n(X,T)= \sqrt{\frac{n}{4\pi T}} e^{-n \frac{(X-2\nu T)^2}{4T}}$. We thus propagates each initial condition on each half-space as
\begin{equation} \label{propagation1}
e^{2\nu n y X}\Theta(-X) \to \int_\R \rmd x' G_n(X-x',T)e^{2\nu n y x'}\Theta(-x') = \frac{1}{2} e^{2 \nu  n y (2 \nu  T (y-1)+X)} \text{Erfc}\left( \sqrt{\frac{n}{4T}} (2 \nu  T (2 y-1)+X)\right)
\end{equation}

and
\begin{equation}
e^{2\nu n y X}\Theta(X) \to \int_\R \rmd x' G_n(X-x',T)e^{2\nu n y x'}\Theta(x') = \frac{1}{2} e^{2 \nu  n y (2 \nu  T (y-1)+X)} \text{Erfc}\left(-\sqrt{\frac{n}{4T}} (2 \nu  T (2 y-1)+X)\right)
\end{equation}

Putting all together we obtain our final result
\begin{equation}
\label{eq:derivative_Psi_alltimes}
\begin{split}
\Psi^{(n)}(0)=& - \frac{\rho _1^n \left(1-\rho _2\right)^n}{4\nu n }\frac{d^{n-1}}{dy^{n-1}} \left( \frac{(y (1-y))^{n-1} e^{2 \nu  n y (2 \nu  T (y-1)+X)}}{  (y-\rho_1)^n }  \text{Erfc}\left(-\sqrt{\frac{n}{4T}} (2 \nu  T (2 y-1)+X)\right)\right)|_{y=\rho_2} \\
&+(-1)^n
\frac{\rho _1^n \left(1-\rho _2\right)^n}{4\nu n }\frac{d^{n-1}}{dy^{n-1}} \left( \frac{(y (1-y))^{n-1}e^{2 \nu  n y (2 \nu  T (y-1)+X)}}{  (\rho_2-y)^n }  \text{Erfc}\left( \sqrt{\frac{n}{4T}} (2 \nu  T (2 y-1)+X)\right)\right)|_{y=\rho_1}
\end{split}
\end{equation}
\begin{remark}
The step initial condition
is recovered as a special case, setting $\rho_1=1$, $\rho_2=0$
\begin{equation}
\begin{split}
\Psi^{(n)}(0)
&= \frac{(-1)^{n-1} (n-1)!}{2\nu n }\int_{\I \mathbb{R} + \delta} \frac{\rmd y}{2\I \pi }\frac{ 1 }{y (1-y) }  e^{- 4\nu^2 n y(1-y)  T + 2\nu n yX }
\end{split}
\end{equation}

At initial time $T=0$ one finds
\be
\Psi^{(n)}(0) \underset{T=0}{=}  \frac{(-1)^{n-1} (n-1)!}{2\nu n } ( \Theta(-X) e^{2 \nu n X} + \Theta(X))
\ee
and for later times $T>0$ one has
\be
\Psi^{(n)}(0) = \frac{ (-1)^{n-1} (n-1)!}{4 \nu n }  \left( e^{2 \nu n X}
{\rm Erfc}\left(\sqrt{\frac{n}{4T}} (2 \nu  T +X)\right) + {\rm Erfc}\left(\sqrt{\frac{n}{4T}} (2 \nu  T -X)\right) \right)
\ee
\end{remark}

\section{Cumulants of the integrated current $J$} \label{sec:cum}

Using the Legendre transform equations \eqref{eqslegendre1}--\eqref{eqslegendre2} we can obtain the cumulants of the time-integrated current $J=-h$ at any order, i.e., the coefficients
$\kappa_n$
\be
\langle J^n \rangle^c =  \varepsilon^{n-1}  \kappa_n
\ee
Note that $J$ here denotes $J(X,T)$ defined in Section~\ref{app:definitions}.
\begin{remark}
Let us recall that the total flux in time $T$ through point $X$ is
$Q_T(X)=J(X,T)-J(X,0)$, with $J(0,0)=0$. Hence $J(0,T)$
is also the total flux $Q_T=Q_T(0)$ through the origin. Knowledge of the
cumulants of $J(X,T)$ alone does not allow to determine the cumulants of $Q_T(X)$
for general $X$, except in two cases (i) for the step initial condition: in that case
$J(X,0)=- X \theta(-X)$ is deterministic and one has
$\langle Q_T(X)^n \rangle^c = \langle J(X,T)^n \rangle^c$ for $n \geq 2$
(ii) for the stationary initial condition $\rho_1=\rho_2$
in which case by translational invariance
$\langle Q_T(X)^n \rangle^c = \langle Q_T(0)^n \rangle^c = \langle J(0,T)^n \rangle^c$.
\end{remark}
To obtain the cumulants of $J$, we first express them in terms of the derivatives $\Psi^{(n)}(0)$ of the rate function $\Psi(u)$. To obtain these relations we applied two equivalent methods.
\begin{enumerate}
\item The first one is to invert, from \eqref{eqslegendre1}, the series of $P= 2 \nu u \Psi'(u)$ which gives $u=u_P$ as
a series in $P$,
and then, from \eqref{eqslegendre2}, to insert it into
\be
\kappa_{n} = [\partial_P^{n-1} \Phi'(P)]_{P=0} = - \frac{1}{2 \nu} \left[\partial_P^{n-1} \left( \log((1-e^{-P})(e^P-\alpha) ) - \log u_P \right) \right]_{P=0}
\ee

\item The second one is to write using the Jacobian of the map $u \to P$
\be \label{cumfromderPsi}
\kappa_n =   - \frac{1}{2\nu} \left[\left(\frac{1}{2\nu \p_u( u \Psi'(u)) }\p_u\right)^{n-1}   \log \left( \frac{(1- e^{-  2\nu u \Psi'(u)}) (e^{  2\nu u \Psi'(u)}-\alpha)}{u} \right) \right]_{u=0}
\ee
\end{enumerate}

We can further work on this expression to express the $n$-th cumulant as
\be
\begin{split}
\kappa_n
&=   - \frac{1}{2\nu} \left[\p_P^{n-1}\log\left(\frac{(1-e^{-P})(e^P-\alpha)}{P}\right)\right]_{P=0} - \frac{1}{(2\nu)^n} \left[ \left(\frac{1}{ \p_u( u \Psi'(u)) }\p_u\right)^{n-1}  \log(2\nu \Psi'(u))\right]_{u=0}\\
\end{split}
\ee
The first part can be computed exactly. Indeed, we can use the following Taylor expansions
\begin{equation}
\log \left(\frac{1-e^{-P}}{P}\right)=-\frac{P}{2}+\sum_{n\geq 1}\frac{B_{2n}}{(2n)! 2n}P^{2n}, \quad \log (e^P-\alpha)=P+\sum_{n \geq 0 } \frac{(-1)^{n+1}}{n!}  {\rm Li}_{1-n}(\alpha) \, P^n
\end{equation}
where $B_n$ are the Bernoulli numbers. This expansion is valid for $\alpha<1$. The limit $\alpha \to 1$
is studied in Section~\ref{sec:stationary}. Hence we obtain
\bea
&& \kappa_n = \kappa_n^0 + \kappa_n^1 \quad , \quad \kappa^0_1= - \frac{1}{2 \nu}  \log(1-\alpha) \\
&& \kappa^0_{n=2 q} = - \frac{1}{2 \nu} ({\rm Li}_{2-n}(\alpha) + \frac{1}{2} \delta_{n,2})
\quad , \quad
\kappa^0_{n=2 q+1 } =  \frac{1}{2 \nu} ({\rm Li}_{2-n}(\alpha) - \frac{B_{n-1}}{n-1}   ) \quad , \quad q \geq 1 \\
&& \kappa^1_n = -
\frac{1}{(2\nu)^n} \left[ \left(\frac{1}{ \p_u( u \Psi'(u)) }\p_u\right)^{n-1}
\log(2\nu \Psi'(u))\right]_{u=0}
\eea

We recall that $\alpha$ is defined in Eq.~\eqref{eq:def_alpha_densities}. Defining the ratios $r_m = \Psi^{(m)}(0)/\Psi'(0)^m$, we obtain the first six cumulants as
\bea \label{cum1234}
&&    \kappa_1 = -  \frac{1}{2 \nu}   \log \left(2\nu (1-\alpha ) \Psi '(0)\right) \quad , \quad      \kappa_2 = \frac{1+\alpha}{4 \nu (\alpha-1)} - \frac{r_2}{4 \nu^2} \\
&&   \kappa_3 = -\frac{\alpha ^2-14 \alpha +1}{24 (\alpha -1)^2 \nu  }+\frac{3 r_2^2-r_3}{8 \nu ^3} \quad , \quad    \kappa_4 =  \frac{\alpha  (\alpha
+1)}{2 (\alpha -1)^3 \nu
}+\frac{-20 r_2^3+12 r_3
r_2-r_4}{16 \nu ^4} \\
&& \kappa_5 = \frac{1}{2 \nu} \left(  \frac{\alpha  (\alpha  (\alpha
+4)+1)}{(\alpha
-1)^4}+\frac{1}{120} \right)
-  \frac{1}{(2 \nu)^5}   \left(
r_5-5 \left(42 r_2^4-36 r_3
r_2^2+4 r_4 r_2+3
r_3^2\right) \right)  \\
&&  \kappa_6 = \frac{1}{2 \nu}
\frac{\alpha  (\alpha +1)
(\alpha  (\alpha
+10)+1)}{(\alpha -1)^5}
- \frac{1}{(2 \nu)^6} \left(
6 \left(504 r_2^5-560 r_3
r_2^3+70 r_4 r_2^2+5
(21 r_3^2-r_5)
r_2-10 r_3 r_4\right)+r_6 \right)  \nn
\eea

We can now insert the explicit formula for the $\Psi^{(n)}(0)$ obtained in the previous section and
obtain the explicit formula for the cumulants for arbitrary $(X,T)$ and $(\rho_1,\rho_2)$. Let us
give some examples

\subsection{First moment}

The average integrated current at the space-time observation point $(X,T)$ reads
\bea \label{Jav}
&& \langle J(X,T) \rangle= \kappa_1 = - \frac{1}{2 \nu} \log \bigg(
\frac{1}{2} e^{2 \nu    \rho _1 (2 \nu
(\rho _1-1)
T+X)}
\text{Erfc}(
\sqrt{\frac{1}{4 T}} (2
\nu  (2 \rho_1-1)
T+X))
\\
&& + \frac{1}{2} e^{2 \nu   \rho _2 (2 \nu   (\rho _2-1) T+X)}
\text{Erfc}(- \sqrt{\frac{1}{4 T}} (2
\nu  (2 \rho_2-1) T+X)) \bigg) \nn
\eea
and we recall that $\langle J(X,0) \rangle=- \rho_1 X \Theta(-X) - \rho_2 X \Theta(X)$. We now obtain the two special cases of step and stationary initial conditions.
\begin{itemize}
\item In the stationary limit $\rho_2=\rho_1=\rho$, it simplifies as
\be \label{averageJ}
\langle J(X,T) \rangle =   2 \nu \rho (1-\rho) T- \rho X
\ee
\item  For the step (determinist) initial condition, $\rho_2=0$ and $\rho_1=1$, one finds
\be
\langle J(X,T) \rangle =  -\frac{\log \left(\frac{1}{2}
\text{Erfc}\left(- \sqrt{\frac{1}{4 T}}
(X-2 \nu  T)\right)+ \frac{1}{2} e^{2 \nu
X}
\text{Erfc}\left(
\sqrt{\frac{1}{4 T}} (2 \nu
T+X)\right)\right)}{2
\nu }
\ee
which becomes for $X=0$
\be
\langle J \rangle =-\frac{\log (\text{Erfc}(\nu \sqrt{T}
))}{2 \nu }  = \begin{cases}  \frac{\sqrt{T}}{\sqrt{\pi
}}+\frac{\nu T}{\pi
}+\mathcal{O}\left(\nu ^2\right) \quad , \quad &\nu \to 0 \\
\frac{\nu T}{2}  + \frac{1}{2 \nu} \log(\nu \sqrt{T} \sqrt{\pi}) + \mathcal{O}(1/\nu^3) \quad , \quad &\nu \to + \infty\\
\frac{\log 2}{2|\nu|}+\mathcal{O}(e^{-T\nu^2}) \quad , \quad &\nu \to - \infty\\
\end{cases}
\ee
which is an increasing function of $\nu$. For $\nu=0$ setting $T=1$ we find
$\langle J \rangle=\frac{1}{\sqrt{\pi}}$ in agreement with the result for the SSEP in
\cite[Eqs.~(1-3)]{derrida2009currentExact}, see also \eqref{defomt}.
Note that for the step initial condition the integrated current $J$ must be positive for $X=0$. Furthermore $J$ vanishes when $\nu \to -\infty$, since in that case the drift forbids the particles
to cross the origin.

\item More generally for arbitrary densities, in the SSEP limit $\nu\to 0$ we obtain
\be
\langle J(X,T) \rangle \underset{\nu = 0}{=}
\left(\rho _1-\rho _2\right)
\sqrt{T} {\cal G}\left(\frac{X}{2
\sqrt{T}}\right)-\frac{1}{2
} \left(\rho _1+\rho_2\right) X
\ee
where here and below we define the function
\be \label{defcalG}
{\cal G}(y) := y {\rm Erf}(y) + \frac{1}{\sqrt{\pi}} e^{-y^2}
\ee
For $X=0$, $T=1$ one finds $\langle J \rangle = \frac{\rho_1-\rho_2}{\sqrt{\pi}}$
in agreement with \cite[Eqs.~(1-3)]{derrida2009currentExact,derrida2009currentMFT}.

\end{itemize}
\begin{remark} \label{remarkferrari}
Equation~\eqref{averageJ} for $X=0$ matches the result
$\langle {\sf J}_{\sf T}(0) \rangle = \frac{1}{\varepsilon} \langle J \rangle = \sqrt{\frac{{\sf T}}{T}}
2 \nu \rho(1-\rho) T  =(R-L) {\sf T} \rho(1-\rho)$ for the ASEP microscopic
integrated current $\langle {\sf J}_{\sf T}(0) \rangle$ in \cite{ferrari1994current}.
\end{remark}
\begin{remark}
The error functions in \eqref{Jav} show diffusive broadening around the space-time rays
$X= v_i T$, where $v_i=\nu \sigma'(\rho_i) = 2 \nu (1-2 \rho_i)$ are the sound velocities
obtained by linearizing \eqref{eq:StochCurrent}, and become discontinuous as a
function of $X/T$ as $T \to +\infty$. In Ref.~\cite{AggarwalBorodin6v} the
large time limit of the ASEP is studied, and a phase transition in the
fluctuations of the current $\langle {\sf J}_t(m) \rangle$ occurs along the corresponding rays
$m={\sf T} (1- 2 \rho)$.
\end{remark}
\begin{remark}
For large WASEP time $T$, and for $\nu>0$, the mean current \eqref{Jav} becomes, denoting $v=X/T$
\bea
\frac{\langle J(X,T) \rangle }{T} \underset{T \gg 1}{\simeq} &  \begin{cases}
2 \nu \rho_1(1-\rho_1) - \rho_1 v \quad ,& \quad v < 2 \nu( 1- 2 \rho_1) \\
\frac{(v-2 \nu)^2}{8 \nu} ~~~~~~~~~~~\quad ,& \quad 2 \nu (1-2 \rho_2) < v < 2 \nu( 1- 2 \rho_1) \\
2 \nu \rho_2(1-\rho_2) - \rho_2 v \quad ,& \quad v > 2 \nu( 1- 2 \rho_2)  \end{cases}
\eea

    The derivative of the profile remains continuous in the limit.
       For $\nu<0$ there is a shock in the limit, and one finds instead
\bea
\frac{\langle J(X,T) \rangle}{T}  \underset{T \gg 1}{\simeq} &  \begin{cases}
2 \nu \rho_1(1-\rho_1) - \rho_1 v
\quad , \quad v < 2 |\nu| (\rho_1+\rho_2-1)  \\
2 \nu \rho_2(1-\rho_2) - \rho_2 v \quad , \quad v > 2 |\nu| (\rho_1+\rho_2-1)  \end{cases}
\eea
    The occurrence of a shock in ASEP or TASEP in that case is well known, see e.g., \cite{zhang2024t} and
       references therein.
\end{remark}

\subsection{Second cumulant}

The expression for arbitrary $\rho_1,\rho_2$ is already quite bulky.
It reads for $X=0$ and $T=1$
\bea
&&  \kappa_2 = \frac{1}{2 \nu (\rho_1-\rho_2) }(  \rho_1 \rho_2 - \frac{\rho_1 + \rho_2}{2} +  \frac{A + B}{C} ) \\
&& A = -4  \nu e^{-2 \nu^2} \sqrt{\frac{2}{\pi}}  (\rho_1-\rho_2)  (\rho_1(\rho_1-1) + \rho_2(\rho_2-1)) \\
&& B = (\rho_1+\rho_2 - 2 \rho_1 \rho_2) (g_\nu(\rho_1) + g_{-\nu}(\rho_2))
+ 8 \nu^2 (\rho_1-\rho_2) (\tilde g_\nu(\rho_1) - \tilde g_{-\nu}(\rho_2) ) \\
&& C =  (g_\frac{\nu}{\sqrt{2}}(\rho_1) + g_{-\frac{\nu}{\sqrt{2}}}(\rho_2))^2  \\
&& g_\nu(\rho)= e^{8 \nu^2 \rho(\rho-1)} {\rm Erfc}(\sqrt{2} \nu (2 \rho-1)) \quad , \quad \tilde g_\nu(\rho)= \rho (\rho-1) (2 \rho-1) g_\nu(\rho)
\eea

We now obtain the three special cases of the SSEP limit and then the step and stationary initial conditions.
\begin{itemize}
\item In the SSEP limit $\nu \to 0$ for arbitrary densities it gives
\be
\kappa_2 \underset{\nu = 0}{=} \frac{-\sqrt{2} \left(\rho
_1-\rho _2\right){}^2+2
\rho _1-4 \rho _1 \rho _2+2
\rho _2}{2 \sqrt{\pi }}
\ee
in agreement with \cite[Eqs.~(1-3)]{derrida2009currentExact,derrida2009currentMFT}.
\item
For the step (determinist) initial condition, $\rho_2=0$ and $\rho_1=1$, one finds
\be
\kappa_2 = \frac{1}{4 \nu} \bigg(
\frac{2
\left(\text{Erfc}\left(\frac
{ 2 \nu
T-X}{\sqrt{ 2 T }}\right)+e^{4
\nu  X}
\text{Erfc}\left(\frac{
2 \nu
T+X}{\sqrt{2 T}}\right)\right
)}{\left(\text{Erfc}\left(
\sqrt{\frac{1}{4 T}} (2 \nu
T-X)\right)+e^{2 \nu  X}
\text{Erfc}\left(
\sqrt{\frac{1}{4 T}} (2 \nu
T+X)\right)\right)^2}-1 \bigg)
\ee
   which for $X=0$, $T=1$ simplifies to
\be
\kappa_2 =
\frac{\frac{\text{Erfc}\left
(\sqrt{2} \nu
\right)}{\text{Erfc}(\nu
)^2} - 1 }{4 \nu }   =
\begin{cases}
\frac{
\left(2-\sqrt{2}\right)
}{2 \sqrt{\pi }}+\mathcal{O}\left(\nu
\right) \\ \frac{1}{4} \sqrt{\frac{\pi }{2}}
+\mathcal{O}\left(\frac{1}{\nu}\right) \quad , \quad &\nu \to + \infty\\
\frac{1}{8|\nu|}+\mathcal{O}(e^{-\nu^2})  \quad , \quad &\nu \to - \infty\\
\end{cases} \label{kappa2stepexpand}
\ee
which is an increasing function of $\nu$.

\item In the stationary limit $\rho_1=\rho_2=\rho$, it simplifies for arbitrary $X,T$ as
\bea  \label{secondcumappendix}
\kappa_2 = \rho (1-\rho)   \left( (2
\nu  (2 \rho-1) T+X)
\text{Erf}\left(\frac{2 \nu
(2 \rho -1) T+X}{2
\sqrt{T}}\right) + \frac{2 \sqrt{T} e^{-\frac{(2
\nu  (2 \rho -1) T+X)^2}{4
T}}}{\sqrt{\pi }}\right)
\eea
which  recovers the expression \eqref{cum2text} given in the text.
As a function of $X$ it has a minimum for $X= 2 \nu (1- 2 \rho) T$.
It has the symmetry $(\rho,X) \to (1-\rho,-X)$.
For $X=0$, $\kappa_2$ is even in $\nu$ and has its minimum for $\nu=0$. Setting $T=1$ one finds the asymptotics
for fixed $\rho \neq 1/2$ and large $\nu$
\be \label{kappa2largenu}
\kappa_2 = \begin{cases} \frac{2 \rho (1-\rho)}{\sqrt{\pi} } + \mathcal{O}(\nu^2) \quad , \quad &\nu \to 0 \\
2 |\nu| \rho(1-\rho) |1- 2 \rho| +\mathcal{O}(e^{-\nu^2 (1-2\varrho)^2})\quad , \quad &\nu \to  \pm  \infty
\end{cases}
\ee
As in Remark~\ref{remarkferrari}, the limit $|\nu| \to +\infty$ matches
the result
$\langle {\sf J}_{\sf T}(0)^2 \rangle \simeq |R-L| \rho(1-\rho) |(1- 2 \rho)| {\sf T}$
for the second cumulant of the ASEP microscopic
integrated current through the origin in \cite{ferrari1994current}. However, note that
$\kappa_2$ simplifies for $\rho=1/2$, in which case $\kappa_2=\frac{1 }{2 \sqrt{\pi} }$ for arbitrary $\nu$. This value does not match \eqref{kappa2largenu} at large $\nu$. Indeed, the correction terms in \eqref{kappa2largenu}
cannot be neglected when $\rho \to 1/2$. As a result, there is a crossover for
$\rho-1/2 \sim 1/\nu$ at large $\nu$. This crossover is precisely the
crossover from stationary to droplet initial conditions in the KPZ limit,
obtained in Eq.~\eqref{kappa2kpz} by increasing $\tilde w$ from zero to $+\infty$,
with $\rho_{1,2}= \frac{1}{2} \pm \frac{\tilde w}{2 \nu}$.
One can check that $\kappa_2$ in Eq.~\eqref{kappa2kpz} indeed varies from
$\kappa_2=\frac{1 }{2 \sqrt{\pi} }$ for $\tilde w=0$,
to $\kappa_2=\frac{1 }{4 } \sqrt{\frac{\pi}{2}}$ for $\tilde w \to + \infty$
which is the result for the step initial condition, see \eqref{kappa2stepexpand}.
\end{itemize}

\subsection{Third cumulant}
The expression for arbitrary $\rho_1,\rho_2$ is very bulky.
\begin{itemize}
\item For the step initial condition, $\rho_2=0$ and $\rho_1=1$, one finds for $X=0$ and $T=1$
\be
\kappa_3 = \frac{f(\nu) -1}{24 \nu} ~,~ f(\nu)=
\frac{9
\text{Erfc}\left(\sqrt{2}
\nu \right)^2-8
\text{Erfc}(\nu )
\text{Erfc}\left(\sqrt{3}
\nu \right)}{\text{Erfc}(\nu
)^4} ~,~ \kappa_3 = \begin{cases}
\frac{6-9 \sqrt{2}+4
\sqrt{3}}{6 \sqrt{\pi}}+\mathcal{O}\left(\nu\right)
\\
\left(\frac{3}{16}-\frac{1}{3
\sqrt{3}}\right) \pi  \nu
+\mathcal{O}\left(\frac{1}{\nu}\right), \quad &\nu \to +\infty\\
\frac{1}{32|\nu|}+\mathcal{O}(e^{-\nu^2}), \quad &\nu \to -\infty
\end{cases} \label{kappa3stepexpand}
\ee

which has a maximum equal to $\kappa_3^{\rm \max}= 0.0239568$ at $\nu = -0.715953$.
\item
In the stationary limit $\rho_1=\rho_2=\rho$ one finds, for $X=0$ and $T=1$
\be
\begin{split}
\kappa_3 &=  2 \nu  (\rho -1) \rho  \left(6
(\rho -1) \rho  \left(-\nu
^2 (1-2 \rho )^2+\frac{e^{-2
\nu ^2 (1-2 \rho )^2}}{\pi
}-1\right)-1\right) \\
& + 12 \nu ^3 (\rho -1)^2 \rho ^2
(2 \rho -1)^2 \text{Erf}(\nu
-2 \nu  \rho )^2 + \frac{24
\nu ^2 (\rho -1)^2 \rho ^2
(1-2 \rho) e^{-\nu ^2 (1-2
\rho )^2} \text{Erf}(\nu -2
\nu  \rho )}{\sqrt{\pi
}}
\end{split}
\ee
which is odd in $\nu$, and invariant by $\rho \to 1-\rho$. It behaves at small asymmetry $\nu$ as
\be
\kappa_3 \underset{|\nu| \ll 1}{=}  -\frac{2 \nu  ((\rho -1) \rho
(6 \pi  (\rho -1) \rho -6
(\rho -1) \rho +\pi ))}{\pi
} + \mathcal{O}(\nu^3)
\ee
and at large asymmetry $\nu \to \pm \infty$, for $\rho \neq 1/2$, as
\be
\kappa_3 \underset{|\nu|\gg 1}{\simeq}
2 \nu  (1-\rho ) \rho  (6
(\rho -1) \rho
+1) + \mathcal{O}( e^{- (2 \rho -1)^2 \nu^2} )  \label{kappa3largenu}
\ee
which in principle could be matched to third cumulant for the ASEP, as above.
Again $\kappa_3$ simplifies for $\rho=1/2$ in which case $\kappa_3=-\frac{(\pi -3) \nu }{4 \pi }$ for arbitrary $\nu$, and again it does not match \eqref{kappa3largenu} at large $\nu$.
The crossover for  $\rho-1/2 \sim 1/\nu$ at large $\nu$
is discussed in the KPZ limit in Eq.~\eqref{eq:third-cum-kpz}
as a function of $\tilde w$. Again one finds a crossover from
the value at $\rho=1/2$ for $\tilde w=0$, to the
$\nu \to +\infty$ asymptotics for the step initial condition displayed in \eqref{kappa3stepexpand},
for $\tilde w \to + \infty$.\\

For fixed $\rho$, $\kappa_3$ has a rich behavior as a function of $\nu$.
Since $\kappa_3$ is odd let us discuss $\nu>0$.
One can already
see from the large $\nu$ asymptotics in \eqref{kappa3largenu} that for
$\rho \in [1-\rho_c,\rho_c]$ the prefactor is negative, and $\kappa_3 \to - \infty$
as $\nu \to +\infty$, while for $\rho \in [0,1-\rho_c] \cup [\rho_c,1]$
one has $\kappa_3 \to - \infty$ as $\nu \to +\infty$. Here one has $\rho_c = \frac{1}{6} (3 + \sqrt{3})
= 0.788675$. In addition, as $\rho$ increases from $1/2$ a global maximum of
$\kappa_3$ for $\nu>0$ appears when
the coefficient of the small $\nu$ asymptotics in \eqref{kappa3largenu}
changes sign, i.e for $\rho=\rho'_c=0.574227$. Beyond this
point $\kappa_3$ becomes non monotonic. This global maximum becomes
local for $\rho>\rho_c$ and completely disappears for some
value of $\rho$ in $[\rho_c,1]$, so that $\kappa_3$ becomes again
monotonic as a function of $\nu$.

\end{itemize}
\begin{remark}
One can compare with the very recent result of \cite{berlioz2024tracer} for the stationary initial condition.
For models with $D(\rho)=1$, they read
\bea \label{secondthirdaurelien}
&&  \kappa_2 = \sigma(\rho) {\cal G}(y= \frac{\nu}{2}  \sigma'(\rho) )
\quad , \quad {\cal G}(y) = y {\rm Erf}(y) + \frac{1}{\sqrt{\pi}} e^{-y^2} \\
&&  \kappa_3 = \frac{\nu}{4} \sigma(\rho) (\sigma'(\rho)^2
+ \sigma(\rho) \sigma''(\rho) (1 + 3 (y^2 - {\cal G}(y)^2))
\eea

Inserting $\sigma(\rho) = 2 \rho (1-\rho) $ one finds agreement with our results for $X=0$ and $T=1$.
\end{remark}

\subsection{Fourth cumulant}
We restrict here to the formula for the step initial condition as well as the stationary one.
\begin{itemize}
\item
For the step initial condition for $X=0$ and $T=1$ we obtain
\be
\kappa_4 = \frac{5
\text{Erfc}\left(\sqrt{2}
\nu \right)^3-8
\text{Erfc}(\nu )
\text{Erfc}\left(\sqrt{3}
\nu \right)
\text{Erfc}\left(\sqrt{2}
\nu \right)+3
\text{Erfc}(\nu )^2
\text{Erfc}(2 \nu )}{4 \nu
\text{Erfc}(\nu )^6}
\ee

It is negative for $\nu < \nu^*= 2.79349$ and positive for $\nu> \nu^*$.
It has a negative minimum for $\nu= 0.762317$
of value $\kappa_4^{\rm min}=-0.0151279$. One finds the asymptotics
\be
\kappa_4 = \begin{cases} \frac{-4-7 \sqrt{2}+8
\sqrt{3}}{2 \sqrt{\pi}}+\mathcal{O}\left(\nu\right) \quad , \quad &\nu \to 0 \\
\frac{1}{48} \left(18+15
\sqrt{2}-16 \sqrt{6}\right)
\pi ^{3/2} \nu^2 + \mathcal{O}(1) \quad , \quad &\nu \to + \infty \end{cases}
\ee
and it vanishes for $\nu \to - \infty$. The
value for $\nu=0$ recovers the one for the SSEP, obtained from
\cite[Eqs.~(1-3)]{derrida2009currentExact,derrida2009currentMFT}.

\item In the stationary limit $\rho_1=\rho_2=\rho$, $X=0$ and $T=1$, with some help from
Mathematica, we find, with the same notations as in \eqref{secondthirdaurelien}
\bea \label{kappa4stat}
&&  \kappa_4 =
-\sqrt{2} {\cal G} \left(\sqrt{2} y\right)
\sigma (\rho )^2 \left(2 \nu ^2
\left(3-\left(4 y^2+9\right)
\sigma (\rho
)\right)+3\right) +20 \nu ^2
{\cal G} (y)^3 \sigma (\rho)^3
\\
&& +{\cal G} (y) \sigma
(\rho ) \left(12 \nu ^2 \sigma
(\rho ) \left(1-3
\left(y^2+1\right) \sigma (\rho
)\right)+1\right)
+\frac{e^{-2 y^2} \sigma (\rho
)^2 \left(-2 \nu ^2+2
y^2+3\right)}{\sqrt{2 \pi }}
\eea
i.e., with $\sigma (\rho) = 2 \rho (1-\rho)$ and $y = \frac{\nu}{2}  \sigma'(\rho)$.
\begin{figure}[t!]
\centering
\includegraphics[width=0.4\linewidth]{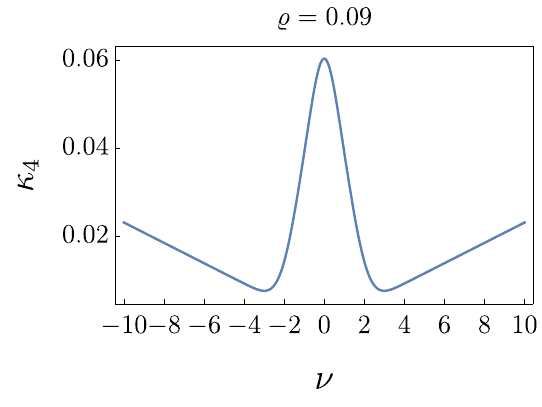}\\
\includegraphics[width=0.4\linewidth]{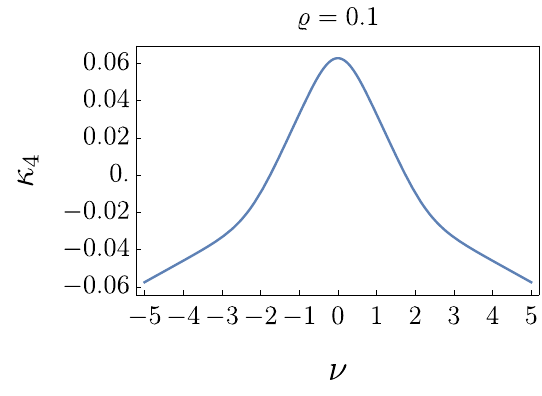}
\includegraphics[width=0.4\linewidth]{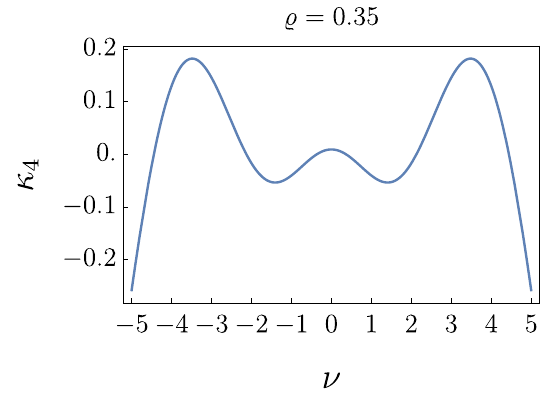}
\caption{Plot of the fourth cumulant $\kappa_4$ in the stationary limit $\varrho_1=\varrho_2=\varrho$ for $X=0$ and $T=1$ as a function of $\nu$ for various values of $\varrho=\{0.09,0.1,0.35\}$.
The behavior at infinity changes at $\rho \approx 0.0917$ and it can exhibit multiple extrema.  }
\label{fig:kappa-4-stat}
\end{figure}
It is an even function of $\nu$ and invariant by $\rho \to 1-\rho$.
Setting $\nu=0$ one finds
\be
\kappa_4 \underset{\nu=0}{=}  \frac{2 (1-\rho ) \rho  \left(3
\sqrt{2} (\rho -1) \rho
+1\right)}{\sqrt{\pi }}  \label{kappa4ssep}
\ee
which agrees with the result for the SSEP obtained in \cite[Eqs.~(1-3)]{derrida2009currentExact,derrida2009currentMFT}. Moreover, it
agrees with the general result for the fourth cumulant for $\nu=0$ given in
\cite[Eq.~(S61) in the Supp. Mat]{grabsch2024tracer}, which reads for
general $\sigma(\rho)$ and $D(\rho)=1$
\bea
\kappa_4 = \frac{2 \sigma(\rho) \sigma'(\rho)^2 + (3 \sqrt{2}-4) \sigma(\rho)^2 \sigma''(\rho)}{8 \sqrt{\pi}}
\eea
which for the SSEP, $\sigma(\rho)=2 \rho(1-\rho)$ reduces to \eqref{kappa4ssep}. We note once again that for $\rho=1/2$ the result \eqref{kappa4stat} simplifies
for all $\nu$
\be
\kappa_4 \underset{\rho=1/2}{=}
\frac{4
\left(5+\left(\sqrt{2}-3\right) \pi \right) \nu
^2+\left(4-3
\sqrt{2}\right) \pi }{8 \pi^{3/2}}
\ee
although now it depends on $\nu$, and behaves as $\mathcal{O}(\nu^2)$ at large $\nu$.
By contrast the asymptotics for $\nu \to + \infty$ and fixed $\rho \neq 1/2$ is quite different
\be
\kappa_4 \simeq 2 \nu  (1-\rho) \rho  |2
\rho -1| (12 (\rho -1) \rho
+1)
\ee
i.e., it is linear in $\nu$. Again that discrepancy is related to the KPZ crossover,
see discussion above and \eqref{eq:fourth-cum-kpz}.
The behavior of $\kappa_4$ as a function of $\nu$ is again quite rich.
Again the sign of the divergence at infinity changes at $\rho=\rho'_c = \frac{1}{6} (3 + \sqrt{6})=0.908248$
and $\rho=1-\rho'_c$.
One observes several extrema for a window
of $\rho$, see Fig.~\ref{fig:kappa-4-stat}.

\end{itemize}
\begin{remark}
For $\rho_1=\rho_2=\rho=1/2$ we conjecture that the cumulants of $J=J(0,T)$ are simply polynomials in $\nu$.
Our results are consistent with $\kappa_{n}$ being a polynomial of degree $n-1$ in $\nu$,
for $n \geq 2$ (even in $\nu$ for $n$ even and odd in $\nu$ for $n$ odd).
\end{remark}
\begin{remark}
In this section we have first obtained the cumulants for $\rho_1 \neq \rho_2$ and on their expressions
we have performed the limit $\rho_1\to \rho_2=\rho$. When performing these limits
we have noted that as $\alpha \to 1$ all divergent parts of the rational functions cancel with the divergent
part of the second term in the cumulant coming for the $r_k$. The $r_k$ have an expansion $r_k = 1/\eta^{k-1} + \dots$ with $\eta=1-\alpha$.
To obtain $\tilde \kappa_2$ one only needs $r_2$ to order $\mathcal{O}(1)$. To obtain $\tilde \kappa_3$ one needs $r_2^2$ and $r_3$ to order $\mathcal{O}(1)$
which means one needs $r_2$ to order $\mathcal{O}(\eta)$. To obtain $\tilde \kappa_4$ one needs $r_2^3$ and $r_3 r_2$ and $r_4$ to order $\mathcal{O}(1)$
which means one needs $r_2$ to order $\mathcal{O}(\eta^2)$ and $r_3$ to order $\mathcal{O}(\eta)$ and so on.
In the next Section we establish a different method which addresses directly the case $\rho_1=\rho_2$
\end{remark}
\section{Stationary limit $\rho_1=\rho_2$} \label{sec:stationary}
In this Section we address directly the stationary limit $\rho_1=\rho_2$. In that limit however
the rate function $\Psi(u)$
is not analytic in $u$ anymore around $u=0$, but admits instead an expansion of powers of $\sqrt{u}$. To see that, let us set formally $\alpha=1$. One has
\begin{equation}
\omega= \omega_{uz} = \frac{u z +\sqrt{u^2 z^2+4 u z}}{2 u
z} = \frac{2}{ -u z+\sqrt{ u^2 z^2 +4 u z}} \underset{u \to 0}{\simeq} \frac{1}{\sqrt{u z} }
\end{equation}

A consequence of this is that from \eqref{eqslegendre1} we have
\be
\Psi'(u)= \frac{1}{2 \nu} \frac{\log (1+u \omega_{uz} z)}{u}  \underset{u \to 0}{\simeq} \frac{1}{2 \nu} \sqrt{\frac{z}{u}}
\ee
where $z=z_u$ has a finite value for $u=0$. Hence we see that $u \Psi'(u)$ now
starts as $\sqrt{u}$. Note that a similar behavior in $\sqrt{u}$ at small $u$ was
obtained for the rate function of the weak noise theory stationary KPZ \cite{krajenbrink2017exact}, so it
is not surprising that it appears here.

\subsection{Variational problem}

We study here the variational problem \eqref{PsiMin}. We have that in the limit $\alpha \to 1$
\begin{equation}
\begin{split}
F(\omega)-\mathrm{Li}_2(-u \omega z) &= \mathrm{Li}_2\left(\frac{\sqrt{(uz)^2+4 uz}-uz}{2}\right)-\mathrm{Li}_2\left(-\frac{\sqrt{(uz)^2+4 uz}+uz}{2}\right)\\
& \underset{u \to 0}{=} 2 \sqrt{uz}-\frac{(uz)^{3/2}}{36}+\frac{3 (uz)^{5/2}}{1600}+\mathcal{O}\left(u^{7/2}\right)\\
& \underset{u \to +\infty}{=}\frac{\log (uz)^2}{2}+\frac{\pi^2}{3}+\mathcal{O}\left(\frac{1}{u}\right)
\end{split}
\end{equation}

The variational problem then formally becomes, recalling that $z=e^{-2 \nu J}$,
\be
\Psi(u) = \min_{J} \left[ \Phi(J) +  \frac{1}{2 \nu} \left[\mathrm{Li}_2\left(\frac{\sqrt{(uz)^2+4 uz}-uz}{2}\right)-\mathrm{Li}_2\left(-\frac{\sqrt{(uz)^2+4 uz}+uz}{2}\right) \right]  \right]   \label{PsiMinStat}
\ee

\subsection{Large deviation function $\psi(v)$}

Consider the case $\rho_1=\rho_2$.
Let us write $\Psi(u)=\psi(v)$ with $v=\sqrt{u}$ and explore the consequences
for the cumulant generating function $\phi(P)$. If $\psi(v)$ starts as $v$ at small $v$ then upon inversion
\be
P =  2 \nu u \Psi'(u) = \nu v \psi'(v) \quad  \Leftrightarrow \quad v=v_P
\ee
one has that $v_P \simeq P/(\nu \psi'(0))$ is linear in $P$ hence one can take the limit in
\be
\phi'(P) = - \frac{1}{2 \nu}  \log\left(\frac{(1-e^{-P})(e^P-\alpha)}{v_P^2} \right) \to
- \frac{1}{2 \nu}  \log\left(\frac{(1-e^{-P})(e^P-1)}{v_P^2} \right)   \label{PhiPv}
\ee
which will remain finite with a finite series in $P^n$ as we want to obtain well defined cumulants. We can {\it conjecture} that this is the correct limit $\rho_1=\rho_2$.
Let us check that it is correct.

Let us write
\be
v \psi'(v) = 2 u \Psi'(u)  =
\frac{1}{\nu} \int_{\I \mathbb{R} + \delta} \frac{\rmd y}{2\I \pi y(1-y)} \log\left(1+ v^2 \frac{\rho _1 \left(1-\rho _2\right) (1-y) y}{\left(\rho_1-y \right)
\left(y-\rho_2\right)} e^{- 4\nu^2 y(1-y) T + 2\nu y X   } \right)
\ee

Let us take another derivative w.r.t. $v$ and note that here we can set $\rho_1=\rho_2$
\be
\begin{split}
v \psi''(v) + \psi'(v) &=
\frac{1}{\nu} \int_{\rho + \I \mathbb{R}  } \frac{\rmd y}{2\I \pi } \frac{2 v}{y(1-y) v^2 - \frac{(y- \rho)^2}{\rho (1-\rho) } e^{ 4\nu^2 y(1-y) T - 2\nu y X }  } \\
& = \frac{1}{\nu} \int_{\mathbb{R}  } \frac{\rmd k}{2 \pi }
\frac{2 v}{(\rho+ \I k)(1-\rho-\I k) v^2 + \frac{k^2}{\rho (1-\rho) }
e^{ 4\nu^2 (\rho+ \I k)(1-\rho-\I k) T - 2\nu (\rho+\I k) X } } \label{sh9}
\end{split}
\ee
where we have set $y = \rho + \I k$. It is easy to see that $\psi'(0)$ is finite.
Indeed using
\be
\lim_{v \to 0^+} \frac{v}{a v^2 + b k^2} \to \frac{\pi}{\sqrt{a b}} \delta(k)
\ee
and taking $a=\rho(1-\rho)$ and $b= \frac{1}{\rho (1-\rho) }
e^{ 4\nu^2 \rho (1-\rho) T - 2 \nu \rho X }$ we obtain
\bea
\psi'(0) = \frac{1}{2 \pi \nu} \frac{2 \pi}{\sqrt{a b}}  = \frac{1}{ \nu} e^{ - 2 \nu^2 \rho (1-\rho) T + \nu \rho X } \label{psifirst}
\eea
which incidentally satisfies the heat equation
$\partial_T = 2 \partial_X^2 - 2 \nu \partial_X$,
corresponding to formally setting $n=1/2$ in \eqref{heat2}.
From $\psi'(0)$ we can recover the first moment of $J$.
Using \eqref{PhiPv} and $v_P \simeq P/(\nu \psi'(0))$ we obtain
\be
\langle J \rangle = \phi'(P)|_{P=0}  =
- \frac{1}{ \nu} \log \nu \psi'(0)
=  2 \nu \rho(1-\rho) T - \rho X
\ee
which is the correct result \eqref{averageJ}.

We will compute explicitly higher derivatives of $\psi(v)$ below and
obtain some cumulants of $J$. Before that, let us give the general relation
between cumulants and derivatives of $\psi(v)$.

\subsection{General relation between the cumulants $\kappa_n$ and the derivatives $\psi^{(n)}(0)$}

The cumulants can be obtained using the formula \eqref{cumfromderPsi}, which becomes
\be
\label{eq:supp-mat-cumulant-stat}
\kappa_n =   - \frac{1}{2\nu} \left[\left(\frac{1}{\nu \p_v( v \psi'(v)) }\p_v\right)^{n-1}
\log \left( \frac{(1- e^{-  \nu v \psi'(v)}) (e^{  \nu v \psi'(v)}-1)}{v^2} \right)\right]_{v=0}
\ee

As in the non-stationary initial condition, we can slightly simplify \eqref{eq:supp-mat-cumulant-stat} as follows.
\be
\begin{split}
\kappa_n &=   - \frac{1}{2\nu} \left[\p_P^{n-1}
\log \left( \frac{(1- e^{-  P}) (e^{  P}-1)}{P^2} \right)\right]_{P=0}    - \frac{1}{\nu^n} \left[\left(\frac{1}{ \p_v( v \psi'(v)) }\p_v\right)^{n-1} \log(\nu\psi'(v)) \right]_{v=0}
\end{split}
\ee

We now can use the Taylor expansion
\begin{equation}
\frac{1}{2} \log \left( \frac{(1- e^{-  P}) (e^{  P}-1)}{P^2} \right) =   \log \left(\frac{\sinh(P/2)}{P/2} \right) =\sum_{n\geq 1}\frac{B_{2n}}{(2n)! 2n}P^{2n}
\end{equation}
to evaluate explicitly the first part of the $n$-th cumulant. One finds
\be \label{kappangen}
\kappa_n = -  \frac{1}{\nu} \frac{B_{n-1}}{n-1} \delta_{n \geq 3,{\rm odd}}
- \frac{1}{\nu^n} \left[\left(\frac{1}{\p_v( v \psi'(v)) }\p_v\right)^{n-1} \log(\nu\psi'(v)) \right]_{v=0}
\ee
and the lowest order formula
\bea
\label{dertocumpsi}
&&  \kappa_1= -  \frac{1}{\nu} \log (\psi'(0)  \nu )
\quad , \quad
\kappa_2 = -\frac{R_2}{\nu^2} \quad , \quad R_n = \frac{\psi^{(n)}(0)}{\psi'(0)^n} \\
&&  \kappa_3 = -\frac{\nu^2-36 R_2^2+12 R_3}{12 \nu^3}
= - \frac{1}{12 \nu} + \frac{1}{\nu^3} (3 R_2^2 - R_3)
\quad , \quad   \kappa_4 = -\frac{20 R_2^3-12 R_3 R_2+R_4}{\nu ^4} \nn
\eea
Note that these combinations are simpler than those appearing in the first method, and they can be computed by recursion: at each order $n$ the higher cumulants can be expressed
as a combination of the $n$-th derivative $R_n$ and some combination
of the (already computed) lower order cumulants, e.g., one has
\be  \label{cumiter}
\kappa_3 = -  \frac{R_3}{\nu^3}  - \frac{1}{12 \nu} + 3 \nu \kappa_2^2
\quad , \quad \kappa_4 = -  \frac{R_4}{\nu^4} + \kappa_2( 1 + 12 \nu \kappa_3 - 16 \nu^2 \kappa_2^3)
\ee

\subsection{Calculation of the derivatives $\psi^{(n)}(0)$ and results for the cumulants}
Now we need to compute the derivatives $\psi^{(n)}(0)$. We will attempt to extend the method used in \cite[Section 3.2]{krajenbrink2017exact}
for the stationary KPZ equation. We return to
\eqref{sh9} and
define a change of integration variable from $k$ to ${\sf z}$

where
\be \label{zk}
{\sf z}= {\sf z}(k) =
\frac{k}{\sqrt{\rho (1-\rho)(\rho+ \I k)(1-\rho-\I k)} }
e^{ 2 \nu^2 (\rho+ \I k)(1-\rho-\I k) T - \nu X (\rho + \I k) }
\ee
and rewrite \eqref{sh9} as
\be
\begin{split}
& v \psi''(v) + \psi'(v) =
\frac{1}{\nu} \int_{\Gamma  } \frac{\rmd {\sf z}}{ \pi } \frac{\rmd k}{\rmd {\sf z}}\frac{1}{(\rho+ \I k({\sf z}))(1-\rho-\I k({\sf z}))}
\frac{v}{ v^2 + {\sf z}^2  }\\
\end{split} \label{newpsi}
\ee
Then in the limit $v=0$, we find, using $\frac{v}{v^2 + {\sf z}^2} \to \pi \delta({\sf z})$
\be
\psi'(0) = \frac{1}{\nu } \left(\frac{\rmd k}{\rmd {\sf z}}\frac{1}{(\rho+ \I k({\sf z}))(1-\rho-\I k({\sf z}))}\right)|_{{\sf z}=0}\\
=
\frac{e^{-2\nu^2\rho(1-\rho) T + \nu \rho X }}{\nu}  \label{psip0}
\ee
which agrees with \eqref{psifirst}. Recall that \cite[Eq.~(65)]{krajenbrink2017exact} for any integer $q \geq 1$
\be \label{identitiesder}
\partial_v^{2 q} \frac{v}{ v^2 + {\sf z}^2  } = (-1)^q \partial_{\sf z}^{2 q} \frac{v}{ v^2 + {\sf z}^2  }
\quad , \quad \partial_v^{2 q-1} \frac{v}{ v^2 + {\sf z}^2  } = (-1)^q \partial_{\sf z}^{2 q-1} \frac{{\sf z}}{ v^2 + {\sf z}^2  }
\ee
and we will also use
\bea
&& \partial_v^{2 q} (v \psi''(v) + \psi'(v) )|_{v=0} = \partial_v^{2 q}  \partial_v (v \psi'(v))|_{v=0} = (2 q+1) \psi^{(2q+1)}(0) \\
&& \partial_v^{2 q-1} (v \psi''(v) + \psi'(v) )|_{v=0} = \partial_v^{2 q-1}  \partial_v (v \psi'(v))|_{v=0} = (2 q) \psi^{(2q)}(0)
\eea

\subsubsection{Odd derivatives}
Let us apply $\partial_v^{2 q}$ to \eqref{newpsi} and set $v \to 0^+$. It leads to
\bea
(2 q+1) \psi^{(2q+1 )}(0) =  \frac{(-1)^q}{\nu}  \bigg[ \partial_{{\sf z}}^{2 q}  \left(\frac{\rmd k}{\rmd {\sf z}}\frac{1}{(\rho+ \I k({\sf z}))(1-\rho-\I k({\sf z}))}\right) \bigg]_{{\sf z}=0}
\eea
where we used \eqref{identitiesder}, performed $2 q$ integrations by part
and sent $v \to 0^+$. This leads to the following formula for the odd derivatives as
\be \label{derpsiodd}
\psi^{(2q+1 )}(0) = \frac{(-1)^q}{(2 q+1) \nu } \left[\left( \frac{1}{{\sf z}'(k)} \partial_k \right)^{2 q} \left( \frac{1}{{\sf z}'(k)}
\frac{1}{(\rho+ \I k)(1-\rho-\I k)} \right)\right]_{k=0}
\ee
where ${\sf z}(k)$ is given by \eqref{zk} and $k$ is set to zero at the end of the calculation. We see that this is a pretty straightforward
algebraic formula, and that the result will not contain error functions.

\subsubsection{Even derivatives}

Unfortunately the calculation of the even
derivatives is much more subtle, and more complicated than for
stationary KPZ in \cite{krajenbrink2017exact}. One writes
\begin{equation} \label{longequation}
\begin{split}
\psi^{(2 q)}(0)= \frac{1}{2 q}  \p_v^{2q}(v\psi'(v))|_{v=0}&= \frac{1}{ 2 q \nu} \int_{\Gamma  } \frac{\rmd {\sf z}}{ \pi } \frac{\rmd k}{\rmd {\sf z}}\frac{1}{(\rho+ \I k({\sf z}))(1-\rho-\I k({\sf z}))}
\p_v^{2q-1}\frac{v}{ v^2 + {\sf z}^2  }\\
&= \frac{(-1)^q}{ 2 q \nu} \int_{\Gamma  } \frac{\rmd {\sf z}}{ \pi } \frac{\rmd k}{\rmd {\sf z}}\frac{1}{(\rho+ \I k({\sf z}))(1-\rho-\I k({\sf z}))}
\p_{{\sf z}}^{2q-1}\frac{{\sf z}}{ v^2 + {\sf z}^2  }\\
&= \frac{(-1)^{q-1}}{ 2 q \nu} \int_{\Gamma  } \frac{\rmd {\sf z}}{ \pi } \frac{{\sf z}}{ v^2 + {\sf z}^2  }
\p_{{\sf z}}^{2q-1}\left(\frac{\rmd k}{\rmd {\sf z}}\frac{1}{(\rho+ \I k({\sf z}))(1-\rho-\I k({\sf z}))}\right)
\\
&\underset{v\to 0}{=} \frac{(-1)^{q-1}}{ 2 q \nu} \dashint_{\Gamma  } \frac{\rmd {\sf z}}{ \pi } \frac{1}{{\sf z}}
\p_{{\sf z}}^{2q-1}\left(\frac{\rmd k}{\rmd {\sf z}}\frac{1}{(\rho+ \I k({\sf z}))(1-\rho-\I k({\sf z}))}\right)
\\
&=  \frac{(-1)^{q-1}(2q-2)!}{2 q \nu} \dashint_{\Gamma  } \frac{\rmd {\sf z}}{ \pi } \frac{1}{{\sf z}^{2q-1}}
\left[ \p_{\sf z}\left(\frac{\rmd k}{\rmd {\sf z}}\frac{1}{(\rho+ \I k({\sf z}))(1-\rho-\I k({\sf z}))}\right)\right]|_{\rm reg,{\sf z}^{2q-1}}
\end{split}
\end{equation}

where we used the identity in \cite[Eqs.~(67-68)]{krajenbrink2017exact}.
Here we define
\be
g(y)|_{\mathrm{reg},y^{m}} = g(y) - \sum_{p=0}^{m-1} \frac{g^{(p)}(0)}{p!} y^p
\ee
so that the regularized function starts at order $y^{m}$ or higher.
Integrating by part once more we find
\be \label{evender1}
\psi^{(2q)}(0)
= \frac{(-1)^{q-1}(2q-1)!}{2q \nu}
\dashint_{\R} \frac{\rmd k}{ \pi }  \left( \frac{1}{{\sf z}'(k) (\rho+ \I k)(1-\rho-\I k)}\right)|_{\rm reg, k^{2q}}
\frac{{\sf z}'(k)}{{\sf z}(k)^{2q}}
\ee
where we have used that if ${\sf z}(k) \sim k$ at small $k$ then
$A({\sf z})_{\rm reg, {\sf z}^{n}} = A({\sf z}(k))_{\rm reg, k^{n}}$. We now compute explicitly some of these derivatives and
obtain the associated cumulants.

\subsubsection{Calculation for $q=1$: second and third cumulant $\kappa_2$, $\kappa_3$}

For $q=1$ one needs to compute the integral
\bea
&& \dashint_{\R} \rmd k  \left( \frac{1}{{\sf z}'(k) (\rho+ \I k)(1-\rho-\I k)}\right)|_{\rm reg,k^2}
\frac{{\sf z}'(k)}{{\sf z}(k)^{2}} \\
&& =  \dashint_{\R} \rmd k  \bigg(
\frac{(\rho (1-\rho))  }{k^{2}}
e^{ - 4  \nu^2 (\rho+ \I k)(1-\rho-\I k) T + 2  \nu X (\rho+ \I k)  }  +
e^{ - 2  \nu^2 \rho (1-\rho) T +   \nu X \rho   } \partial_k \frac{1}{{\sf z}(k) } \bigg)
\eea
One can check that the counterterm (second term) cancels the $1/k^2$ divergence at $k=0$
but for generic values of the parameters the principal part remains necessary to
regularize the $\mathcal{O}(1/k)$ term. We have checked numerically that this integral
gives the same result as the fourth line in \eqref{longequation}.
Note that for $T=0$ the integrand oscillates and does not decay to zero at infinity because of
the last term, hence $T>0$ is needed for convergence. An important property is that
we can
check that this integral obeys the heat equation
$(\partial_T - ( \partial_X^2 - 2 \nu \partial_X) ) \psi''(0)=0$,
corresponding to formally setting $n=1$ in \eqref{heat2}.\\

The analytical calculation of the above integral for $T>0$ is not easy. So we will resort to a
trick by evaluating it at $T=0^+$ and then propagating it using the heat equation.
We first note that the first term can be integrated at $T=0$ so that the total integral
can be written as $\dashint_{\R} \rmd k  \partial_k f(k)$ with
\be
f(k) =  f_1(k) + f_2(k) \quad , \quad
f_1(k) = \frac{(\rho -1) \rho  e^{2 \nu X (\rho +\I k)}}{k}-2 \I \nu (\rho -1) \rho  X e^{2 \nu\rho  X} \text{Ei}(2 \I k X\nu ) \quad , \quad  f_2(k)=
e^{  \nu X \rho   }  \frac{1}{{\sf z}(k) }
\ee
Plotting $f(k)$ we see that it does not diverge at $k=0$
but that its imaginary part has a $\log$ divergence
and its real part has a jump at $k=0$. Hence we must evaluate the principal part with care and the total
integral is thus
\be
\dashint_{\R} \rmd k  \partial_k f(k) =
f(+\infty) - f(0^+) + f(0^-) - f(-\infty)
\ee
The contribution at infinity can be evaluated from the first term only
\be
f(+\infty) - f(-\infty) = f_1(+\infty) - f_1(-\infty) = - 4 \pi \nu (1-\rho) \rho   e^{2 \nu\rho  X}  |X|  \quad \text{with} \; T=0^+ 
\ee
using
\begin{equation}
\lim_{k\to \pm \infty} \mathrm{Ei}(\I X k ) = \I \pi \, \mathrm{sign}(k X)  \quad , \quad
(\lim_{k\to + \infty}-\lim_{k\to - \infty}) \text{Ei}(\I X k ) = 2 \I \pi  \mathrm{sign}(X)
\end{equation}
Indeed we assume $T=0^+$, i.e., a slightly positive $T$,
so that the second term does not contribute (since it contains a $e^{- k^2 T}$ factor).
Next we obtain the jump at $k=0$
\be  \label{jump}
f(0^+) - f(0^-)= - 2 \pi \nu (1-\rho) \rho   e^{2 \nu\rho  X}  |X|
\ee
which we note is $1/2$ of the jump at infinity.
To obtain the jump of $f$ we cannot simply set $f=f_1$ because it diverges at $k=0$, but
we can add to it the divergent part of $f_2$, i.e., consider the jump of
$f_1(k) + \rho(1-\rho) e^{2 \nu \rho X}/k$. We have further checked \eqref{jump}
numerically. We also find that the log divergence of the imaginary part cancels out.
In total we thus find
\bea
&& \dashint_{\R} \rmd k  \partial_k f(k) \underset{T=0^+}{=}
- 2 \pi \nu (1-\rho) \rho   e^{2 \nu\rho  X}  |X| \\
&& \psi^{(2)}(0) \underset{T=0}{=} \frac{1}{2 \pi \nu} \dashint_{\R} \rmd k  \partial_k f(k) |_{T=0^+}
= - (1-\rho) \rho   e^{2 \nu\rho  X}  |X|
\eea

One finally obtains (by propagating from $T=0$ using the heat kernel $G_n(X,T)= \sqrt{\frac{n}{4\pi T}} e^{-n \frac{(X-2\nu T)^2}{4T}}$
associated to \eqref{heat2}, setting $n=1$, as we do in \eqref{propagation1})
\be
\psi^{(2)}(0)=
(\rho -1) \rho
e^{-\frac{(X-2 \nu  T)^2}{4 T}}
\left( e^{\frac{(2 \nu
(2 \rho -1) T+X)^2}{4 T}} (2 \nu
(2 \rho -1) T+X)
\text{Erf}\left(\frac{2 \nu  (2
\rho -1) T+X}{2
\sqrt{T}}\right)+ \frac{2
\sqrt{T}}{\sqrt{\pi }} \right) \nn
\ee
Using $\psi'(0)  = e^{-2\nu^2\rho(1-\rho) T + \nu \rho X }/\nu$ from
\eqref{psip0} and $ \kappa_2 = -\frac{r_2}{\nu^2} = - \psi''(0)/( \nu^2 \psi'(0)^2)$
from \eqref{dertocumpsi} one obtains exactly the result \ref{secondcumappendix}
for the second cumulant $\kappa_2$ derived by the first method. We
recall that it can be written in a compact form for any $X,T$, as
given in the main text
\be
\kappa_2 = \kappa_2(X,T)=\sqrt{T} \sigma(\rho) {\cal G}(y) , \quad y = - \frac{\nu (2 \rho-1) T + X/2}{\sqrt{T}}
\ee
with $\sigma(\rho)=2 \rho(1-\rho)$ and ${\cal G}(y) =  y {\rm Erf}(y) + \frac{1}{\sqrt{\pi}} e^{-y^2}$
is an even function. The nice feature of the present calculation is that once the even cumulant
is known the next odd one is immediate to compute. Using the formula \eqref{derpsiodd}
for $\psi^{2q+1}(0)$ for $q=1$ we easily obtain the third cumulant in a compact form
for arbitrary $X,T$
\be
\kappa_3 = \kappa_3(X,T) = \frac{\nu T}{4} \sigma(\rho)^2  \sigma''(\rho) (1 + 3 (y^2 - {\cal G}(y)^2))
+ \frac{1}{2} \sqrt{T} y \sigma(\rho) \sigma'(\rho)
\ee
and one can check that for $X=0$ and $T=1$ it is equivalent to the formula
\eqref{secondthirdaurelien}. In the limit $\nu \to 0$ it gives (setting $T=1$)
\be
\kappa_3 \underset{\nu \to 0}{=} \kappa_3(X)= \rho (1-\rho) (2 \rho-1) X
\ee
which, one can check, agrees with the result for the SSEP extracted from
\eqref{resssep} and the first line of
\eqref{Psi0Mallick} (in agreement with \cite[Eq.~(6.35)]{mallick2024exact}).

\subsubsection{Calculation for $q=2$: fourth and fifth cumulant $\kappa_4$, $\kappa_5$}

For $q=2$, from \eqref{evender1} we need to compute
\bea \label{psi400}
&& \psi^{(4)}(0)  = - \frac{3!}{4 \nu}
\dashint_{\R} \frac{\rmd k}{ \pi } \bigg(
\frac{1}{ (\rho+ \I k)(1-\rho-\I k) {\sf z}(k)^4}  - A_0 \frac{{\sf z}'(k)}{{\sf z}(k)^{4}} - A_1 \frac{{\sf z}'(k)}{{\sf z}(k)^{3}} - A_2  \frac{{\sf z}'(k)}{{\sf z}(k)^2}  \bigg)
\\
&& = - \frac{3!}{4 \nu \pi }  \dashint_{\R} {\rmd k} \partial_k f(k)
\eea
where
\be
\frac{1}{{\sf z}'(k)  (\rho+ \I k)(1-\rho-\I k) } = A_0 + A_1 {\sf z}(k) + A_2 {\sf z}(k)^2  + o(k^2)
\ee
One finds at $T=0$
\be
A_0 = e^{\nu
\rho  X}  ~,~ A_1=  -2 \I
\nu  (\rho -1) \rho  X e^{2
\nu  \rho  X}
~,~ A_2=
-\frac{1}{8} e^{3 \nu  \rho
X} \left(36 \nu ^2 (\rho
-1)^2 \rho ^2 X^2+12 \nu
\left(2 \rho ^2-3 \rho
+1\right) \rho
X-1\right)
\ee
One can also obtain these coefficients at $T>0$ (not shown).
One can check using Mathematica that the terms $1/k^4, 1/k^3, 1/k^2$ near $k=0$ do cancel in the
integrand at $T=0$ and also at $T>0$, so that the integral is well defined.

One can rewrite the first term in the integrand in \eqref{psi400} as
\be
\frac{1}{ (\rho+ \I k)(1-\rho-\I k) {\sf z}(k)^4} = \frac{1}{k^4} \Delta_4(k) \quad , \quad \Delta_4(k) =
(\rho -1)^2 \rho ^2 (\rho+ \I k)(1-\rho-\I k)
e^{4  \nu  X (\rho +\I k)}
\ee
At $T=0$ this term can be integrated, and the primitive $f_1(k)$ has the same structure as for $q=1$,
with one ${\rm Ei}$ function. It is then easy to extract the jump at infinity.
One finds that it can be written as
\bea
f_1(+\infty) - f_1(-\infty) = \frac{\I \pi}{3}  {\rm sgn}(X) \, \left[\partial_k^3 \Delta_4(k)\right]_{k=0}
\eea
which we double checked with Mathematica.
Returning to $q=1$ we see that in that case $f_1(+\infty) - f_1(-\infty) = 2 \I \pi {\rm sgn}(X) \, \partial_k|_{k=0} \Delta_2(k)$.
Hence we conjecture that for general $q$
\bea \label{conjecturef1}
f_1(+\infty) - f_1(-\infty) = \frac{2\I \pi}{(2q-1)!}  {\rm sgn}(X) \, \left[\partial_k^{2q-1} \Delta_{2q}(k) \right]_{k=0}
\eea
The primitive of the second term is
\bea
f_2(k)=  \frac{1}{3} A_0 \frac{1}{{\sf z}(k)^{3}} +\frac{1}{2} A'_1 \frac{1}{{\sf z}(k)^{2}} + A'_2  \frac{1}{{\sf z}(k)}
\eea
By adding its singular part at $k=0$ to $f_1(k)$ we can compute with Mathematica
the jump of $f(k)$ at $k=0$ and, remarkably, we find
\be \label{relationff1}
f(0^+)-f(0^-) =  \frac{1}{2} (f_1(+\infty) - f_1(-\infty) )
\ee
i.e., there is the same factor of $1/2$ as noted for $q=1$. Hence we find
\bea
\psi^{(4)}(0)  &\underset{T=0}{=}& - \frac{3!}{4 \nu \pi }  \dashint_{\R} {\rmd k} \partial_k f(k)
= - \frac{3!}{8 \nu \pi }  (f_1(+\infty) - f_1(-\infty) ) \\
&=& - \frac{\I}{4 \nu  }   {\rm sgn}(X) \, \left[\partial_k^3 \Delta_4(k)\right]_{k=0} \\
& = & 2 (\rho -1)^2 \rho ^2 |X|
e^{4 \nu
\rho  X} (2 \nu  X (6 \rho
+4 \nu  (\rho -1) \rho
X-3)+3)
\eea
To obtain $\psi^{(4)}(0)$ for general $T$ we simply propagate its value at $T=0$ as we did in \eqref{propagation1} using
the heat kernel $G_n(X,T)= \sqrt{\frac{n}{4\pi T}} e^{-n \frac{(X-2\nu T)^2}{4T}}$
setting $n=2$. From \eqref{cumiter} we then obtain the fourth cumulant for arbitrary $X,T$ in the form
\be
\begin{split}
\kappa_4 =  &\kappa_2( 1 + 12 \nu \kappa_3 - 16 \nu^2 \kappa_2^3) +
\frac{1}{2} \sqrt{T} \sigma
(\rho )^2
\bigg(\sqrt{\frac{2}{\pi }}
e^{-2 y^2} \left(8 \nu ^2 T
\left(2 y^2+1\right) \sigma
(\rho )+12 \nu  (2 \rho -1)
\sqrt{T} y-3\right)  \\
& -  2
\text{Erf}\left(\sqrt{2}
y\right) \left(-4 \nu ^2 T
y \left(4 y^2+3\right)
\sigma (\rho )-3 \nu  (2
\rho -1) \sqrt{T} \left(4
y^2+1\right)+3 y\right)\bigg)
\end{split}
\ee
where $y = - \frac{\nu (2 \rho-1) T + X/2}{\sqrt{T}} $ and $\kappa_2$, $\kappa_3$, $\sigma$ and ${\cal G}$
have been given above. In the limit $\nu \to 0$ it becomes
\be
\kappa_4 \underset{\nu =0}{=} 2 \rho (1-\rho)   \left(3
\sqrt{2} (\rho -1) \rho \,
{\cal G}\left(\sqrt{2}
y\right)+ {\cal G}(y)\right)  \quad , \quad y = - \frac{X}{2}
\ee
which, one can check, agrees with the result for the SSEP extracted from
\eqref{resssep} and the first line of
\eqref{Psi0Mallick} (in agreement with \cite[Eq.~(6.35)]{mallick2024exact}).

\subsubsection{A formula for the general stationary cumulant $\kappa_n$ }

Conjecturing the formula for the even derivatives of $\psi(v)$ leads to
a formula for the general cumulant for $\rho_1=\rho_2$. From
the conjecture \eqref{conjecturef1} and assuming
 \eqref{relationff1} holds for any $q$,
 we obtain
\bea \label{conjecturef1-2}
\psi^{(2q)}(0) \underset{T=0}{=} \frac{\I (-1)^{q-1}(\rho (1- \rho))^q}{2 q \nu}  \, \left[\partial_k^{2q-1}
((\rho + \I k) (1- \rho - \I k))^{q-1} e^{2 q \nu X (\rho + \I k) } {\rm sgn}(X)
\right]_{k=0}
\eea

We can commute the evolution by the heat kernel and the derivatives so we obtain, setting $p = \I k$
\be \label{conjecturef1-3}
\psi^{(2q)}(0) = - \frac{(\rho (1- \rho))^q}{2 q \nu}
\left[\partial_p^{2q-1} \left(
((\rho + p) (1- \rho - p))^{q-1}
e^{- 4 q \nu \sqrt{T} (y (\rho+ p) + \nu \sqrt{T} (\rho^2 - p^2) ) }
{\rm Erf}(\sqrt{q} (2 p \nu \sqrt{T} - y) ) \right)
\right]_{p=0}
\ee
with $y = - \frac{\nu (2 \rho-1) T + X/2}{\sqrt{T}} $.\\

The odd derivatives are given by \eqref{derpsiodd}, which setting $p= \I k$ can be rewritten
as
\bea  \label{derpsiodd2}
&& \psi^{(2q+1 )}(0) = \frac{1}{(2 q+1) \nu } \left[\left( \frac{1}{\hat {\sf z}'(p)} \partial_p \right)^{2 q} \left( \frac{1}{\hat {\sf z}'(p)}
\frac{1}{(\rho+ p)(1-\rho-p)} \right)\right]_{p=0}
\\
&&
\hat {\sf z}(p)
= \frac{p e^{2 \nu  \sqrt{T}
(\rho+p ) \left(\nu
\sqrt{T} (\rho-p)+y\right)}}{\sqrt{\rho (\rho-1) (\rho+p )  (\rho+p -1)}}
\eea
An equivalent way to compute it is to expand $\hat {\sf z}$ in series of $p$
and inverse the series to get $p$ in series of $\hat {\sf z}$, and compute
\bea
&& \psi^{(2q+1 )}(0) = \frac{1}{(2 q+1) \nu }
\left[\left( \frac{d}{d \hat {\sf z}}  \right)^{2 q} \left( \frac{dp}{d \hat {\sf z}}
\frac{1}{(\rho+ p(\hat{{\sf z}})(1-\rho-p(\hat{{\sf z}}))} \right)\right]_{\hat {\sf z}=0}
\eea

These formula allow to obtain the higher cumulants using
\eqref{kappangen} and \eqref{dertocumpsi}. We have obtained
for instance the fifth and sixth cumulants, which we give below
in some special case. We have checked that they reduce for $\nu=0$
to their SSEP limit which we also computed from
\eqref{resssep} and the first line of
\eqref{Psi0Mallick} (in agreement with \cite[Eq.~(6.35)]{mallick2024exact}).

\paragraph{Special case.}
In the special case $\rho=1/2$, $X=0$, $T=1$, which implies $y=0$, we find for the even and odd derivatives
simplify a little as
\bea \label{evenodd}
&&  \psi^{(2q)}(0) = - \frac{1}{2^{2 q+1} q \nu}  e^{- \nu^2 q}
\left[\partial_p^{2q-1} \left(
( \frac{1}{4} - p^2)^{q-1} e^{4 \nu^2 p^2 q} {\rm Erf}(2 \nu p \sqrt{q}) \right)
\right]_{p=0}  \\
&& \psi^{(2q+1 )}(0) = \frac{1}{(2 q+1) \nu }
\left[\left( \frac{d}{d \hat {\sf z}}  \right)^{2 q} \left( \frac{dp}{d \hat {\sf z}}
\frac{1}{\frac{1}{4} - p^2} \right)\right]_{\hat {\sf z}=0}  \quad , \quad \hat {\sf z} =
\frac{2p e^{ 2 \nu^2 (\frac{1}{4}-p^2)   }}{\sqrt{(\frac{1}{4} - p^2)} }
\eea
As a result all the cumulants are polynomials in $\nu$. We give here the
explicit forms
\bea
&& \kappa_5 = \frac{5 \left(21+2 \left(4
\sqrt{2}-9\right) \pi
\right) \nu ^3+2 \pi
\left(15-15 \sqrt{2}+2 \pi
\right) \nu }{8 \pi ^2} \\
&& \kappa_6 = \frac{1}{8
\pi ^{5/2}} \bigg( \left(4-15 \sqrt{2}+10
\sqrt{3}\right) \pi ^2
+5 \pi
\left(56-63
\sqrt{2}+\left(6+13
\sqrt{2}-8 \sqrt{3}\right)
\pi \right) \nu
^2 \\
&& +    \left(756+420
\left(\sqrt{2}-2\right) \pi
+\left(45-60 \sqrt{2}+24
\sqrt{3}\right) \pi
^2\right) \nu ^4 \bigg)
\eea

\paragraph{KPZ limit.} Consider now the further limit to stationary KPZ where
in addition we take $\nu \to +\infty$. From Eqs.~\eqref{rescu}, \eqref{PsiKPZgeneral} we expect that in that limit
\be
\psi(v) \to  \frac{1}{\nu^2} \psi_{\rm KPZ}(\tilde v) \quad , \quad  v=\frac{2\tilde{v}}{\nu}e^{\nu^2/2}
\ee
hence we expect that in that limit
\be \label{limiteven}
\psi^{(2q)}(0) \underset{\nu \to \infty}\simeq \frac{1}{\nu^2} \left(\frac{\nu}{2}e^{-\nu^2/2}\right)^{2q} \times
\psi_{\rm KPZ}^{(2q)}(0) \quad , \quad \psi_{\rm KPZ}^{(2q)}(0)  =
\frac{2 (-1)^q q^{q-\frac{3}{2}} \Gamma
\left(\frac{3}{2}-q\right) \Gamma (2
q-1)}{\pi }
\ee
where the values of $\psi_{\rm KPZ}^{(2q)}(0)$ were obtained in \cite[Eq.~(69)]{krajenbrink2017exact}.
To check that our result agrees with that limit, we rescale $p=\tilde{p}/(2\nu)$
in \eqref{evenodd} and, taking $\nu$ large, we replace $1/4 -p^2 \to 1/4$ to leading order.
We use that $e^{a^2} {\rm Erf}(a)= \sum_{q' \geq 1} \frac{1}{\Gamma( \frac{1}{2} + q')} a^{2 q'-1}$
and the identity $\frac{(2q-1)!}{\Gamma( \frac{1}{2} + q)} = (-1)^{q-1}
\frac{2}{\pi} \Gamma(\frac{3}{2} - q) \Gamma(2 q-1)$, and we find that
\eqref{limiteven} holds as required. \\

For the odd derivatives, one defines $p=\tilde{p}/\nu$ in \eqref{evenodd} so that
in the large $\nu$ limit one has $\hat {\sf z}=\tilde z/\nu$ with $\tilde z=
 4\tilde{p}e^{ -2 \tilde{p}^2 } $ and one finds
\bea  \label{derpsiodd2-x03}
\psi^{(2q+1 )}(0)&& \underset{\nu \to \infty}{\simeq}  4 \frac{\nu^{2q} e^{-(2q+1)\nu^2/2}}{(2 q+1) \nu }
\left[\left( \frac{d}{d \tilde z}  \right)^{2 q+1} \tilde p \right]_{\tilde z=0}
\eea
The relation between $\tilde{z}$ and $\tilde{p}$ can be explicitly inverted as $\tilde{p}^2=-\frac{1}{4}W(-\tilde{z}^2/4)$, where $W$ represents the main branch of the Lambert function. Using the following series $ \sqrt{W(y^2)/y^2}=\frac{1}{2}\sum_{n=0}^{\infty} \frac{(n+1/2)^{n-1}}{n!}(-y^2)^n$,
we obtain $\tilde p = \frac{1}{8} \sum_{q=0}^{\infty} \frac{(q+1/2)^{n-1}}{4^q q!} z^{2 q+1}$.
This leads to
\be \label{limitodd}
\psi^{(2q+1)}(0) \underset{\nu \to \infty}\simeq \frac{1}{\nu^2} \left(\frac{\nu}{2}e^{-\nu^2/2}\right)^{2q+1} \times
\psi_{\rm KPZ}^{(2q+1)}(0) \quad , \quad \psi_{\rm KPZ}^{(2q+1)}(0)  =
\frac{(2q)!}{q!}(q+\frac{1}{2})^{q-1}
\ee
where the values of $\psi_{\rm KPZ}^{(2q)}(0)$ were obtained in \cite[Eq.~(66)]{krajenbrink2017exact}.
The agreement with the stationary KPZ limit provides a test on our formula for general $q$.

\begin{remark}
The above formula for the cumulants are displayed (in a slightly simpler form) in terms of the $R_n$
in Appendix~\ref{app:statcum}. Note that performing the limit $\nu \to 0$ 
to the SSEP directly on these formula is delicate due to cancellations 
of various orders between the $R_n$ when computing the cumulants. 
The SSEP limit is studied in more details in Section~\ref{sec:ssep}.
\end{remark}

\section{Cumulants of the position of a tracer}

If we denote $M_{\sf t}$ the position of the tracer in ASEP, i.e., a tagged particle of the ASEP, its position in the continuum model is $Y_t = \varepsilon M_{{\sf t}=t/\varepsilon^2}$
where $1/\varepsilon=\sqrt{{\sf T}/T}$. Consider now the position of a tracer $Y_t$ starting at the initial position $Y_0=0$. It is defined by the conservation of particle number to the right of the tracer
\be
\int_{Y_t}^{\infty} dx \rho(x,t) = \int_{Y_0}^{+\infty} \rmd x \rho(x,0) \quad  \Leftrightarrow  \quad
h(Y_t,t)= 0 \quad  \Leftrightarrow  \quad   z(Y_t,t)= 1 \quad \text{for} \,\, Y_0=0
\ee
so that one has, as in Ref.~\cite{imamura2017large}, the equality in probability
\begin{equation}
\mathbb{P}( Y_t < x) = \mathbb{P}(h(x,t) >0)
\end{equation}
(with $N$ there being $h$ here). Note that if $Y_0 \neq 0$ the above condition becomes $h(Y_t,t)= h(Y_0,0)$
and $z(Y_t,t)= e^{2 \nu h(Y_0,0)}$ (we recall that $h(0,0)=0$).
\\

We want to compute the CGF and the PDF of the tracer's position $Y_T$ at time $T$, which take the large deviation forms
\be
\langle e^{ \frac{1}{\varepsilon} \lambda Y_T } \rangle \sim  e^{  \frac{1}{\varepsilon} C(\lambda)}  \quad , \quad P( Y_T=X) \sim e^{- \frac{1}{\varepsilon} \chi(X) }
\ee
where the rate functions are related by $C(\lambda) = \max_{X \in \mathbb{R}} [\lambda X - \chi(X)   ]$.
In most formula below the dependence in $T$ is implicit.
Once $C(\lambda)$
is known, its Taylor expansion coefficients $c_n=c_n(T)$ give the cumulants of the tracer position
\be
C(\lambda) = \sum_{n \geq 1} \frac{1}{n!} c_n(T) \lambda^n  \quad , \quad \langle Y_T^n \rangle^c \simeq \varepsilon^{n-1} c_n(T) \quad , \quad \langle M_{\sf T}^n \rangle^c \simeq c_n(T=1) \sqrt{{\sf T}}
\ee

We can now use our main result for the CGF of $h(X,T)=-J(X,T)$ for
arbitrary value of $X$.
Let us now make the $X$ dependence explicit
and denote $\phi(p)=\phi(p;X)$ and $\Phi(J)=\Phi(J;X)$.
Now, since $J=-h$, we have
\be
P( Y_T = X) = P(h(X,T)=0 ) \sim e^{ - \frac{1}{\varepsilon}  \Phi(0;X) }
\ee
which implies $\chi(X)= \Phi(0;X)$. From Legendre inversion of \eqref{Lagrangephi} one obtains
\be
\Phi(J;X) = \max_P [ P J -  \phi(P;X)   ]
\ee
The value of $J$ which realizes the maximum is thus such that
\be
P =  \partial_J \Phi(J;X) \quad , \quad \partial_P \phi(P;X) = J
\ee
where the derivatives act on the first argument. The constraint that we are following the tracer can be implemented
by setting $J=0$ in the above equations. This gives a relation
between $p$, $u$ and $X$ which we denote $p^*=p^*(X)$, $P^*=P^*(X)$ and $u^*=u^*(X)$.
One then obtains the rate function of the PDF of the tracer as
\bea
&& \chi(X)= \Phi(0;X)
= -  \phi(P^*;X) \\
&&
\p_P  \phi(P;X)|_{P=P*}=0 \label{conditionder}
\eea
where the last equation determines
$P^*=P^*(X)$ if $\phi$ and $\Phi$ are known.
We then obtain the CGF for the tracer's position from
\bea  \label{Cprime}
&& C'(\lambda) =  X \\
&& \lambda = \chi'(X)
=  - \partial_X  \phi(P;X) |_{P=P^*(X)}
\eea
where we have used the condition \eqref{conditionder} to replace $d/dX$ by $\partial_X$.

\subsection{First method using the CGF and cumulants of the current}
In summary, once the CGF and PDF rate functions of the current, $\phi$ and $\Phi$, are known, i.e., once the cumulants $\kappa_n=\kappa_n(X)$
computed in a previous section are known, we can obtain the cumulants of the tracer $c_n$ by eliminating
$P^*$ and $X$ in the system of {\it three} equations
\bea
&& \p_P  \phi(P;X)|_{P=P*} = \kappa_1(X) + \sum_{n \geq 1} \frac{1}{n!} \kappa_{n+1}(X) (P^*)^n = 0 \label{sysser1} \\
&& - \partial_X  \phi(P;X) |_{P=P^*} =  - \sum_{n \geq 1} \frac{1}{n!} (\partial_X \kappa_n(X)) (P^*)^n = \lambda \label{sysser2} \\
&& C'(\lambda) = c_1  +  \sum_{n \geq 1} \frac{1}{n!} c_{n+1} \lambda^n  = X \label{sysser3}
\eea
To understand the structure let us write the lowest order terms. One has
\bea
&& X = c_1 + c_2 \lambda + \mathcal{O}(\lambda^2) \\
&& \lambda = - \kappa_1'(X) P^* + \mathcal{O}((P^*)^2) \\
&& 0 = \kappa_1(X) + \kappa_2(X) P^* + \mathcal{O}((P^*)^2)
\eea
The elimination problem can be solved in a systematic expansion in $\lambda$, $P^*=\mathcal{O}(\lambda)$ and $X-c_1= \mathcal{O}(\lambda)$.
One first obtains that the first moment $c_1$ is the root of the equation
\be
\kappa_1(c_1) = 0
\ee
Next one obtains for the second cumulant
\be c_2 = \frac{\kappa _2}{\left(\kappa
_1'\right){}^2} : = \frac{\kappa
_2\left(c_1\right)}{\kappa
_1'\left(c_1\right){}^2} \label{c2}
\ee
To pursue further, one way is to invert the series \eqref{sysser2} to obtain $P^*$ as a function of $\lambda$, then insert in \eqref{sysser1}
and replace $X$ by its series \eqref{sysser3} and expand all in $\lambda$ to obtain recursive equations
for the $c_n$. The higher cumulants become quickly quite complicated. They can be written in a more economical form by introducing
\be
\tilde \kappa_n(X) := \frac{\kappa_n(X)}{\kappa'_1(c_1)^n}
\ee
In these notations they read
\bea  \label{allcn}
&& c_2 = \tilde \kappa _2 \quad , \quad c_3 =
3 \tilde \kappa _2 \left(\tilde \kappa
_2'-\tilde \kappa _2 \tilde \kappa
_1''\right)-\tilde \kappa
_3 \\
&& c_4 =  6 \tilde \kappa _2
\left(2 \tilde \kappa _3-7 \tilde \kappa
_2 \tilde \kappa _2'\right) \tilde \kappa
_1''+15 \tilde \kappa _2
\left(\tilde \kappa
_2'\right){}^2-4 \tilde \kappa _2
\tilde \kappa _3'-6 \tilde \kappa _3
\tilde \kappa _2'+24 \tilde \kappa _2^3
\left(\tilde \kappa
_1''\right){}^2+6 \tilde \kappa
_2^2 \tilde \kappa _2''-4 \tilde \kappa
_1{}^{(3)} \tilde \kappa
_2^3+\tilde \kappa _4 \nn
\eea
where the derivatives are taken at $X=c_1$, i.e., as in \eqref{c2}.\\

The cumulants of the tracer can thus be obtained from the cumulants of the current at $X,T$
although the combinatorics is not simple. This combinatorics is general and valid for any model,
since it is only a consequence of the definitions of the rate functions.\\

Let us apply it in some simple cases. For the first moment the equation
$\kappa_1(c_1)=0$ reads, more explicitly
\be
e^{2 \nu    \rho _1 (2 \nu
(\rho _1-1)
T+c_1)}
\text{Erfc}( \frac{\nu  (2 \rho_1-1)+\frac{c_1}{2}}{\sqrt{T}} )
+ e^{2 \nu   \rho _2 (2 \nu   (\rho _2-1) T+c_1)}
\text{Erfc}( - \frac{\nu  (2 \rho_1-1)+\frac{c_1}{2}}{\sqrt{T}} )
=2
\ee
In the SSEP limit $\nu \to 0$,
one must cancel the term $\mathcal{O}(\nu)$ and
one obtains the equation
\be
2 \rho_2 \xi = (\rho_1-\rho_2)  f(\xi) \quad , \quad f(\xi)= \int_{\xi}^{+\infty} du {\rm Erfc}(u) =  \frac{1}{\sqrt{\pi}} e^{- \xi^2} - \xi {\rm Erfc}(\xi)
\quad , \quad \xi= \frac{X}{\sqrt{4 T}}
\ee
It agrees with Ref.~\cite[Eq.~(13)]{imamura2017large} (with the matching $\rho_-=\rho_2$
and $\rho_+=\rho_1$).

Let us consider now the higher cumulants and focus on the stationary case $\rho_1=\rho_2$. Then one has from \eqref{averageJ},
i.e., $\kappa_1(X,T)= 2 \nu \rho(1-\rho) T - \rho X$, and we obtain the mean position of the tracer as
\be
\langle Y_T \rangle = c_1= 2 \nu (1-\rho) T \label{c1stat}
\ee
In that case one has $\kappa'_1(c_1)= - \rho$ and from the explicit expression
\eqref{secondcumappendix} for $\tilde \kappa_2(X,T)$
we obtain
\be  \label{c2explicit}
c_2= \frac{1}{\rho^2} \kappa_2(c_1) = \frac{2 (1-\rho)}{\rho}  \sqrt{T}   \left(  \nu  \rho \sqrt{T}
\text{Erf}\left( \nu  \rho \sqrt{T} \right)
+ \frac{  e^{-(  \nu  \rho \sqrt{T})^2}}{\sqrt{\pi }}\right)
\ee
which setting our $T=1$ recovers \cite[Eq.~(S135)]{berlioz2024tracer} (see also \cite{de1985self}).
It is an even function of $\nu$, which behaves as $c_2 \simeq 2 (1- \rho) \nu + O(e^{- \rho^2 \nu^2})$ at large $|\nu|$.
For the third cumulant we find (for $T=1$)
\bea
&& c_3 = \frac{1-\rho}{\rho} \bigg(
-6 \pi  \nu  (\rho -1) (2
\nu ^2 \rho ^2+1)
\text{Erf}(\nu  \rho
)^2-\frac{6 \sqrt{\pi }
(\rho -1) e^{-\nu ^2 \rho
^2} \left(4 \nu ^2 \rho
^2+1\right) \text{Erf}(\nu
\rho )}{\rho } \\
&& +2 \nu
(\pi  (2 \rho
(3 \nu ^2 (\rho -1)
\rho +2)-3)-6
(\rho -1) e^{-2 \nu ^2 \rho
^2}) \bigg)
\eea
which can be checked, agrees with \cite[Eq.~(S136)]{berlioz2024tracer}.
It is an odd function of $\nu$, which behaves as $c_3 \simeq 2 (1- \rho) \nu + O(e^{- \rho^2 \nu^2})$ at large $|\nu|$,
and can be non monotonic as a function of $\nu$ in some range of $\rho$. For the fourth
cumulant we find (setting $T=1$)
{\small
\bea
&& c_4 = \nu  (\rho -1)^2 \rho
\left(80 \nu ^4 \rho ^4+84
\nu ^2 \rho ^2+15\right)
\text{Erf}(\nu  \rho
)^3+\frac{15 (\rho -1)^2
e^{-\nu ^2 \rho ^2}
\left(16 \nu ^4 \rho ^4+12
\nu ^2 \rho ^2+1\right)
\text{Erf}(\nu  \rho
)^2}{\sqrt{\pi }}
\\
&&  +\nu  \rho
\left(\frac{12 (\rho -1)^2
e^{-2 \nu ^2 \rho ^2}
\left(20 \nu ^2 \rho
^2+9\right)}{\pi }+2 \rho
\left(27-2 \rho  \left(3
\nu ^2 (\rho -1) \left(3
\rho  \left(4 \nu ^2 (\rho
-1) \rho
+5\right)-13\right)+8\right
)\right)-21\right)
\text{Erf}(\nu  \rho ) \nn \\
&& +2
\nu  (\rho -1) \rho
\left(\rho  \left(4 \nu ^2
\rho  \left(4 \rho  \left(2
\nu ^2 (\rho -1) \rho
+3\right)-9\right)+9\right)
-3\right)
\text{Erf}\left(\sqrt{2}
\nu  \rho
\right) \\
&& +\frac{e^{-3 \nu ^2
\rho ^2} }{
\pi ^{3/2}}
\bigg(4 (\rho
-1)^2 \left(20 \nu ^2 \rho
^2+3\right)+\sqrt{2} \pi
\rho  (\rho -1) e^{\nu ^2
\rho ^2} \left(4 \nu ^2
\rho  \left(2 \rho  \left(4
\nu ^2 (\rho -1) \rho
+5\right)-7\right)+3\right)
\\
&& -\pi  e^{2 \nu ^2 \rho ^2}
\left(4 \rho  \left(6 \nu
^2 (\rho -1) \rho  \left(6
\rho  \left(\nu ^2 (\rho
-1) \rho
+1\right)-5\right)+2 \rho
-3\right)+3\right)\bigg)
\eea }
This bulky expression simplifies in the limit $\nu=0$ and reproduces the known result for
the SSEP \cite{krapivsky2014large} (see also (9) and Section III.A.2 in
\cite{grabsch2024tracer}). Remarkably, for large $\nu$ it again
simplifies to $c_4 \simeq 2 (1- \rho) \nu T $.
\begin{remark}
It is reasonable to expect that for the WASEP (see the discussion in \cite{berlioz2024tracer})
in the so-called ballistic regime $\nu \gg 1$ with $\rho$ fixed, all cumulants $c_n \simeq 2 (1- \rho) \nu T$, i.e.,
the tracer position becomes a Poisson random variable. Indeed, this was
shown for the ASEP in \cite{Ferrari_Fontes_1996}.
Our results for the first four
cumulants are
compatible this result. In terms of the rate functions defined above it would mean that
in the limit $c_1= 2 (1-\rho) \nu T \gg 1$
\bea
&& \chi(Y) \simeq c_1 \, \chi_{\rm Poisson}(Y/c_1) \quad , \quad C(\lambda) \simeq c_1 (e^{\lambda}-1) \\
&& \chi_{\rm Poisson}(y) = y \log y - y + 1
\eea

where $\chi_{\rm Poisson}$ is the large deviation rate function
of a Poisson random variable of unit mean.
\end{remark}
\begin{remark}
In the case of the step initial condition one finds $c_1=+\infty$. This feature also holds for $\nu=0$, i.e.
for the SSEP, see
Eq.(13) in \cite{imamura2017large}.
It may thus be a degenerate case, as the MFT does not allow to study the edge.
One would need to study the case $Y_0<0$ where the tracer starts in the bulk.
\end{remark}
\subsection{Second method using the rate function $\Psi(u)$}
In the second approach one obtains the tracer
CGF directly from the function $\Psi(u)$ that we have computed.
Setting $J=0$ in \eqref{eqslegendre1} leads to a relation
between $u$ and $X$, i.e., we define $u=u^*(X)$ the solution of
\bea
2\nu u^*(X) (\p_u\Psi)(u,X)|_{u=u^*(X)}
=  \log( 1 + u^*(X) \omega_{u^*(X)}) \label{liste}
\eea

From \eqref{ParamPhi}, setting $J=0$ and
making explicit the $X$ dependence we obtain
\begin{equation}
\chi(X)=   \Phi(0;X)=\Psi(u^*(X),X)+\frac{1}{2\nu}[\mathrm{Li}_2(-u^* \omega_{u^*})-F(\omega_{u^*})]|_{u^*=u^*(X)}
\end{equation}
 Differentiating with respect to $X$, we obtain
\begin{equation}
\begin{split}
\chi'(X) =    \p_X \Phi(0,X) &= u^{*\prime}(X) \left(\p_u \Psi(u^*,X)-\frac{\log(1+\omega_{u^*} u^*)}{2\nu u^*}\right)+(\p_X \Psi)(u^*,X)
= (\p_X \Psi)(u^*,X)
\end{split}
\end{equation}
since the first part is zero by construction. Hence we obtain from \eqref{Cprime}
\begin{equation}
\lambda=  (\p_X \Psi)(u^*(X),X)
\end{equation}

In summary, to obtain the cumulants $c_n$ of the tracer's position we must eliminate $u$ and $X$
in the system of {\it three} equations
\bea
&& \p_u \Psi(u,X) = \sum_{n\geq 0}\frac{ \Psi^{(n+1)}(0,X) }{n!} u^n
= \frac{\log(1+u \omega_{u})}{2\nu u} = \sum_{n \geq 0}  \frac{a_n}{n!}  u^n  \\
&& \p_X \Psi(u,X)= \sum_{n\geq 1}\frac{ \Psi^{(n,1)}(0,X) }{n!} u^n = \lambda \quad , \quad
\Psi^{(n,m)}(u,X) := \partial_u^n \partial_X^m \Psi(u,X) \\
&&  C'(\lambda)= c_1 + \sum_{n \geq 1} \frac{1}{n!} c_{n+1} \lambda^n = X
\eea
where we recall that $\omega_u$ is given in \eqref{omegauz}.
An equivalent expression for the
coefficients $a_n$ can be read from Eqs.~\eqref{smallnu}, \eqref{psidernu0}
\be
a_n = - \frac{\rho _1^{n+1} \left(1-\rho _2\right)^{n+1}}{2\nu (n+1) } \frac{d^{n}}{dy^{n}} \left( \frac{(y (1-y))^{n}}{  (y-\rho_1)^{n+1} } \right)|_{y=\rho_2}
\ee

To order zero we obtain that $c_1$ is the root of the equation
\begin{equation}
\Psi ^{(1,0)}(0;c_1)=\frac{1}{2 \nu(1-\alpha )  }   \quad , \quad \alpha=\frac{\left(1-\rho _1\right) \rho _2}{\rho _1 \left(1-\rho _2\right)}
\end{equation}
which, as expected from \eqref{cum1234} and \eqref{Jav} is equivalent to the condition $\kappa_1(c_1) = 0$ obtained above.

Further expansion can be performed in terms of the derivatives $\Psi^{(n,m)}(0;c_1) $. We obtain for
the second cumulant
\begin{equation} \label{c2second}
c_2=\frac{\frac{\alpha +1}{4 (\alpha -1)^3 \nu } - \Psi ^{(2,0)}}{\left(\Psi ^{(1,1)}\right)^2}
\end{equation}
and more complicated expressions for the higher cumulants.
Using the expressions for the $\Psi^{n}(0;X) $ obtained in Section~\ref{sec:derPsi}
yields the $c_n$ for arbitrary $\rho_1$, $\rho_2$. We have checked that
taking the limit $\rho_2=\rho_1=\rho$ of \eqref{c2second} recovers
the result \eqref{c2explicit} (taking into account the first order correction
of $c_1$ in $\rho_1-\rho_2$).

\section{Macroscopic fluctuation theory approach to the large deviations}

\subsection{Framework and boundary conditions}

The fluctuating hydrodynamics equation for the WASEP read (setting $D_0=1$ and $\sigma(\rho)=2 \rho (1-\rho)$)
\begin{equation}
\partial_t \rho=\partial^2_{x} \rho - \p_x(  2\nu  \rho (1-\rho) + \sqrt{2 \varepsilon \rho (1-\rho) }  \eta)
\end{equation}
In addition we introduce the fields
\be
z(x,t) = e^{2\nu h(x,t)} \quad , \quad  h(x,t)= h(0,t) + \int_0^{x} dy  \rho(y,t)  \quad , \quad h(0,t)=-Q_t = - \int_0^{+\infty} dx (\rho(x,t)-\rho(x,0))
\ee
where $h(x,t)=-J(x,t)$ and
$Q_t$ is the number of particles
which have crossed the origin from left to right minus right to left during time $t$,
and is also the integrated current $Q_t = J(0,T)= \int_0^t dt' j(0,t')$. \\

The expectation value of any observable of the form $\exp( \frac{1}{\varepsilon} {\cal O})$ can be represented as a MSR path integral over the field $\rho$ and the response field
$\tilde \rho/\varepsilon$ as
\be   \label{eq:intro-mft-observable2}
\begin{split}
&  \big\langle e^{\frac{1}{\varepsilon} {\cal O}}  \big\rangle =
\iiint {\cal D} \varrho {\cal D} \tilde{\varrho}  e^{-\frac{1}{\varepsilon}\left[S[\varrho,\tilde{\varrho}]+\mathcal{F}(\varrho)- {\cal O} \right]}
\end{split}
\ee
with the dynamical action (after averaging over the noise and integration by part)
\begin{equation}
\label{eq:action-mft-rho-tilderho}
S[\varrho,\tilde{\varrho}]=\iint \rmd x \rmd t \left(\tilde{\varrho}  (\p_t \varrho-\p_x^2 \varrho )-\varrho(1-\varrho)(\p_x \tilde{\varrho})^2-2\nu \varrho(1-\varrho)\p_x \tilde{\varrho}\right)
\end{equation}
where we introduced the measure on the initial condition for the density field, parametrized \cite{derrida2009currentMFT} by
\begin{equation}
\mathcal{F}(\varrho) = \int_\R \rmd x  \int_{\bar{\varrho}(x)}^{\varrho(x,0)}\rmd r \frac{\varrho(x,0)-r}{r(1-r)}
\end{equation}
where $\bar{\varrho}(x)=\varrho_1 \Theta(-x)+\varrho_2 \Theta(x)$. We recall
that $\langle \cdot \rangle$ denotes the expectation value over
the noise $\eta$ and over the initial density. We will only
consider here observables ${\cal O}$ which depends on the density field
only at the final and initial times.\\

As $\varepsilon \to 0$ the path integral is dominated by a saddle point.
The saddle point equations give back the standard MFT equations of the WASEP for the fields $(\rho, \tilde \rho)$,
which at the saddle point will be denoted $(\rho, \tilde \rho)|_{SP}=(q,p)$ (and for simplicity we keep the same
notation for $z$ and $h$). They read
\be
\label{eq:MFTbulkWASEP-suppmat}
\begin{split}
\partial_t q  &       = \partial_x \left[
\partial_x q - 2 q(1-q) ( \partial_x p + \nu)
\right]
\:,
\\
- \partial_t p & =   \partial_x^2 p +  (1-2 q) \p_x p (   \partial_x p
+  2\nu  )
\:,
\end{split}
\ee
These equations have to be supplemented by boundary conditions at the final and initial time. These
conditions depend on the chosen observable ${\cal O}$. Let us discuss two choices of observables.
\bea
&& {\cal O} = {\cal O}_1 = P \, J(X,T) = - P \, h(X,T)  \\
&& {\cal O} = {\cal O}_2 = \frac{1}{2\nu} [{\rm Li}_2(- u \omega_{u z(X,T)} z(X,T)  ) -F(\omega_{u z(X,T)})]
\eea
The first observable corresponds to the CGF of the current in \eqref{defphi} while
the second corresponds to the (more complicated) observable which appears in \eqref{observableB},
where in the limit $\varepsilon \to 0$ the additional variable
$\omega$ can be taken at its saddle point value $\omega= \omega_{u z(X,T)}$.

To derive the boundary condition associated to each observable we
express them in terms of the density field, and take the functional
derivative. Recalling that
\begin{equation}
h(X,T)=\int_0^{+\infty} dy  \rho(y,0) -\int_X^{+\infty} dy  \rho(y,T)
\end{equation}
One finds
\bea
&& \frac{\delta {\cal O}_1}{\delta \rho(x,t)} = - P  ( \delta(t) \Theta(x)
- \delta(t-T) \Theta(x- X) ) \\
&&  \frac{\delta {\cal O}_2}{\delta \rho(x,t)} =  - \log(1+u \omega_{uz(X,T)} z(X,T) )
( \delta(t) \Theta(x)
- \delta(t-T) \Theta(x- X) )
\eea
To obtain the second equation we note that the
functional derivative applied to $\omega_{u z(X,T)}$ vanishes from the saddle point
condition.

The causality property of the response field imposes that $p(x,0^-)=0$ and $p(x,T+0^+)=0$.
For the two-sided Bernoulli initial condition one thus obtains that one
must solve the saddle point equations \eqref{eq:MFTbulkWASEP-suppmat}
with the following initial and terminal conditions on the fields $p,q$: for the observable ${\cal O}_1$
one has
\be
\label{eq:mft-initial-terminal-conditions}
\begin{split}
& p(x,T) =  P  \, \Theta(x- X) \\
& p(x,0) =  P \, \Theta(x) + {\cal F}'(q(x,0)) = P  \, \Theta(x)+\log \frac{q(x,0)(1-\bar{q}(x))}{\bar{q}(x)(1-q(x,0))}
\end{split}
\ee
where $\bar{q}(x)=\bar{\varrho}(x)$.
For the observable ${\cal O}_2$ one has the same conditions, with to the substitution
\be
P = \log(1+u \omega_{u z(X,T)} z(X,T) )    \label{Pversus}
\ee
This is now a self-consistent equation, since
where $z(X,T)=\exp( 2 \nu (\int_0^{+\infty} dy  q(y,0) -\int_X^{+\infty} dy  q(y,T)))$
itself depends on the solution. However we see that \eqref{Pversus} is precisely
the relation obtained in \eqref{eqslegendre1} from Lagrange inversion at the saddle point. Thus
the calculation is in fact the same for the two observables, but expressed in different
ensembles.

In practice the solution to the above system can be studied in perturbation theory in $P$.
One writes $q(x,t)=\bar \rho(x) + \sum_{n \geq 1} P^n \rho_n(x,t) $ and $p(x,t)= \sum_{n \geq 1} P^n
p_n(x,t)$, and computes iteratively the functions $q_n,p_n$. Once $q_n,p_n$ is known
one can obtain from them the $n$ order cumulant of the integrated current $J$, and
the coefficients $\kappa_n$. The procedure becomes quickly extremely tedious as
$n$ increases. For the general MFT model with driving, including the WASEP as
a special case,
it was performed very recently in \cite{berlioz2024tracer}, up to and including $n=3$.
\begin{remark}
For the step initial condition, the boundary conditions become
\begin{equation}
\begin{split}
p(x,T) &= P  \, \Theta(x- X)  \quad , \quad P = \log(1+u  z(X,T) ) \\
q(x,0)& = \Theta(-x)
\end{split}
\end{equation}

with $h(x,0)=x \Theta(-x)$ and $z(x,0)=\Theta(x) + e^{ 2\nu x} \Theta(-x)$
\end{remark}

\subsection{MFT equations and change of variables}

We explain in this Section how to transform the MFT equations \eqref{eq:MFTbulkWASEP-suppmat} to reveal their integrability. We expect that a similar derivation will unveil the integrability of all MFT with an asymmetry parameter, a constant diffusion parameter and a quadratic mobility, i.e., $\sigma(\varrho)$ being a second order polynomial in the density. Note that we have shown in Ref.~\cite{UsWNTCrossover} that all MFT with quadratic mobility for the symmetric case $\nu=0$ were integrable by a direct mapping to the imaginary-time DNLS system, which is itself gauge equivalent to the imaginary-time NLS system.\\

We start by first performing the transformation $r=\p_x p$, leading to
\be
\begin{split}
\partial_t q         &= \partial_x \left[
\partial_x q - 2 q(1-q) ( r + \nu)
\right] \\
-\p_t r &=  \partial_x^2 r +  \p_x\left(r(1-2 q)  (   r + 2\nu ) \right)
\end{split}
\ee
The Cole-Hopf transformation $q=\frac{1}{2\nu}\p_x \log z$ then yields
\begin{equation} \label{CH}
\begin{split}
\p_t z &= \p_x^2 z -\p_x z \left(2\nu +2r-\frac{1}{\nu} r\p_x \log z\right)\\
-\p_t r &=  \partial_x^2 r +  \p_x\left(r(1-\frac{1}{\nu} \p_x \log z) (   r + 2\nu ) \right)
\end{split}
\end{equation}

These equations seem highly nonsymmetric and we thus propose the following change of variable to make them symmetric \footnote{To provide some insight on this change, we note that we originally tried an ansatz $r=f(\tilde{r}z)$, obtained an equation for the time derivative $\p_t \tilde{r}$ in which we have canceled the prefactor of $\p_x^2 z$ which gave the explicit expression of $f$.}. Let
\begin{equation}
\tilde{r}(x,t)=-\frac{r(x,t)}{(2\nu+r(x,t))z(x,t)} \quad \Longleftrightarrow r=-\frac{2   \nu  \tilde{r} z}{1+  \tilde{r} z}
\end{equation}
we then have
\begin{equation}
\begin{split}
\p_t z &= \p_x^2 z -2\nu \p_x z -2  \frac{  \tilde{r} \p_x z \left(\p_x z-2 \nu  z\right)}{1+  \tilde{r} z}\\
-\p_t \tilde{r} &=  \partial_x^2 \tilde{r}+2\nu \p_x \tilde{r} - 2  \frac{ z \p_x \tilde{r}  \left(\p_x \tilde{r}+2 \nu  \tilde{r}\right)}{1+  \tilde{r} z}
\end{split}
\end{equation}
where to establish the second equation one needs to use also the first one.
We propose to bias the variables as follows to symmetrize the interaction term and eliminate the drift
\be
z(x,t) = Z(x,t) e^{\nu x - \nu^2 t} \quad , \quad \tilde{r}(x,t) =  R(x,t) e^{- \nu x + \nu^2 t}
\ee
and we finally obtain a pair of non-linear equations
\begin{equation}
\label{eq:MFT-system-coherent-ZR}
\begin{split}
\p_t Z &= \p_x^2 Z - \frac{  2   R  }{1+  R Z}\left((\p_x Z)^2- (\nu  Z)^2\right)\\
-\p_t R &=  \partial_x^2 R-\frac{   2   Z }{1+  RZ}\left((\p_x R)^2- (\nu  R)^2\right)
\end{split}
\end{equation}
which we have found to be convenient for multiple reasons that we now expose. The extreme case $\nu \to (0, \infty)$ allow to obtain the NLS and DNLS systems which correspond to the KPZ/SSEP limits as follows
\begin{enumerate}
\item To obtain DNLS, one formally takes $\nu=0$ and then applies the stereographic change of variable
\begin{equation}
\begin{split}
Q&=\frac{Z}{1+ZR } \\
P&= \p_x R
\end{split}
\end{equation}
This leads to
\begin{equation}
\label{eq:MFT-system-coherent-ZR-DNLS-limit}
\begin{split}
\p_t Q &= \p_x^2 Q +   2 \p_x (Q^2 P)  \\
-\p_t P &=  \partial_x^2 P-2\p_x (Q P^2)
\end{split}
\end{equation}
We can additionally map DNLS to NLS using the following non-local change of variable
\begin{equation}
u=(Q^2P+\p_x Q)e^{2\int^x \rmd y QP}, \quad v=-P e^{-2\int^x \rmd y QP}
\end{equation}
as explained in \cite[Appendix~L]{UsWNTCrossover} in the context of the MFT and originally derived in \cite{Wadati_1983}.

\item To obtain NLS, one rescales $R = \tilde{R}/\nu^2$ and then take the limit $\nu \to \infty$. This leads to the system
\begin{equation}
\label{eq:MFT-system-coherent-ZR-NLS-limit}
\begin{split}
\p_t Z &= \p_x^2 Z +   2  Z^2 \tilde{R}  \\
-\p_t \tilde{R} &=  \partial_x^2 \tilde{R}+  2   Z \tilde{R}^2
\end{split}
\end{equation}
\end{enumerate}

Before introducing the last change of variable required to unveil the integrability of the MFT of the ASEP, we briefly comment the structure of the action under the new variables. Performing the change of
variable in the action \eqref{eq:action-mft-rho-tilderho} (discarding boundary terms as well as the Jacobian since we study here the large deviations hence the saddle point)
we find that the action in the new variables
can be represented as
\begin{equation}
\label{eq:action-mft-coherent}
\begin{split}
S[Z,R]&= \iint \rmd t \rmd x \left[-\frac{  R
}{1+  R Z }\left(\p_t Z-\p_x^2 Z\right)+\frac{ R^2 }{(1+ R Z)^2}\left(\nu ^2 Z^2-(\p_x Z)^2\right)\right]\\
&= \iint \rmd t \rmd x \left[\p_Z \mathrm{Li}_1(- R Z)\p_t Z-[\p_Z \mathrm{Li}_1(- R Z)\p_x^2 Z+\p^2_Z \mathrm{Li}_1(- R Z)(\p_x Z)^2]+\frac{\nu^2  R^2 Z^2 }{(1+  R Z)^2} \right]\\
&= \iint \rmd t \rmd x \left[-  \frac{R   \p_t Z}{1+  R Z }-\frac{ \p_x R \p_x Z-\nu^2 Z^2 R^2}{(1+  R Z)^2}\right]\\
\end{split}
\end{equation}
It can be checked that the saddle point equations associated to this action indeed reproduces
the system \eqref{eq:MFT-system-coherent-ZR}.
\begin{remark}
The Hamiltonian appearing in the third line of \eqref{eq:action-mft-coherent} resembles the non-linear Schrodinger Hamiltonian with a coherent dressing, see \cite[Eq.~(3.6)]{Stone_2000}. The equation~\eqref{eq:action-mft-coherent} for $\nu=0$ previously appeared in \cite[Eqs.~(35-36)]{BernardNahumEntanglementCoherent} in the context of a SU(2) Heisenberg chain.
\end{remark}
We now show that this action is symmetric in the variables $(Z,R)$ up to the time reversal $t \to -t$. Only the time derivative term seems nontrivially symmetric. Using
\begin{equation}
\p_t \mathrm{Li}_1(-RZ) = \p_R \mathrm{Li}_1(- RZ) \p_t R + \p_Z \mathrm{Li}_1(\gamma RZ) \p_t Z
\end{equation}

we have up to boundary terms
\begin{equation}
\int \rmd t \p_Z \mathrm{Li}_1(-RZ) \p_t z = -\int \rmd t  \p_R \mathrm{Li}_1(- RZ) \p_t R
\end{equation}
The symmetry of the space derivative term expressed in terms of $\mathrm{Li}_1$ arises from the fact that
\begin{equation}
\begin{split}
&\p_x \left(\p_Z \mathrm{Li}_1(-RZ) \p_x Z -\p_R \mathrm{Li}_1(-RZ) \p_x R \right) \\
&= (\p_Z \mathrm{Li}_1(-RZ) \p^2_x z + \p_Z^2 \mathrm{Li}_1(-RZ) (\p_x Z)^2) - (\p_R \mathrm{Li}_1(-RZ) \p_x^2 R + \p_R^2 \mathrm{Li}_1(-RZ) (\p_x R)^2)
\end{split}
\end{equation}
and thus discarding any boundary term we have
\begin{equation}
\int \rmd x (\p_Z \mathrm{Li}_1(-RZ) \p^2_x Z + \p_Z^2 \mathrm{Li}_1(-RZ) (\p_x Z)^2) =  \int \rmd x (\p_R \mathrm{Li}_1(-RZ) \p_x^2 R + \p_R^2 \mathrm{Li}_1(-RZ) (\p_x R)^2)
\end{equation}

Additionally, we guess the first two conserved quantities of the dynamics with the new variables
\begin{equation}
\p_t \left(\int_\R \rmd x \frac{Z R}{1+Z R}\right) = 0, \quad \p_t \left(\int_\R \rmd x \frac{Z \p_x R}{1+ Z R}\right) = 0
\end{equation}
We can indeed find the associated currents and check that
\bea
&& \p_t \left(\frac{Z R}{1+Z R}\right) = \partial_x \left( \frac{- Z \partial_x R + R \partial_x Z}{(1+ R Z )^2}\right)  \\
&& \p_t \left( \frac{-Z \p_x R}{1+ Z R} \right)
= \partial_x \left( \frac{Z}{1+ R Z} \partial_x^2 R + \frac{\partial_x R (2 Z^2 \partial_x R + \partial_x Z)}{(1+ R Z)^2} + \nu^2 \frac{1 + 2 R Z}{(1 + R Z)^2} \right)
\eea

\subsection{Mapping the WASEP to the anisotropic Landau-Lifshitz model}

 The last change of variable we now introduce allows us to represent \eqref{eq:MFT-system-coherent-ZR} as a complex extension of the dynamics of the classical anisotropic Landau-Lifshitz model in its stereographic frame \cite{LandauLifschitzGilbert,LakshmananFascinating}.
 Let us define the matrix spin $\mathcal{S}$
\begin{equation}
\label{eq:stereographic-representation-spin-coherent}
\begin{split}
\mathcal{S}&=\frac{1}{{1+RZ}}\begin{pmatrix}
1-RZ & 2 R \\
2 Z & -(1-RZ) \\
\end{pmatrix} := \begin{pmatrix}
S_z & S_+\\
S_- & - S_z \\
\end{pmatrix}
\end{split}
\end{equation}
The direct properties of this representation are $\mathcal{S}^2=I_d$, $\det \mathcal{S}=-1$ and $S_- S_+ + S_z^2 =1$. Discarding again the Jacobian and any boundary term (total derivative), the action \eqref{eq:action-mft-coherent} becomes, in the new variables
\begin{equation}
S[\mathcal{S}]= \iint \rmd t \rmd x \left[\frac{S_-\p_t S_+-S_+\p_tS_-}{4(1+S_z)} -\frac{1}{4}(\p_x S_+ \p_x S_-+(\p_x S_z)^2-\nu^2(S_z-1)^2) \right]
\end{equation}
The anisotropy of the WASEP is then mapped exactly to the anisotropy of the Landau-Lifshitz model.\\

To make more precise the connection let us recall that
in \cite{LakshmananFascinating} the anisotropic Landau-Lifshitz model in a field was
studied. We note that the spin representation in Eq.~(4.b) there is consistent with identifying $\omega=R$ and $\omega^*=Z$.
With this identification, the dynamical equation (13) there (together with the remark below that about adding
a longitudinal field $B^L$) is exactly equivalent to our system \eqref{eq:MFT-system-coherent-ZR}
under the identification
\be \label{identificationspin}
2 A = \mu B^L = - \nu^2 \quad , \quad R(x,t)= \omega(x,\tau)|_{\tau \to \I t}
\quad , \quad Z(x,t)= \omega^*(x,\tau)|_{\tau \to \I t}
\ee
With this choice of parameter the Hamiltonian studied in \cite{LakshmananFascinating} reads
\begin{equation}
\mathcal{H}_{LL}=\frac{1}{2}\int \rmd x\, (\p_x S_+ \p_x S_-+(\p_x S_z)^2-\nu^2(S_z-1)^2)
\end{equation}
for a classical spin $\vec S=(S_x,S_y,S_z)$ on the sphere $\vec S^2=1$,
with $S_\pm=S_x \pm \I S_y$. The complex variable $\omega$ is then the
stereographic projection of the spin $\vec S$ (from the south pole).
This Hamiltonian, using $\frac{d}{d\tau} S_i=\{ S_i , H_{LL}\} $
and $\{ S_i(x), S_j(x') \}= \sum_k \epsilon_{ijk} S_k(x) \delta(x-x')$,
leads to the LL equations (we recall that
$\{ A, B \} = \sum_{i,j,k} \epsilon_{ijk} (\partial_{S_i} A) (\partial_{S_j} B)   S_k$)
\be
\partial_\tau \vec S = \vec S \wedge ( \partial_x^2 \vec S + \nu^2 (S_z - 1) \vec e_z )
\ee
equivalently
\bea
&& \partial_\tau S_\pm
= \pm  \I \partial_x (S_z \partial_x S_\pm - S_\pm \partial_x S_z)  \pm\I \nu^2 S_{\pm} (1-S_z) \\
&& \partial_\tau S_z
= \frac{\I}{2} \partial_x (S_+ \partial_x S_- - S_- \partial_x S_+ )
\eea
Changing $\tau=\I t$ and using the correspondence \eqref{eq:stereographic-representation-spin-coherent} one can verify that indeed these equations
are equivalent to \eqref{eq:MFT-system-coherent-ZR}.
\begin{remark}[Spin representation with the MFT variables and separation of variables]
We can also represent the spin variable as a function of the original MFT variables $(z,r)$. In this case, we observe a separation of variables where the spin matrix can be factorised into the form $\mathcal{S}=g_1 g g_1^{-1}$ where $g_1$ solely depends on $z$ while $g$ depends on $r$.
\begin{equation}
\begin{split}
\mathcal{S} &=    \begin{pmatrix}
\frac{\nu +r}{\nu } & -\frac{r e^{\nu  (x-\nu  t)}}{\nu  z} \\
\frac{z (2 \nu +r) e^{\nu  (\nu  t-x)}}{\nu } & -\frac{\nu +r}{\nu } \\
\end{pmatrix}\\
&=
\underbrace{\begin{pmatrix}
\frac{1}{\sqrt{z}}e^{\frac{\nu}{2}x-\frac{\nu^2}{2} t}& 0 \\
0 &  \sqrt{z} e^{-\frac{\nu}{2}x+\frac{\nu^2}{2} t}  \\
\end{pmatrix}}_{g_1}
\underbrace{\begin{pmatrix}
\frac{\nu +r}{\nu } & -\frac{r}{\nu  } \\
\frac{ (2 \nu +r)}{\nu } & -\frac{\nu +r}{\nu } \\
\end{pmatrix}}_g
\underbrace{\begin{pmatrix}
\sqrt{z} e^{-\frac{\nu}{2}x+\frac{\nu^2}{2} t}  & 0 \\
0 &  \frac{1}{\sqrt{z}} e^{\frac{\nu}{2}x-\frac{\nu^2}{2} t}\\
\end{pmatrix}}_{g_1^{-1}}
\end{split}
\end{equation}

Additionally, we can factorise the inner matrix as $g=g_2 \sigma_3 g_2^{-1}$, i.e.
\begin{equation}
\begin{pmatrix}
\frac{\nu +r}{\nu } & -\frac{r}{\nu  } \\
\frac{ (2 \nu +r)}{\nu } & -\frac{\nu +r}{\nu } \\
\end{pmatrix} =
\underbrace{\begin{pmatrix}
\frac{r}{2 \nu } & \frac{r}{2 \nu } \\
\frac{r}{2 \nu } & \frac{r}{2 \nu }+1 \\
\end{pmatrix} }_{g_2}
\sigma_3
\underbrace{\begin{pmatrix}
\frac{r}{2 \nu } & \frac{r}{2 \nu } \\
\frac{r}{2 \nu } & \frac{r}{2 \nu }+1 \\
\end{pmatrix}^{-1}}_{g_2^{-1}}
\end{equation}

Thus, we overall have $\mathcal{S}=g_1g_2\sigma_3(g_1g_2)^{-1}$.
We comment on this separation of variable later in Section~\ref{subsec:comment-gauge-equivalence}. Note that for $\nu \to 0$, one needs to proceed to a change of variable $r \to \nu r$ so that the spin remains well defined. This spin representation is a priori different from the one considered for the SSEP in \cite{Tailleur_2008}
although they may differ by a gauge transformation.

\end{remark}

\subsection{Lax pair representation of the MFT}

The Lax pair for the anisotropic Landau-Lifshitz magnet has been known for a few decades, see e.g., Ref.~\cite{Nakamura_1982}. To obtain a Lax pair for the MFT of the WASEP we
use the conventions of this reference, with minor changes: we set $\lambda=-k/2$ and since the time there,
which we denote by $\tau$, is $\tau = \I  t$ (see above), we set $M=- \I M_1 -\frac{\nu^2}{2} \sigma_3 $,
with an additional term, see below.
Let $\vec{v}(x,t,k)$ be a two-dimensional vector depending on space, time and a spectral parameter $k$. We define the linear problem
\begin{equation}
\p_x \vec{v}=L \vec{v}, \quad \p_t \vec{v}=M \vec{v}
\end{equation}
where $L,M$ are two $2 \times 2$ matrices. The compatibility of this system $\p_{xt}=\p_{tx} $ provides a zero-curvature condition
\be \label{zerocurvature}
\p_t L - \p_x M +[L,M]=0
\ee
where $[L,M]=LM-ML$. A nonlinear system is said to be integrable if there exist a pair of Lax matrices $(L,M)$ so that their compatibility yields back the original system. In the present case, we define
\begin{equation}
L= -\frac{\I k}{2}\mathcal{S}+\mu [\sigma_3,\mathcal{S}]
\end{equation}
and
\begin{equation}
M = \frac{k^2}{2}\mathcal{S}+\frac{\I k}{2}\mathcal{S} \p_x \mathcal{S}+\I \mu k [\sigma_3,\mathcal{S}]-\mu [\sigma_3,\mathcal{S} \p_x \mathcal{S}]+4\mu^2 \{\sigma_3,\mathcal{S} \}\sigma_3-\frac{\nu^2}{2} \sigma_3
\end{equation}
where $\{\sigma_3,\mathcal{S} \}=\sigma_3\mathcal{S}+\mathcal{S}\sigma_3$ and
$\mu = \sqrt{-2 A}/4 = \frac{\nu}{4}$ from \cite{Nakamura_1982} and \eqref{identificationspin}
(although $\mu=-\nu/4$ also leads to a Lax pair). Note that we found that it was necessary to add
the term $-\frac{\nu^2}{2} \sigma_3$ in the presence of the magnetic field.
Choosing $\mu= \frac{\nu}{4}$,
we write explicitly the Lax pair in terms of the variables $(Z,R)$ as
\begin{equation}
L = \frac{1}{1+RZ}
\begin{pmatrix}
-\frac{\I k}{2}  \left(1-RZ \right) & (\nu -\I k) R \\
-(\nu+\I k ) Z & \frac{\I k}{2}  \left(1-RZ \right) \\
\end{pmatrix}
\end{equation}
and
\begin{equation}
\begin{split}
&M = \frac{1}{(1+RZ)^2}\\
&\begin{pmatrix}
\frac{k^2}{2}- \nu ^2   ZR-\I k (Z \p_x R-R\p_x Z)-R^2 Z^2 \left(\frac{k^2}{2}+\nu ^2\right) & (k+\I\nu ) \left(R \left(k+R \left(k Z+\I\p_x Z\right)\right)+\I\p_x R\right) \\
(k-\I\nu ) \left(Z \left(k+Z \left(k R-\I\p_x R\right)\right)-\I\p_x Z\right) &-\frac{k^2}{2}+ \nu ^2   ZR+\I k (Z \p_x R-R\p_x Z)+R^2 Z^2 \left(\frac{k^2}{2}+\nu ^2\right) \\
\end{pmatrix}
\end{split}
\end{equation}
One can now show that the zero curvature condition of these two matrices is equivalent to the system \eqref{eq:MFT-system-coherent-ZR}. This is
checked by explicit calculation, inserting these matrices in the l.h.s of \eqref{zerocurvature},
and replacing the time derivatives of $Z$ and $R$ by their expressions from
\eqref{eq:MFT-system-coherent-ZR}. A massive cancellation occurs and one finds zero.
These matrices thus form a good Lax pair for the system \eqref{eq:MFT-system-coherent-ZR}. This open the way to study the scattering and inverse scattering problem associated with this Lax pair,
which is deferred to a subsequent work.
\begin{remark}
It was noted in \cite{Rizkallah_2023} that a number of dualities do exist in the macroscopic fluctuation theory. It would be interesting to study the implications of such dualities on the mappings we have proposed in this Section.
\end{remark}
\subsection{Comment on gauge equivalences}
\label{subsec:comment-gauge-equivalence}
There is no uniqueness of the Lax pair to represent \eqref{eq:MFT-system-coherent-ZR} and thus we cannot exclude that other gauge equivalent representations of this pair can be convenient to solve a scattering problem. A gauge transformation on the Lax matrices is defined as the map $(L,M) \mapsto (\tilde{L},\tilde{M})$ involving an invertible gauge $\mathcal{G}$ so that
\begin{equation}
\tilde{L}=\mathcal{G}^{-1}L\mathcal{G}-\mathcal{G}^{-1}\p_x \mathcal{G}, \quad \tilde{M}=\mathcal{G}^{-1}M\mathcal{G}-\mathcal{G}^{-1}\p_t \mathcal{G}
\end{equation}

Since we have shown above that the spin matrix can get factorised as $\mathcal{S}=g_1g_2\sigma_3(g_1g_2)^{-1}$, proposing the following gauge $\mathcal{G}=g_1g_2$ would ensure that the coefficient in front of the spectral parameter $k$ in the space Lax matrix $\tilde{L}$ would be independent of the space-time variables $(x,t)$. While this consideration emerges from an algebraic standpoint, gauge equivalences have been historically paramount from a physics standpoint to relate different models together, as we now recall.\\

We have mapped in this work the MFT of the WASEP to the anisotropic Landau-Lifshitz model. We refer the reader to various references for additional context, \cite{LakshmananFascinating} for a review on the Landau-Lifshitz model, \cite{LandauLifschitzGilbert} for the analysis of the Landau-Lifshitz model in the stereographic frame. The mappings between symmetric exclusion processes (including the SSEP) and isotropic spin chains was
considered in \cite{Tailleur_2008}, and  \cite{derrida2009currentExact,derrida2009currentMFT}.
For a recent appearance of the Landau-Lifshitz model in the context of an entanglement entropy calculations in random unitary circuits, see \cite{BernardNahumEntanglementCoherent}. \\

The anisotropic Landau-Lifshitz model has other remarkable mappings in the context of integrability
\begin{itemize}
\item The anisotropic Landau-Lifshitz model is gauge equivalent to the isotropic Landau-Lifshitz model and the nonlinear Schrodinger equation, see \cite{Nakamura_1982,Kundu_1983,zakharov1979equivalence}.
\item The nonlinear Schrodinger equation has been shown to be gauge equivalent to the derivative nonlinear Schrodinger equation, see \cite{Wadati_1983}, through a triangular gauge transformation.
\end{itemize}
These mappings have recently been used to obtain exact results in the context of the study of the large deviations in a number of stochastic integrable models.
\begin{itemize}
\item The weak noise theory of the KPZ equation was solved using its mapping to the imaginary-time NLS system in Ref.~\cite{UsWNTDroplet2021} for droplet initial condition and Ref.~\cite{UsWNTFlat2021} for the flat and Brownian initial condition.
\item The MFT equations for the SSEP and KMP with quenched initial condition were solved in Refs.~\cite{UsWNTCrossover,NaftaliKMP1,NaftaliKMP2} using its mapping to the imaginary-time DNLS equation.
\item The MFT equations for the SSEP with annealed initial condition were solved in Ref.~\cite{SasamotoExactSSEP} using the gauge transformation of Wadati and Sogo \cite{Wadati_1983}. The derivation of \cite{SasamotoExactSSEP} showed that the annealed initial condition of SSEP maps to the initial condition studied in \cite{UsWNTDroplet2021} in the context of the KPZ equation, indicating that the scattering results from \cite{UsWNTDroplet2021} can be used directly. This points out to the existence of additional duality results between particle models.
\end{itemize}
We believe more gauge equivalences do exist, both in the continuous and semi-discrete or discrete settings which will lead to a variety of new results in the large deviation study of driven diffusive systems.
\subsection{Extension to the MFT of asymmetric models with quadratic mobility}
\label{sec:extension-quadratic}
We now extend the integrability argument we have derived to the MFT of asymmetric models with quadratic mobility. We recall that the general MFT depends on two functions $D(q)$ and $\sigma(q)$. With the choice
$\sigma(q)= \sigma_{A,B}(q) := 2Aq(B-q)$, $D(q)=1$, the MFT reads
\begin{subequations}
\begin{align}
\partial_t q
&= \partial_x \left[
\partial_x q - 2 A q(B-q) (\partial_x p + \nu )
\right]
\:,
\\
- \partial_t p
&=   \partial_x^2 p + A(B-2q) \partial_x p (\partial_x p + 2 \nu)
\:,
\end{align}
\end{subequations}
This system maps, under the change of variable
\begin{equation}
\begin{split}
q(x,t)&=\frac{1}{2A\nu} \p_x \log [Z(x,t)e^{AB \nu x-(AB\nu)^2 t}],\\
\p_x p(x,t)&=-\frac{2\nu R(x,t) Z(x,t)}{1+R(x,t)Z(x,t)}
\end{split}
\end{equation}
exactly to \eqref{eq:MFT-system-coherent-ZR} with $\nu$ replaced by $A B \nu$. Hence the
original MFT maps to the complex extension of the classical anisotropic Landau-Lifshitz model with anisotropy $AB \nu$. This mapping allows to solve the MFT for other models
such as the Weakly Asymmetric Simple Inclusion Process (WASIP), setting
$A=-1$, $B=-1$, and the driven continuum KMP model (WAKMP) in the limit $A=-1$, $B=0$,
as well as their dual models, see Section~\ref{sec:duality-aurelien}. \\

This remark will be most useful to address general initial conditions.
For stationary initial conditions however, i.e., for an initial density distributed as
\be
{\cal P}(\rho) = e^{- \frac{1}{\varepsilon} {\cal F}(\rho,\bar \rho,\sigma_{A,B}) }
\quad , \quad
{\cal F}(\rho,\bar \rho,\sigma)= 2 \int_\R \rmd x \int_{\bar \rho(x)}^{\rho(x)} \rmd r\, \frac{\rho(x)- r}{\sigma(r)}
\ee
one can infer the result from the one we have obtained for the WASEP,
by simply extending to the driven case the mapping between quadratic models
introduced in \cite[Eq. (33)]{derrida2009currentMFT}.
In our notations it amounts to note that the action and the free energy transform as
\bea
&& S[\tilde \rho,\rho,\sigma_{A,B},\nu] = \frac{1}{A} S[A B \tilde \rho  ,\frac{\rho}{B},\sigma_{1,1},
A B \nu]
\quad , \quad {\cal F}(\rho,\bar \rho,\sigma_{A,B}) =
\frac{1}{A} {\cal F}(\frac{\rho}{B},\frac{\bar \rho}{B},\sigma_{1,1})
\eea
Consider the CGF of the integrated current $J$ defined in
\eqref{defphi}, which we denote $\phi_{A,B}(P)$ for the model with $\sigma_{A,B}$.
Since the observable $J$ is linear in the density field,
we need to rescale $P \to A B P$ so that all terms in the
exponential scale as $1/A$ under the
change $\rho \to \rho/B$. Hence one obtains
\be
\phi_{A,B}(P, \bar \rho(x), \nu)= \frac{1}{A} \phi_{\rm WASEP}(AB P , \frac{\bar \rho(x)}{B}  , \nu A B)
\ee
where $\phi_{\rm WASEP}=\phi_{1,1}$.

Hence we obtain the CGF of the integrated current
for the WASIP with $\sigma(\rho)=2 \rho (1+\rho)$ for the two sided
stationary initial condition for that model, with densities $\rho_1,\rho_2$
\bea
&& \phi_{\rm WASIP}(P,\rho_1,\rho_2,\nu)= - \phi_{\rm WASEP}(P , - \rho_1 , -\rho_2 , \nu)
\\
&&
\kappa_n^{\rm WASIP}(\rho_1,\rho_2,\nu)= - \kappa_n^{\rm WASEP}(- \rho_1 , -\rho_2 , \nu)
\eea
This relation must be understood as an analytical continuation of the parameters (which, order
by order in a perturbative calculation is well defined).
Note that this model was studied recently for a different initial condition
\cite{MeersonInclusion}. \\

Similarly one obtains for the weakly asymmetric KPM model
with $\sigma(\rho)=2 \rho^2$
\bea
&& \phi_{\rm WAKMP}(P)= - \lim_{B \to 0}  \phi_{\rm WASEP}(- B P , \frac{\rho_1}{B} , \frac{\rho_2}{B} , - B \nu)
\\
&&
\kappa_n^{\rm WAKMP}(\rho_1,\rho_2,\nu)= (-)^{n+1} \lim_{B \to 0} B^n
\kappa_n^{\rm WASEP}(\frac{\rho_1}{B} , \frac{\rho_2}{B} , - B \nu)
\eea

Having the $\kappa_n$ for the WASIP and the WAKMP we also obtain cumulants $c_n$ of the
tracer position for these models from the relations \eqref{allcn}.

\subsection{Duality in the MFT}
\label{sec:duality-aurelien}

There is a general duality property of the MFT models, see \cite{berlioz2024tracer,Rizkallah_2023}
for more details.
Consider a MFT model with density field $\rho(x,t)$ and
parameters $D(\rho), \sigma(\rho), \nu$. The dual MFT model
describes a density field $\tilde \rho(x,t)$ and
parameters $\tilde D(\rho), \tilde \sigma(\rho), \tilde \nu$,
with the following involutive relations
\bea
&& \rho(x,t) = \frac{1}{\tilde{\rho}(k(x,t),t)} \quad , \quad \partial_x k(x,t) = \rho(x,t) \\
&&
\tilde{D}(\rho) = \frac{1}{\rho^2}D\left(\frac{1}{\rho}\right), \quad
\tilde{\sigma}(\rho) = \rho\,\sigma\left(\frac{1}{\rho}\right), \quad
\tilde{\nu} = -\nu.
\eea
In the first equation line one should eliminate the "height field" $k(x,t)$.
Denoting $(J(0,T), Y_T)$ the integrated current and position of tracer in
the original MFT model, this duality maps it onto $(-Y_T, -J(0,T))$ in the
dual MFT model \cite{berlioz2024tracer,Rizkallah_2023}.
This duality in some cases originates from a duality of the microscopic
models, for instance the ASEP is dual to a zero range process (ZRP)
Indeed the evolution of the gaps in the ASEP is described by a ZRP.
In the weakly asymmetric limit this becomes the duality between
the WASEP $D=1$, $\sigma=2 \rho(1-\rho)$ and a ZRP with
$D=1/\rho^2$ and $\sigma = 2 (1 - \frac{1}{\rho})$. \\

This duality thus allows us to immediately obtain the integrated current
and tracer position large deviation functions for the ZRP,
with the corresponding double sided stationary initial condition
$\tilde \rho_1=1/\rho_1$, $\tilde \rho_2=1/\rho_2$. from Section~\ref{sec:extension-quadratic},
it also yields the exact results for the dual models to all quadratic models,
which have
\begin{equation}
\tilde{D}(\rho)= \frac{1}{\rho^2}, \quad \tilde{\sigma}(\rho) = 2A(B-\frac{1}{\rho})
\end{equation}
in particular the dual of the WASIP and the dual of the WAKMP (an asymmetric version of the random average
process, see Table~1 in \cite{Rizkallah_2023}).

\subsection{Optimal density at initial time $T=0$}

At time $T=0$ the density field $\rho(x)$ of the WASEP is the coarse-grained version of the density field of the
ASEP with a double sided i.i.d. Bernoulli initial condition, i.e., for each $x$ one can write
$\rho(x) = \sum_{i=1}^{1/\varepsilon} n_i$ where $n_i=0,1$ with probabilities $1-\bar \rho(x)$
and $\bar \rho(x)$, respectively. Its probability distribution thus decouples
in space and takes the form
\be
{\cal P}(\rho) \sim e^{- \frac{1}{\epsilon} \int dx \hat f(\rho(x),\bar \rho(x)) }
\ee
where $\hat f(\rho,\bar \rho)$ is the large deviation rate function of sums of i.i.d Bernoulli variables.
To obtain it one computes its CGF
\be \label{CGFBernoulli}
\langle e^{\frac{p}{\epsilon} \rho  } \rangle
= \langle e^{\frac{p}{\epsilon} \sum_{i=1}^{1/\varepsilon} n_i  } \rangle
= e^{\frac{1}{\epsilon} \log(1 - \bar \rho + \bar \rho e^{p}) }
\ee
The Legendre inversion of
\be \label{inversionBernoulli}
\min_\rho (p \rho - \hat f(\rho,\bar \rho) ) = \log(1 - \bar \rho + \bar \rho e^{p})
\ee
then gives the "free energy density"
\be
\hat f(\rho,\bar \rho) = \rho \log \frac{\rho}{\bar \rho}
+ (1- \rho ) \log \frac{1-\rho}{1-\bar \rho}  = \int_{\bar{\varrho}}^{\varrho}\rmd z \frac{\varrho-z}{z(1-z)}
\ee
for $0<\rho<1$,
leading to the well known form \eqref{initialP} in the main text, see e.g \cite{derrida2009currentMFT,bodineau2005current}.

Once can then study the large deviations for any given observable at initial time, by computing the associated optimal particle density. In the MSR action, the only term which remains at time $t=0$ is the free energy describing the stationary fluctuations of the density compared to the imposed density profile.
The optimal density is found by balancing this free energy with the observable of interest.

Let us illustrate this on the simplest example, and obtain the exact rate function $\Phi(J)$ at initial time $T=0$.
The observable is $J(X,0)= - \int_0^X dy \rho(y,0)$. We first compute its CGF,
$\big\langle e^{\frac{1}{  \varepsilon} P J } \big\rangle \sim e^{ \frac{1}{ \varepsilon}  \phi(P) }$.
One has
\be
\label{eq:optimal-density-initial-time}
\phi(P) =  \max_\varrho \left[  - P \int_0^X  \rmd y\rho(y)  -  \int dx \hat f(\rho(x),\bar \rho(x))
\right]
\ee
Taking a functional derivative w.r.t. $\rho(x)$ we find that the
optimum density is the solution of
\begin{equation}
\label{eq:initial_time_optimal_density_derivative}
-  P \, {\rm sign}(X) \Theta_X(x) (x \in [0,X] ) - \log \frac{\rho(x)(1-\bar{\varrho}(x))}{\bar{\varrho}(x)(1-\rho(x))}=0
\end{equation}
where $\Theta_X(x)=\Theta(x \in [0,X] ) $ for $X>0$ and
$\Theta_X(x)=\Theta(x \in [X,0] ) $ for $X<0$. We thus obtain the
optimal density as
\be
\label{eq:initial_time_optimal_density_derivative_solution}
\varrho(x,0) = \varrho_P(x) = \frac{e^{- P \, {\rm sign}(X) \Theta_X(x) }\bar{\rho}(x)}{1-\bar{\rho}(x)+e^{- P \, {\rm sign}(X) \Theta_X(x) }\bar{\rho}(x)}
\ee
Using that $\phi'(P) = - \int_0^X dy \varrho_P(y)$ from taking a derivative of \eqref{eq:optimal-density-initial-time},
we obtain
\be
\phi(P)= |X| \log(1 + \bar \rho(X) (e^{- {\rm sign}(X) P}-1))
\ee
Comparing with \eqref{CGFBernoulli} we see that it is consistent with $J = - {\rm sign}(X) R$
where $R$ is a sum of $|X|/\epsilon$ Bernoulli variables, hence one finds (by the
same inversion as in \eqref{inversionBernoulli})
\bea
\Phi(J=J(X,0)) &=& |X| f(  \frac{ - {\rm sign}(X) J }{|X|}, \bar \rho(X) )
\\
& = & - {\rm sgn}(X) J \log \frac{J}{- X \bar \rho(X)}
+ |X| (1 + \frac{J}{X} ) \log \frac{1+ \frac{J}{X}}{1-\bar \rho(X)}
\eea
This gives the large deviation form of the PDF of $J$
at time $T=0$, i.e., ${\cal P}(J) \sim e^{- \frac{1}{\varepsilon} \Phi(J)}$, and we note that for $X>0$, $J \in [-X,0]$,
while for $X<0$, $J \in [0,X]$. At $X=0$ one has $J(0,0)=0$.
\\

One can ask how to match these variational calculations with setting $T=0$ in our result for $\Psi(u)$ in
\eqref{eq:rate-function-wasep-general}. Taking $X>0$ here, the variational problem then become
\be
\Psi(u) \underset{T=0}{=} \max_\varrho \left[ - f(u e^{ 2 \nu \int_0^X  \rmd y\rho(y)}) -  \int_\R dx f(\rho(x),\bar \rho(x))) \right]
\ee
where the function $f(a)$ was defined below \eqref{complicated}. The optimum density obeys the same equation as \eqref{eq:initial_time_optimal_density_derivative} setting
$P= \log(1+u z \omega_{u z})$, where $z=e^{ 2 \nu \int_0^X \rho(y)}$ and its
solution is given by \eqref{eq:initial_time_optimal_density_derivative_solution}.

One has $\varrho(x,0)=\bar \rho(x)$ for $x$ outside of $[0,X]$. For $x\in [0,X]$, the optimum density is given by
\be
\varrho(x,0) = \frac{1}{1 + \frac{1- \bar \rho(x)}{\bar \rho(x)} (1+u \omega e^{ 2 \nu \int_0^X \rho(y,0)} ) }
\ee
with $\bar{\varrho}(x)=\varrho_1 \Theta(-x)+\varrho_2 \Theta(x)$ and thus the optimal density is constant on the interval $[0,X]$ (we will denote $\varrho(x,0)=\varrho(0)$ in this interval). By integration, one can find a self-consistent equation for $\int_0^X \rmd y \rho(y,0)$ which can be evaluated numerically. We should now verify that to what extend this result should be consistent with the large deviation result
\be
\begin{split}
&\int_{1}^{+\infty} \rmd w \big\langle e^{\frac{1}{2\nu \varepsilon} [{\rm Li}_2(- u \omega z(X,T=0)  ) -F(\omega)]}\big\rangle \sim e^{- \frac{1}{ \varepsilon} \Psi(u) }
\end{split}
\ee
From the MSR representation \eqref{eq:intro-mft-observable2}, this is equivalent to
\be
\begin{split}
\mathcal{F}(\varrho)-\frac{1}{2\nu} [{\rm Li}_2(- u \omega_{uz} z  ) -F(\omega_{uz})] = \Psi(u)  = - \frac{1}{2\nu} \int_{\I \mathbb{R} + \delta}
\frac{\rmd y}{2\I \pi y(1-y)}  \mathrm{Li}_2\left(-u\frac{\rho _1 \left(1-\rho _2\right) (1-y) y}{\left(\rho_1-y\right)
\left(y-\rho _2\right)} e^{ 2\nu y X } \right)
\end{split}
\ee

The left hand side can be evaluated using
\begin{equation}
\begin{split}
& z=e^{ 2 \nu  X \varrho(0)}, \quad \quad \omega_{uz} = \frac{\alpha-1+u z  +\sqrt{\left(\alpha +uz -1\right)^2+4 uz }}{2 uz } , \quad \quad
\varrho(0) = \frac{1}{1 + \frac{1- \varrho_2}{\varrho_2} (1+u z\omega_{uz})}\\
&\mathcal{F}(\varrho)=X \left(\left(1-\rho _2\right) \log (\frac{1-\rho (0)}{1-\varrho_2})+\rho _2 \log (\frac{\rho(0)}{\varrho_2})\right)\\
\end{split}
\end{equation}

\section{Cumulants of the partition function $z$ and perturbative expansion of $z(u)$}

The aim of this section is two-fold. First, since the field
$z= e^h= e^{-2 \nu J}$ plays the role of a partition sum (particularly so in
the KPZ limit) it is interesting to be able to compute
its cumulants at the final point, i.e., the cumulants of the random variable $z=z(X,T)$,
solution of the WASEP-MFT stochastic equation.
These cumulants are of the form
\be \label{cumzeps}
\langle z^n \rangle_c \simeq z_n \varepsilon^{n-1}
\ee
to leading order in $\varepsilon$. Furthermore, since $z$ also denotes (and one should not confuse these two
objects, abusively denoted by the same letter) the field which enters in the MFT saddle point equations
\eqref{CH} (after a Cole-Hopf transform),
our second aim is to obtain the solution $z=z(X,T)$ of these equations
at the final time, and at the special point $X$ (which is also
a function of $u$, which enters the boundary condition
of these equations, see \eqref{eq:mft-initial-terminal-conditions}
\eqref{Pversus} and the line below). We will denote $z=z(u)$ this latter quantity.\\

We will show how to
obtain both quantities from the rate function $\Psi(u)$ and
its derivatives at $u=0$, for which we have explicit
expressions. To do so we will use our saddle point method, together with Legendre transforms
as in \eqref{eqslegendre1}, \eqref{eqslegendre2}.

Let us define $\hat \Psi(u)$ to be the CGF of the random variable $z=z(X,T)$, from which
we can obtain the cumulants \eqref{cumzeps}. One has
\be
\langle e^{ - \frac{1}{\varepsilon} u z  } \rangle \sim  e^{- \frac{1}{\varepsilon} \hat \Psi(u) }
\quad , \quad z_n =  (-1)^{n-1} \hat \Psi^{(n)}(0)
\ee
By contrast, from the knowledge of $\Psi(u)$, the formula \eqref{observableB} gives us
only access to the expectation value of a more complicated observable
\be \label{complicated}
\langle e^{ - \frac{1}{\varepsilon} f(u z)  } \rangle \sim  e^{- \frac{1}{\varepsilon}  \Psi(u) }
\ee
where $f(a) := - \frac{1}{2\nu}  [ {\rm Li}_2(- a \omega_{a}  ) - F(\omega_{a})]$
and $F(\omega)=\mathrm{Li}_2\left(\frac{1}{\omega}\right)-\log \omega \log \alpha +\mathrm{Li}_2(\alpha)-\mathrm{Li}_2(1)$, where $\omega_u$ is the value at the saddle point given in \eqref{omegauz}. The derivative of the function $f$ has a simpler
expression,
$g(a) := 2 \nu a f'(a) =  \log(1 + a \omega_{a})$, and we have $f(0)=0$.
We recall that the inverse function $g^{-1}$ has a simple form, $g^{-1}(P) = (1-e^{-P})(e^P-\alpha)$,
see \eqref{inversionuz}.\\

One way to obtain the cumulants of the random variable $z=z(X,T)$ is by expanding in powers of $u$ both sides
of \eqref{complicated}, identifying and matching the powers of $\varepsilon$
using \eqref{cumzeps}. It is not very convenient however, and a more powerful method
is to use Legendre transforms.

One defines the rate function $\hat \Phi(z)$ for the PDF of $z$, i.e.,${\cal P}(z) \sim e^{- \frac{1}{\varepsilon} \hat \Phi(z)}$ and write
\bea  \label{2eq}
&& \Psi(u) = \min_{z \in \mathbb{R}^+ } [f(u z) + \hat \Phi(z)] \\
&& \hat \Psi(u) = \min_{z \in \mathbb{R}^+ } [u z + \hat \Phi(z)]
\eea
Let us first consider the first equation, which yields the pair of equations
\be
\Psi'(u)= z f'(u z) \quad , \quad \hat \Phi'(z) = - u f'(u z)
\ee
The first equation define $z=z(u)$. It can be rewritten
as $2 \nu u \Psi'(u) = g(u z)$ which can be inverted, see above and
\eqref{inversionuz}, leading to
\begin{equation} \label{zdeu}
z=z(u)=\frac{(1- e^{-  2\nu u \Psi'(u)}) (e^{  2\nu u \Psi'(u)}-\alpha)}{u}
\end{equation}
This expression is valid for the main branch of $\Psi(u)$,  see discussion in Section~\ref{sec:domain}. We can Taylor expand this relation around $u=0$ which should match the solution of the MFT equations obtained perturbatively term by term in powers of $u$, hence providing a check of the
solution of these equations
\begin{equation}
\begin{split}
z(u)=&2 (1-\alpha ) \nu  \Psi '(0)+2 \nu  u \left((\alpha +1) \nu  \Psi '(0)^2+(1-\alpha ) \Psi ''(0)\right)\\
&+u^2 \left(\frac{4}{3} (1-\alpha ) \nu ^3
\Psi '(0)^3+4 (\alpha +1) \nu ^2 \Psi '(0) \Psi ''(0)+(1-\alpha ) \nu  \Psi ^{(3)}(0)\right)+\mathcal{O}(u^3)
\end{split}
\end{equation}
\begin{remark}
Note that one cannot obtain $q(X,T)$ from $z(u)$ as we only have access to a single space-time point $(X,T)$.
\end{remark}
Pursuing the calculation to obtain the cumulants the second equation in \eqref{2eq}
leads to the pair of equations
\be
\hat \Psi'(v)= z \quad , \quad \hat \Phi'(z) = - v
\ee
where we changed the name of one of the variable, but chose to
identify the other one with $z$. Since $z=z(u)$ from the first
system, this leads to a relation between $u$ and $v$. It is obtained by
noting that
\be
u \Psi'(u) = - z \hat \Phi'(z) = z v   \quad \Rightarrow \quad v = \frac{u \Psi'(u)}{z(u)}
\ee
so that finally we obtain
\be
\hat \Psi'( \frac{u \Psi'(u)}{z(u)} ) = z(u)
\ee
Expanding in powers of $u$ using \eqref{zdeu} gives the relation between the
derivatives of $\hat \Psi$ and those of $\Psi$. We obtain the first three cumulants as
\bea
&& z_1= 2 (1-\alpha) \nu
\Psi '(0) \quad , \quad
z_2= - 4 (1-\alpha) \nu^2 \left( (\alpha +1) \nu \Psi '(0)^2
+ (1-\alpha)
\Psi''(0)\right) \\
&& z_3 = \frac{8}{3}
(\alpha -1) \nu ^3 \left(18
\left(\alpha ^2-1\right)
\nu  \Psi '(0) \Psi ''(0)-2
(\alpha  (5 \alpha +2)+5)
\nu ^2 \Psi '(0)^3-3
(\alpha -1)^2 \Psi^{(3)}(0)\right) \nn
\eea
The first moment is simply $z_1= e^{-2 \nu \kappa_1}$ as expected (since
there are typical values). The second cumulant reads, for $\rho_1=\rho_2$,
\be
z_2 = 8 \nu^2 \rho(1-\rho) z_1^2 \sqrt{T}  {\cal G}(y) \quad , \quad
z_1= e^{- 4 \nu^2 \rho(1-\rho) T + 2 \nu \rho X} \quad , \quad
y = - \frac{1}{\sqrt{T}}(\nu (2 \rho-1) T + X/2)
\ee
\\
\begin{remark}[Cumulants of $z^{1/2}$ in the stationary case $\varrho_1=\varrho_2$]
Interestingly the stationary case $\rho_1=\rho_2=\rho$, the cumulants which can be extracted from $\psi(v)$
are those of $z^{1/2}$. Indeed expanding on both sides of the large deviation identity \eqref{PsiMinStat}
in powers of $v=\sqrt{u}$
\bea
&&  \big\langle  e^{-\frac{1}{2\nu \varepsilon} \left[\mathrm{Li}_2\left(\frac{\sqrt{(uz)^2+4 uz}-uz}{2}\right)-\mathrm{Li}_2\left(-\frac{\sqrt{(uz)^2+4 uz}+uz}{2}\right)\right]}\big\rangle \sim e^{- \frac{1}{\varepsilon} \psi(v) }
\eea

we obtain
\begin{equation}
\label{eq:cum-z-taylor-psi-2}
\begin{split}
\psi(v)=&\frac{v  }{\nu }\langle z^{\frac{1}{2}}\rangle +\frac{v^2
}{2 \nu ^2 \varepsilon } \left(\langle z^{\frac{1}{2}}\rangle^2-\langle z\rangle \right)+\frac{v^{3} }{72 \nu ^3 \varepsilon
^2}\left(\langle z^{\frac{3}{2}} \rangle \left(12-\nu ^2 \varepsilon ^2\right)+24
\langle z^{\frac{1}{2}}\rangle^3-36 \langle z \rangle  \langle z^{\frac{1}{2}}\rangle \right)+\mathcal{O}(v^4)\\
\end{split}
\end{equation}
\end{remark}

\section{Limit to the weak-noise KPZ equation $\nu \to \infty$}

We study in this Section the $\nu \to \infty$ limit of our results and show a perfect matching with the results previously obtained for the weak noise regime of the KPZ equation
for the two-sided Brownian initial condition in \cite{krajenbrink2017exact,UsWNTFlat2021}, and for the droplet
initial condition in \cite{le2016exact,UsWNTDroplet2021}. It is known that the ASEP converges under a weak asymmetry $\mathcal{O}(1/\sqrt{N})$,
to the KPZ equation, see
\cite{bertini1997stochastic,corwin2018open}.
Here we start from the WASEP, which is defined with an even weaker asymmetry $\mathcal{O}(\nu/N)$.
The large $\nu$ limit again brings us
to the KPZ equation, but in its short time/weak noise regime.\\

In this Section we first parametrize the initial densities as
\begin{equation} \label{param12}
\varrho_1=\frac{1}{2}+\frac{\tilde{\varrho}_1}{2\nu}, \; \varrho_2=\frac{1}{2}+\frac{\tilde{\varrho}_2}{2\nu}
\end{equation}
and in a second stage we study the limit of large $\nu$ with $\tilde \rho_{1,2}$ fixed,
which amounts to rescale the densities around $1/2$. With these definitions, one has at leading order
\begin{equation}
\alpha = 1-2\frac{\tilde{\varrho}_1-\tilde{\varrho}_2}{\nu}+\mathcal{O}\left(\frac{1}{\nu^2}\right)
\end{equation}

We first study the KPZ limit of the large deviation rate function $\Psi(u)$, then of the cumulants of the current, and finally of the MFT equations.

\subsection{KPZ limit of the large deviation rate function}
\label{sec:kpzlimitlargedev}

In the KPZ limit, we rescale the generalized Laplace parameter $u$ and the partition function $z$ as
\begin{equation} \label{rescu}
u=\frac{4 \tilde{u} e^{\nu ^2 T-\nu  X}}{\nu ^2}, \; z=Ze^{-\nu ^2 T+\nu  X}
\end{equation}
and then take the limit $\nu\to \infty$. We proceed to a change of variable in the integrand of the rate function \eqref{eq:rate-function-wasep-general}
\begin{equation}
y = \frac{1}{2} + \frac{\tilde y}{2 \nu}, \quad \tilde y = \tilde \delta + \I k
\end{equation}
and one finds that at leading order the rate function takes the scaling form
\bea \label{PsiKPZgeneral}
&&   \Psi(u)  \simeq \frac{1}{\nu^2}\tilde{\Psi}(\tilde{u})
\quad , \quad
\tilde{\Psi}(\tilde{u}) =
-  \int_{\tilde{\delta}+\I \mathbb{R}}
\frac{\rmd \tilde y}{2\I \pi}  \mathrm{Li}_2\left(- \frac{\tilde{u} e^{ \tilde y^2 T+ \tilde y X   }}{(\tilde{\varrho}_1-\tilde y)(\tilde y-\tilde{\varrho}_2) }  \right)
\eea
The integration contour is now chosen so that $\tilde{\varrho}_2<\tilde{\delta}<\tilde{\varrho}_1$.
Consider for simplicity the case $X=0$, $T=1$ and $\tilde \rho_1=- \tilde \rho_2=\tilde w$.
Then we can choose $\tilde \delta=0$ and $\tilde y = \I k$. This leads to
\be  \label{finalPsiKPZ}
\tilde{\Psi}(\tilde{u}) = -  \int_{\mathbb{R}}
\frac{\rmd k}{2 \pi}  \mathrm{Li}_2\left(- \frac{\tilde{u} e^{ - k^2   }}{k^2 + \tilde w^2}\right)
\ee
which is exactly the rate function for the KPZ equation with a double-sided Brownian initial condition
obtained in Ref.~\cite{krajenbrink2017exact}. There is was obtained in another
form which is equivalent to \eqref{finalPsiKPZ} as shown in
\cite[Eq.~(156)]{ProlhacKrajenbrink} (see also \cite{krajenbrink2019beyond}).

Hence we obtain that in the large $\nu$ limit the r.h.s. of \eqref{observableB} takes the
form
\begin{equation}
\exp\left(- \frac{\Psi(u)}{ \varepsilon}  \right)  \to \exp\left(- \frac{\tilde{\Psi}(\tilde{u})}{ \nu^2 \varepsilon}  \right)
\end{equation}
This corresponds to a large deviation form with a speed $1/(\nu^2 \varepsilon)$,
which we can precisely identify with the speed $1/\sqrt{T_{\rm KPZ}}$
of the short-time weak noise KPZ equation, by defining the KPZ time as $T_{\rm KPZ}= \nu^4 \varepsilon^2  \ll 1$.\\

Let us examine now the limit of the l.h.s. of \eqref{observableB}.
The auxiliary quantities admit the following limit
\begin{equation} \label{omegatilde}
\omega =   \frac{\nu\tilde{\omega}}{2} +\mathcal{O}(1)
, \quad \tilde{\omega} = \frac{  -\tilde{\varrho}_1+\tilde{\varrho}_2+\sqrt{\left(\tilde{\varrho}_1-\tilde{\varrho}_2\right)^2+4
\tilde{u} Z}}{2 \tilde{u} Z}
\end{equation}
\bea
&&    F(\omega)=\frac{2}{\nu} \tilde{F}(\tilde{\omega})  +\mathcal{O}\left(\frac{1}{\nu^2}\right), \quad   \tilde{F}(\tilde{\omega})     =\left(\tilde{\varrho}_1-\tilde{\varrho}_2\right)
\log \frac{\tilde{\omega}\left(\tilde{\varrho}_1-\tilde{\varrho}_2\right)}{e}
+\frac{1}{\tilde{\omega} }
\eea
The auxiliary function $ \tilde{F}(\tilde{\omega}) $ matches exactly the one in \cite[Eq.~(71)]{krajenbrink2017exact} for the large deviation of stationary KPZ with the mapping $\tilde{\varrho}_1-\tilde{\varrho}_2=2\tilde{w}$ and $\tilde{\omega} = e^{\chi'}$. Note that these quantities solely depend on the difference of the rescaled densities, which corresponds to the local slope difference of the height. Upon our rescaling the observable is linearized as
\begin{equation}
\mathrm{Li}_2(-u \omega z) = -\frac{2\tilde{u}\tilde{\omega}Z}{\nu}+\mathcal{O}\left(\frac{1}{\nu^2}\right)
\end{equation}
so that the generalised generating function in the l.h.s. of \eqref{observableB}
becomes (since preexponential factors can be neglected)
\begin{equation}
\label{eq:generating-func-convergence-kpz}
\int_1^{+\infty} d\omega  \, \langle e^{\frac{1}{2\nu \varepsilon} [{\rm Li}_2(- u \omega z(X,T)  ) -F(\omega)]} \rangle   \to
\int_{0}^{+\infty} d\tilde \omega \, \langle e^{-\frac{1}{\nu^2 \varepsilon} [\tilde{u}\tilde{w}Z +\tilde{F}(\tilde{\omega})]} \rangle
\end{equation}

In terms of the KPZ time $T_{\rm KPZ}= \nu^4 \varepsilon^2  \ll 1$ we finally obtain the
large $\nu$ limit of Eq.~\eqref{observableB} in the form
\begin{equation}
\int_{0}^{+\infty} d\tilde \omega \, \bigg\langle \exp\left(-\frac{ \tilde{u}\tilde{w}Z +\tilde{F}(\tilde{\omega})}{\sqrt{T_{\rm KPZ}}}\right) \bigg\rangle  \sim \exp\left(- \frac{\tilde{\Psi}(\tilde{u})}{ \sqrt{T_{\rm KPZ}}}  \right)
\end{equation}

which is exactly the same form as obtained in \cite{krajenbrink2017exact}.
There the partition sum $Z$ is such that
\be \label{logZ1}
\log Z = H_{\rm KPZ} = [ h_{\rm KPZ}(x,t) + t/12 + x^2/(4 t)]_{t= T_{\rm KPZ}}
\ee
where $h_{\rm KPZ}(x,t)$ satisfies the KPZ equation with unit space time white noise,
and with initial condition $h_{\rm KPZ}(x,0) = B(x) - w |x|$, where $B(x)$ is a two-sided standard Brownian.
Let us recall that $\tilde w = w T_{\rm KPZ}^{1/2}$ is the rescaled slope \cite{krajenbrink2017exact} (the same
variable as appears here). The limit $\tilde w \gg 1 $ then corresponds to the limit
of droplet initial conditions.
Here, in the
text we have defined $h = -J = \frac{H}{2 \nu}$, $z=e^H$, hence here we have
\be \label{logZ2}
\log Z = H + \nu^2 T - \nu X
\ee
where \eqref{logZ1} and \eqref{logZ2} allow for a precise identification
in the large $\nu$ limit.

\subsection{KPZ limit of the cumulants of the current}

We now check the KPZ limit on the cumulants of the current.
We consider the height field $H=-2 \nu J$ and use the parameterization
\eqref{param12}. The average of $H$ reads, from \eqref{Jav}
\begin{equation}
\langle H \rangle = \nu X - \nu^2 T + \log  \left(\frac{e^{\tilde{\varrho}_1 \left(\tilde{\varrho}_1 T+X\right)} \text{Erfc}\left(\frac{2 \tilde{\varrho}_1 T+X}{2
\sqrt{T}}\right)+e^{\tilde{\varrho}_2 \left(\tilde{\varrho}_2 T+X\right)} \text{Erfc}\left(-\frac{2 \tilde{\varrho}_2 T+X}{2 \sqrt{T}}\right)}{2}\right) \label{Hb}
\end{equation}
which is exact and does not involve any limit. From \eqref{logZ1} and \eqref{logZ2}
we see that the first two terms are a simple shift, and that
$\langle H_{\rm KPZ} \rangle$ is given only by the last term in \eqref{Hb}.\\

To compare with previous results, we set from now on $\tilde \rho_1=-\tilde \rho_2=\tilde w$
as well as $X=0$ and $T=1$. Then \eqref{Hb} reproduces exactly the result of Ref.~\cite[Eq.~(107)]{krajenbrink2017exact}, taking into account the shift identified above.
For the second cumulant,
taking the large $\nu$ result of our result displayed in
\eqref{secondcumappendix},
we obtain
\be \label{kappa2kpz}
\kappa_2 = \frac{1}{8 \tilde{w}} \left(
\frac{\left(1-4 \tilde{w}^2\right)
\text{Erfc}\left(\sqrt{2}
\tilde{w}\right)+2 \sqrt{\frac{2}{\pi }}
e^{-2 \tilde{w}^2}
\tilde{w}}{\text{Erfc}(\tilde{w})^2}-1 \right) + \mathcal{O}(\frac{1}{\nu^2})
\ee

This predicts that
\bea
\langle H^2 \rangle_c
= 4 \nu^2 \langle J^2 \rangle_c
= 4 \nu^2 \varepsilon  \kappa_2 = 4  \kappa_2 \varepsilon_{\rm KPZ}
\eea
which, thanks to the factor of $4$ is exactly the result in \cite[Eq.~(108)]{krajenbrink2017exact} (with $t^{1/2}=T_{\rm KPZ}^{1/2}=\varepsilon_{\rm KPZ}$).  We have also checked the third and fourth cumulant. The calculation
uses the method of Section~\ref{sec:cum} and was performed using Mathematica.
The prediction from the limit of our result is
\be
\label{eq:third-cum-kpz}
\langle H^3 \rangle_c =  - 8 \nu^3 \langle J^3 \rangle_c
= - 8 \nu^3  \kappa_3 \varepsilon^2 = - \frac{8}{\nu}  \kappa_3 \epsilon_{\rm KPZ}^2
\ee
We find indeed that $- \frac{8}{\nu}  \kappa_3 $ converges to the result of \cite[Eqs.~(110), (116)]{krajenbrink2017exact}. Finally we computed
\be \label{eq:fourth-cum-kpz}
\langle H^4 \rangle_c =  16 \nu^4 \langle J^4 \rangle_c
= 16 \nu^4 \kappa_4 \varepsilon^3 =  \frac{16}{\nu^2}  \kappa_4 \epsilon_{\rm KPZ}^3
\ee
and again we find perfect agreement with \cite[Eqs.~(111) and (117)]{krajenbrink2017exact}.

\subsection{KPZ limit of the MFT equations}
Finally, we
obtain the KPZ limit of the MFT equations by two methods: first directly on the stochastic version of the MFT equation (as in the text) and then on the nonlinear dynamical saddle point equations.
\begin{enumerate}
\item The stochastic MFT equation of the WASEP reads
\be \
\partial_t \rho =  \partial_x (  \partial_x \rho - \nu \sigma(\rho) + \sqrt{\varepsilon \sigma(\rho)}  \eta )
\ee
We first rescale the density around $1/2$, i.e., also the maximum of $\sigma(\varrho)$, as
\be
\rho = \frac{1}{2} + \frac{\tilde \rho}{2\nu}   \quad , \quad \sigma(\rho) = 2 \rho(1-\rho) = \frac{1}{2} \left(1 - \frac{\tilde \rho^2}{\nu^2}\right)
\ee
At large $\nu$, the dynamics of $\tilde \rho$ is governed by the Burgers equation
\be
\partial_t \tilde \rho =   \partial^2_x \tilde \rho  + \partial_x \tilde \rho^2 + 2 \nu \partial_x \sqrt{\frac{\varepsilon}{2}   }  \eta
\ee
Seeing the density as the slope of the height, i.e., $\tilde \rho=\partial_x H$, which is the same as
\be
\rho=  \partial_x h = \frac{1}{2 \nu} \partial_x H \quad , \quad h = \frac{H}{2 \nu} \quad , \quad z = e^{H}
\ee
(we recall $h=-J$)
it leads to the KPZ equation for $H$
\be
\partial_t H =   \partial^2_x H  + (\partial_x H)^2 +  2 \nu \sqrt{\frac{\varepsilon}{2}   }  \eta
\ee
and thus $\varepsilon_{\rm KPZ}= \nu^2 \varepsilon$ (recall that the KPZ case has noise with the following convention $\sqrt{2 \varepsilon_{\rm KPZ}} \eta$). The regime $\varepsilon_{\rm KPZ} \ll 1$ corresponds to the weak noise theory of KPZ, which is also its short-time regime.
\item The MFT equations directly describe the weak noise regime of the stochastic hydrodynamic equation of the WASEP. We now show that they simply converge to the weak noise equations of KPZ. One starts from the pair
\be
\begin{split}
\partial_t q  &       = \partial_x \left[
\partial_x q - 2 q(1-q) ( \partial_x p + \nu)
\right]
\:,
\\
- \partial_t p & =   \partial_x^2 p +  (1-2 q) \p_x p (   \partial_x p
+  2\nu  )
\:,
\end{split}
\ee
and we set the following change of variable
\be
q= \frac{1}{2}  + \frac{\tilde q}{2\nu} \quad , \quad p = \frac{2 \tilde p}{\nu}
\ee
One obtains at large $\nu$ to leading order
\be
\begin{split}
\partial_t \tilde q &= \partial_x^2 \tilde q + \partial_x \tilde q^2 - 2 \partial_x^2 \tilde p \\
- \partial_t \tilde p &= \partial_x^2 \tilde p - 2 \tilde q \partial_x \tilde p
\end{split}
\ee
Setting $\tilde q = \partial_x H$, $\tilde P = - \partial_x \tilde p$ one obtains
\be
\begin{split}
\partial_t H &= \partial_x^2 H + (\partial_x  H)^2 + 2 \tilde P   \\
- \partial_t \tilde P &= \partial_x^2 \tilde P - \partial_x ( 2  \tilde P \partial_x  H )
\end{split}
\ee
which are exactly the nonlinear equations associated to the weak noise theory of the KPZ equation, see Ref.~\cite[Eqs.~(S42)--(S43)]{UsWNTDroplet2021}.
\end{enumerate}
\begin{remark}
The limit from the MFT equation to the weak noise KPZ equation
can be performed around any mean density $\bar \rho$, by going to
a moving frame, and writing $\rho(x,t)=\bar \rho + \frac{\tilde \rho(y,t)}{2\nu}$
where $y= x - 2 \nu(1- 2 \bar \rho) t$. In the large $\nu$ limit
the MFT equation of the WASEP becomes the following stirred Burgers equation
\be
\partial_t \tilde \rho =   \partial^2_y \tilde \rho  + \partial_y (\tilde \rho^2) +  \partial_y ( \sqrt{8 \nu^2 \varepsilon  \bar \rho(1- \bar \rho)     }  \, \eta(y,t) )
\ee

which gives the KPZ equation \eqref{KPZeq} for $H$, with $\tilde \rho = \partial_y H$
and noise $\varepsilon_{\rm KPZ}= 4 \nu^2 \varepsilon \bar \rho(1- \bar \rho)$. Since $\sigma(\rho)$ is quadratic, the WASEP MFT equation is exactly a Burgers equation,
but with a noise of amplitude $\sqrt{4 \nu^2 \varepsilon \sigma(\rho)}$,
depending non-linearly on the density field. It is only for large $\nu$
and with the above scaling that the noise amplitude can be considered as constant
$\sqrt{4 \nu^2 \varepsilon \sigma(\bar \rho)}$. Note that the same rescaling can be performed for the general MFT equation,
with arbitrary $D(\rho)$ and $\sigma(\rho)$, with a boost $y= x - \nu \sigma'(\bar \rho) t$,
diffusion coefficient $D(\bar \rho)$, a non linear term $\frac{1}{4} \sigma''(\bar \rho) \tilde \rho^2$,
and a noise amplitude $\sqrt{4 \nu^2 \varepsilon \sigma(\bar \rho)}$.

\end{remark}

\section{Limit to the SSEP $\nu \to 0$} \label{sec:ssep} 

Here we show that if the asymmetry becomes zero, i.e., $\nu \to 0$, the result of this work matches the known results for the SSEP \cite{derrida2009currentExact,derrida2009currentMFT} for $X=0$, as well as those for $X \neq 0$
in \cite{imamura2017large,Imamura2021,mallick2024exact}. Since $z=e^{-2\nu J}$, the $\nu \to 0$ limit has to be taken carefully. First of all, denoting $\omega_u = \omega_{u,z=1}$, we have that
\begin{equation}
\frac{1}{2\nu}[{\rm Li}_2(- u \omega_{uz} z  ) -F(\omega_{uz})] \underset{\nu \to 0}{\simeq} - \frac{1}{2 \nu} \int_0^u  \frac{\rmd u'}{u'} \log(1+u' \omega_{u'}) +  \log(1+u\omega_{u} )  J +\mathcal{O}(\nu)
\end{equation}
Note that at this order we only need $\omega_{u,z=1}$ since $\omega_{u,z}$ is by definition the
extremum of the l.h.s.  In the remainder of this section we will prove that
\be \label{smallnu}
\Psi(u) = \frac{1}{2 \nu} \int_0^u  \frac{\rmd u'}{u'} \log(1+u' \omega_{u'}) + \Psi_0(u) +\mathcal{O}(\nu)
\ee
and obtain $\Psi_0(u)$. Reporting both results in either sides of Eq.~\eqref{observableB}
we see that the leading term $\mathcal{O}(1/\nu)$ cancels
on both sides, and we are left with the following large deviation principle for the SSEP, as the limit of the one of the WASEP
\begin{equation}
{ \langle }  e^{  \frac{\log(1+u\omega_{u} )  J}{\varepsilon}} { \rangle }  \sim e^{-\frac{\Psi_0(u)}{\varepsilon}}
\end{equation}
This large deviation result is more elegantly written using the reduced variable
\begin{equation}
P = \log(1+u\omega_{u} ) \Longleftrightarrow u=  \left(1-e^{-P}\right) \left(e^P-\alpha \right)
\end{equation}
This implies that in the SSEP limit $\nu \to 0$, the cumulant generating function is
\be \label{resssep}
\phi(P)|_{\nu =0}  = - \Psi_0(u)|_{u= \left(1-e^{-P}\right) \left(e^P-\alpha \right)}
\ee

Let us give here our result for $\Psi_0(u)$, which is proved below. In the case $X=0$
(setting $T=1$) we obtain
\be
\Psi_0(u)= \frac{1}{\sqrt{\pi}} \sum_{n \geq 1} \frac{(- \Omega)^n}{n^{3/2} } \quad , \quad
\Omega = u \rho_1 (1-\rho_2)
\ee
Now let us recall here the result of Derrida and Gershenfeld in \cite[Eqs.~(1-3)]{derrida2009currentExact} .
In our notations it reads
\be
\phi(P)= - \frac{1}{\sqrt{\pi}} \sum_{n \geq 1} \frac{(- \tilde \omega)^n}{n^{3/2}}
=  \int_\R \frac{\rmd k}{\pi} \log(1 + \tilde \omega e^{-k^2})  \quad , \quad
\tilde \omega= \rho_1(1-\rho_2) (e^P-1)+ \rho_2(1-\rho_1) (e^{-P}-1) \label{defomt}
\ee
for $\tilde \omega > -1$.
It is easy to
check that $\tilde \omega=\Omega$ and that our results are identical.

For $X \neq 0$ our result for $\Psi_0(u)$ reads
\begin{equation} \label{Psi0Apps}
\Psi_0(u)= \frac{2\xi\sqrt{T} }{\sqrt{\pi}} \int_{0}^\xi \rmd x \,  {\rm Li}_{1/2}(-\Omega e^{-x^2})+\sqrt{\frac{T}{\pi}}{\rm Li}_{3/2}(-\Omega e^{-\xi^2})+ \sqrt{T}  \xi \log \left( \frac{\left(1+u \omega _u\right) \left(1+\rho _1 u \omega _u\right)}{ \left(1+\left(1-\rho _2\right) u \omega
_u\right)}\right)
\end{equation}
where here and below we use the notations $\Omega = u \rho_1 (1-\rho_2)$ and $\xi=X/\sqrt{(4 T)}$.
We give below an equivalent form which allows to compare with the results of Refs.~\cite{imamura2017large,Imamura2021,mallick2024exact}.\\

In the rest of this Section, we first characterise the expansion of the derivatives of $\Psi(u)$ in the first two orders in $\nu$, then obtain the leading and subleading orders of $\Psi(u)$.
\begin{remark}
Note that our result \eqref{resssep} for $\nu \to 0$
is in agreement with the parametric representation of $\phi(P)$ obtained in \eqref{eq:param-repn-phi-P}
(valid for any $\nu$). Indeed, we have that
\begin{equation} \label{remarkssep}
\int_0^{u(P)}  \frac{\rmd u'}{u'} \log(1+u' \omega_{u'})= P \log u(P)  - \frac{P^2}{2} -\left( {\rm Li}_2\left(\alpha e^{-P}\right) - {\rm Li}_2\left(\alpha\right) \right)  - \left( {\rm Li}_2\left(e^{-P}\right)  - {\rm Li}_2\left(1\right) \right)
\end{equation}

when evaluated at $u(P)= \left(1-e^{-P}\right) \left(e^P-\alpha \right)$. This is easily checked by taking a derivative on both sides of \eqref{remarkssep}
and using the identification $P = \log(1+u\omega_{u} )|_{u=u(P)}$ which is valid for $\nu \to 0$.
\end{remark}

\subsection{Derivatives of $\Psi(u)$}
Starting from the expression of the derivatives of $\Psi(u)$ \eqref{eq:derivative_Psi_alltimes}, we obtain the lowest two orders as $\nu \to 0$.
\begin{equation}
\begin{split}
\Psi^{(n)}(0)=& - \frac{\rho _1^n \left(1-\rho _2\right)^n}{4\nu n }\bigg[\frac{d^{n-1}}{dy^{n-1}} \left( \frac{(y (1-y))^{n-1} e^{2 \nu  n y (2 \nu  T (y-1)+X)}}{  (y-\rho_1)^n }  \text{Erfc}\left(-\sqrt{\frac{n}{4T}} (2 \nu  T (2 y-1)+X)\right)\right)|_{y=\rho_2} \\
&+(-1)^{n-1}
\frac{d^{n-1}}{dy^{n-1}} \left( \frac{(y (1-y))^{n-1}e^{2 \nu  n y (2 \nu  T (y-1)+X)}}{  (\rho_2-y)^n }  \text{Erfc}\left( \sqrt{\frac{n}{4T}} (2 \nu  T (2 y-1)+X)\right)\right)|_{y=\rho_1} \bigg]
\end{split}
\end{equation}

In order to set  $\nu \to 0$, one needs to expand the error and exponential functions as
\begin{equation}
\begin{split}
&e^{2 \nu  n y (2 \nu  T (y-1)+X)}\text{Erfc}\left(\pm \sqrt{\frac{n}{4T}} (2 \nu  T (2 y-1)+X)\right) \\
&= \text{Erfc}\left(\pm X\sqrt{\frac{n}{4T}} \right)+ \left(  2 n X y \text{Erfc}\left(\pm  X \sqrt{\frac{n}{4T}}\right)\pm \sqrt{\frac{4nT}{\pi }}   (1-2 y)  e^{-\frac{n X^2}{4 T}}\right)  \nu +\mathcal{O}(\nu^2)
\end{split}
\end{equation}

Using that
\begin{equation}
\label{eq:magic-identity1}
\begin{split}
\frac{d^{n-1}}{dy^{n-1}} \left( \frac{(y (1-y))^{n-1}}{  (y-\rho_1)^n } \right)|_{y=\rho_2} =(-1)^{n-1}
\frac{d^{n-1}}{dy^{n-1}} \left( \frac{(y (1-y))^{n-1}}{  (\rho_2-y)^n } \right)|_{y=\rho_1}
\end{split}
\end{equation}
and $\text{Erfc}(x)+\text{Erfc}(-x)=2$, we find that the leading order is independent of $(X,T)$ and reads
\begin{equation} \label{psidernu0}
\Psi^{(n)}(0) = - \frac{\rho _1^n \left(1-\rho _2\right)^n}{2\nu n } \frac{d^{n-1}}{dy^{n-1}} \left( \frac{(y (1-y))^{n-1}}{  (y-\rho_1)^n } \right)|_{y=\rho_2}+ \Psi_0^{(n)}(0) +\mathcal{O}(\nu)
\end{equation}
To get the next order, we use that
\begin{equation}
\label{eq:magic-identity2}
\begin{split}
\underbrace{\frac{d^{n-1}}{dy^{n-1}} \left( \frac{ y^n(1-y)^{n-1}}{  (y-\rho_1)^n } \right)|_{y=\rho_2} }_{I_1}=\underbrace{(-1)^{n-1}
\frac{d^{n-1}}{dy^{n-1}} \left( \frac{y^n  (1-y)^{n-1}}{  (\rho_2-y)^n } \right)|_{y=\rho_1}}_{I_2}+(-1)^{n-1}(n-1)!
\end{split}
\end{equation}

Denoting
\begin{equation}
\begin{split}
\Upsilon_1 &=   2 n X  \text{Erfc}\left(  X \sqrt{\frac{n}{4T}}\right)- \sqrt{\frac{16nT}{\pi }}     e^{-\frac{n X^2}{4 T}}= -2n \int_X^{+\infty} \rmd x \, \text{Erfc}\left(x \sqrt{\frac{n}{4T}}\right)\\
\Upsilon_2 &= 2n X
\end{split}
\end{equation}
we have that the next order is given by
\begin{equation}
\label{eq:derivative-ssep-next-to-leading-order-Psi0}
\begin{split}
\Psi_0^{(n)}(0)&=  - \frac{\rho _1^n \left(1-\rho _2\right)^n}{4 n } \left( (2\Upsilon_2-\Upsilon_1)I_1+\Upsilon_1 I_2 \right) \\
&=  - \frac{\rho _1^n \left(1-\rho _2\right)^n}{4 n } \left( 2\Upsilon_2 I_1+(-1)^n (n-1)!\Upsilon_1  \right) \\
&=  -\frac{\rho _1^n \left(1-\rho _2\right)^n }{2}\left(2X \frac{d^{n-1}}{dy^{n-1}} \left( \frac{ y^n(1-y)^{n-1}}{  (y-\rho_1)^n } \right)|_{y=\rho_2}+(-1)^{n-1}(n-1)!\int_X^{+\infty} \rmd x \, \text{Erfc}\left(x \sqrt{\frac{n}{4T}}\right) \right)
\end{split}
\end{equation}
\begin{remark}  The identities \eqref{eq:magic-identity1} and \eqref{eq:magic-identity2} can be proved by considering the residue at infinity (zero in the first case
and $(-1)^n$ in the second case) of the contour integral analog to \eqref{eq:supp-mat-large-dev-ssep-limit-leading-order} below where the logarithm is expanded as a series.
\end{remark}
\subsection{Determination of the leading order of $\Psi(u)$}
\label{subsec:leading-order-ssep-identity}
As we now show, to obtain the leading order of the $\nu \to 0$ limit of $\Psi(u)$, it is sufficient to set $\nu=0$ inside the integrand in \eqref{eq:rate-function-wasep-general}. To see this, first note that from \eqref{uPsi} we have the following relation at the saddle point
\be
\Psi'(u)= \frac{1}{2 \nu} \frac{\log (1+u \omega_{u,z} z)}{u}   \quad , \quad  z=z(u)=e^{-2 \nu J(u)}
\ee

As $\nu\to 0$, the right hand side of this equation can be treated as if $z=1$ at leading order
\begin{equation}
u  \Psi'(u) =  \frac{1}{2 \nu}  \log(1+u \omega_{u}) +\mathcal{O}(\nu^0)
\end{equation}
On the other side, by setting $\nu =0$ in the integrand of \eqref{eq:rate-function-wasep-general}, we obtain that
\be
\label{eq:supp-mat-large-dev-ssep-limit-leading-order}
u  \Psi'(u)  = \frac{1}{2\nu} \int_{\I \mathbb{R} + \delta}
\frac{\rmd y}{2\I \pi y(1-y)}  \log\left( 1 + u\frac{\rho _1 \left(1-\rho _2\right) (y-1) y}{\left(y-\rho_1\right)
\left(y-\rho _2\right)}  \right)      +\mathcal{O}(\nu^0)
\ee

We now show that these two expressions are indeed compatible. The integral \eqref{eq:supp-mat-large-dev-ssep-limit-leading-order} can be performed by integration by part and the boundary term vanishes at infinity.
One replaces $u= \frac{1 + (\alpha-1) \omega}{\omega (\omega-1) } $ where $\omega=\omega_u$. The integrand becomes a rational fraction of $y$ which behaves as $1/y^2$ at infinity and with the following poles
$y_i$
\begin{equation}
y=\left\{ \rho _1, \frac{\rho _1 (\omega
-1)}{\omega -\rho _1} , \rho
_2, \frac{\rho _2 \omega }{\rho _2+\omega
-1}\right\}
\end{equation}
and associated residues $R_i$
\begin{equation}
\begin{split}
& R= \left\{\log \left(\frac{\rho _1}{1-\rho _1}\right),\log
\left(\frac{\left(1-\rho _1\right) \omega }{\rho _1 (\omega
-1)}\right),    \log \left(\frac{\rho _2}{1-\rho
_2}\right),\log \left(\frac{\left(1-\rho _2\right) (\omega
-1)}{\rho _2 \omega }\right)\right\}
\end{split}
\end{equation}
so that up to terms of order $\mathcal{O}(\nu)$ we obtain
\be
2 \nu  u  \Psi'(u)  =  R_2 + R_3 = - (R_1 + R_4)   = \log\left( \frac{\alpha \omega}{\omega-1} \right) =
\log(1+u \omega_u)
\ee

\subsection{Determination of the subleading order of $\Psi(u)$}
To obtain the expression of $\Psi_0(u)$, we proceed to a resummation of its derivatives obtained in \eqref{eq:derivative-ssep-next-to-leading-order-Psi0}. We will show a perfect matching with \cite[Eq.~(6.35)]{mallick2024exact} for any $X$.
We recall that we use
the notations $\Omega = u \rho _1 (1-\rho _2)$, $\xi=\sqrt{X/(4 T)}$ and $P=\log (1+u \omega_u)$.
Note that it implies that $\Omega = \tilde \omega$ defined in \eqref{defomt}.
We first introduce the identity
\begin{equation}
\begin{split}
\sum_{n\geq 1}\frac{\rho _1^n \left(1-\rho _2\right)^n u^n }{n!}\frac{d^{n-1}}{dy^{n-1}} \left( \frac{ y^n(1-y)^{n-1}}{  (y-\rho_1)^n } \right)|_{y=\rho_2} &= -\frac{1}{2}\left( -\log(1+u \varrho_1(1-\varrho_2))+\log \frac{1+(e^{P}-1)\varrho_1}{1+(e^{-P}-1)\varrho_2}\right)\\
&= -\frac{1}{2}\log \left( \frac{\left(1+u \omega _u\right) \left(1+\rho _1 u \omega _u\right)}{\left(1+\rho _1 \left(1-\rho _2\right) u\right) \left(1+\left(1-\rho _2\right) u \omega
_u\right)}\right)
\end{split}
\end{equation}
\begin{remark}
We checked this identity with Mathematica to high orders. It is possible that it can be proven using the Lagrange inversion theorem.
\end{remark}
Part of the resummation of the derivatives of $\Psi_0(u)$ involve manipulating error functions as follows
\bea
&& -\frac{1}{2} \sum_{n \geq 1}
\frac{( u \rho _1 (1-\rho _2))^n}{n!}
(-1)^{n-1}(n-1)!\int_X^{+\infty} \rmd x \,
\text{Erfc}\left(x \sqrt{\frac{n}{4T}}\right)  \\
&& = \sqrt{T}   \sum_{n \geq 1} \frac{(-\Omega)^n}{n^{3/2} } \left( \frac{e^{- n \xi^2}}{\sqrt{\pi}}   - \sqrt{n} \xi {\rm Erfc}(\sqrt{n} \xi) \right) \\
&& = \sqrt{T}   \sum_{n \geq 1} \frac{(-\Omega)^n}{n^{3/2} } \left( \frac{e^{- n \xi^2}}{\sqrt{\pi}}   + \sqrt{n} \xi {\rm Erf}(\sqrt{n} \xi) \right)
- \sqrt{T} \xi   \sum_{n \geq 1} \frac{(-\Omega)^n}{n }
\eea

where we recall that we introduced the notation $\xi=X/\sqrt{4 T}$.  Combining these intermediate identities allows to sum the derivatives of $\Psi_0(u)$ to reconstruct the function as
\be
\begin{split}
\Psi_0(u) &= \sqrt{T}   \sum_{n \geq 1} \frac{(-\Omega)^n}{n } \left( \frac{e^{- n \xi^2}}{\sqrt{n\pi}}   +  \xi {\rm Erf}(\sqrt{n} \xi) \right)
+ \sqrt{T}  \xi \log \frac{1+(e^{P}-1)\varrho_1}{1+(e^{-P}-1)\varrho_2} \\
&=  \sqrt{T}   \sum_{n \geq 1} \frac{(-\Omega)^n}{n } \left(\int_{0}^\xi \rmd x {\rm Erf}(\sqrt{n}x)+\frac{1}{\sqrt{n \pi}}\right)
+ \sqrt{T}  \xi \log \left( \frac{\left(1+u \omega _u\right) \left(1+\rho _1 u \omega _u\right)}{ \left(1+\left(1-\rho _2\right) u \omega
_u\right)}\right) \\
&=  \sqrt{T}   \sum_{n \geq 1} \frac{(-\Omega)^n}{n } \int_{0}^\xi \rmd x {\rm Erf}(\sqrt{n}x)+\sqrt{\frac{T}{\pi}}{\rm Li}_{3/2}(-\Omega)
+ \sqrt{T}  \xi \log \left( \frac{\left(1+u \omega _u\right) \left(1+\rho _1 u \omega _u\right)}{ \left(1+\left(1-\rho _2\right) u \omega
_u\right)}\right) \\
\end{split} \label{Psi0Mallick}
\ee

The first line is exactly \cite[Eq.~(6.35)]{mallick2024exact}
with the conventions matched as follows
$\Psi_0(u) \equiv - \sqrt{T} \mu(\lambda)$, $\Omega =u \rho_1 (1-\rho_2) \equiv \omega $,  $P \equiv \lambda$, $\rho_1 \equiv \rho_-$, $\rho_2 \equiv \rho_+$ and $\xi=X/\sqrt{4 T}$
. One can further resum the first term using the integral definition of the error function ${\rm Erf}(\sqrt{n}x)=\frac{2\sqrt{n}}{\sqrt{\pi}}\int_0^x\rmd t e^{-nt^2}$ to obtain
\begin{equation}
\begin{split}
\sum_{n \geq 1} \frac{(-\Omega)^n}{n } \int_{0}^\xi \rmd x {\rm Erf}(\sqrt{n}x) &=   \frac{2}{\sqrt{\pi}} \int_{0}^\xi \rmd x \int_0^{x} \rmd t \, {\rm Li}_{1/2}(-\Omega e^{-t^2}) =  \frac{2}{\sqrt{\pi}} \int_{0}^\xi \rmd x \, (\xi-x) {\rm Li}_{1/2}(-\Omega e^{-x^2})\\
&=  \frac{2\xi}{\sqrt{\pi}} \int_{0}^\xi \rmd x \,  {\rm Li}_{1/2}(-\Omega e^{-x^2})+\frac{1}{\sqrt{\pi}}({\rm Li}_{3/2}(-\Omega e^{-\xi^2})-{\rm Li}_{3/2}(-\Omega ))
\end{split}
\end{equation}

and thus we obtain our final result displayed in \eqref{Psi0Apps}.
\begin{remark}
In the stationary limit $\varrho_1=\varrho_2$, the last term will admit an expansion in powers of $\sqrt{u}$.
\end{remark}
\section{Tail $J \to +\infty$ of the distribution of the integrated current  }
Here we study the tail of the distribution of the integrated current for $J \to +\infty$,
i.e.,the asymptotic behavior of $\Phi(J)$ for $J \to +\infty$. It is obtained from
the asymptotics of the rate function $\Psi(u)$ in
the limit $u \to +\infty$. It is simple to extract, as it involves only the main branch of the large deviation function,
i.e.,the formula \eqref{eq:rate-function-wasep-general} for $\Psi(u)$. This tail corresponds to the {\it lower tail} for the height field $h=-J$. \\

This tail is known for the SSEP, in which case it is
cubic $\Phi(J) \propto J^{3}$ \cite{derrida2009currentExact,derrida2009currentMFT}, as well as for the KPZ equation,
in which case it exhibits a $5/2$ exponent instead, i.e.,$\Phi(J) \propto J^{5/2}$ \cite{le2016exact,krajenbrink2017exact,JanasDynamical,meerson2016large,kolokolov2007optimal}.
In that case it is called the lower tail since the KPZ height field is $H = - 2 \nu J$.  We first compute the tail for the WASEP, i.e., for a fixed $\nu>0$, and then we will study these two limits,
respectively $\nu \to 0$ and $\nu \to +\infty$. This can be done from
our general formula
for $\Phi(J)$ (at arbitrary $J$ and finite $\nu$) given in Section~\ref{sec:parametricphi}.
Since we find that the tail for the
WASEP is cubic, and with the same amplitude as for the SSEP,
the crossover to the SSEP is simple at leading order (it may
be more delicate at subleading orders).
We will see however that in the
case of the KPZ limit, the crossover from the leading orders $J^3$ to $J^{5/2}$ is nontrivial, and we will
compute the full crossover function which describes it.
\subsection{$J \to +\infty$ tail for the WASEP}
\label{subsec:lower-tail-wasep}
For general $\nu>0$, let us examine the equations \eqref{ParamPhi} as $u \to +\infty$.
In that limit one has $u \Psi'(u) \to +\infty$ logarithmically in $u$ and $\omega_{u}=1+ \frac{\alpha}{u}+ \mathcal{O}(1/u^2)$, thus the various expansions read
\bea \label{legendreasymptotic}
&& \zeta(u) = e^{  2\nu u \Psi'(u)} - (1 + \alpha) + \alpha e^{  - 2\nu u \Psi'(u)} \\
&& J  =  -  u \Psi'(u) + \frac{1}{2\nu} \log u + \frac{1+\alpha}{2 \nu} e^{  - 2\nu u \Psi'(u)} (1 + o(1))  \nn  \\
&& \Phi'(J) = 2\nu u \Psi'(u) \nn
\eea

Note that we have neglected additional $1/u$ corrections since the final expansion will be in $1/\log u$.
It turns out that the terms $e^{  - 2\nu u \Psi'(u)}$ will also be of order $1/u$
and will be neglected later on.\\

Let us now evaluate $u \Psi'(u)$ at large $u$ starting from the general expression
\be
u \Psi'(u) =  \frac{1}{2\nu} \int_{\I \mathbb{R} + \delta} \frac{\rmd y}{2\I \pi y(1-y)} \log\left(1+ u\frac{\rho _1 \left(1-\rho _2\right) (1-y) y}{\left(y-\rho _1\right)
\left(\rho_2-y\right)} e^{- 4\nu^2 y(1-y)  T + 2\nu yX } \right)
\ee

To simplify the analysis we choose $X=0$, $\rho_1= \frac{1}{2} + w$, $\rho_2= \frac{1}{2} - w$, where $ 0 \leq w \leq 1/2$, and $\delta=1/2$, which leads to
\bea \label{startingpoint}
u \Psi'(u) = \int_{ \mathbb{R}} \frac{\rmd k \, }{\pi \nu \left(4
k^2+1\right)  }
\log \left(1 + u \frac{\left(4 k^2+1\right)  (2 w+1)^2
e^{- (4 k^2+1) \nu^2 T }}{16
\left(k^2+w^2\right)}\right)
\eea

Let us define $k_0>0$ such that
\begin{equation}
u=\frac{16\left(k_0^2+w^2\right)}{\left(4 k_0^2+1\right)  (2 w+1)^2 e^{- (4 k_0^2+1) \nu^2 T }} := f(k_0)
\end{equation}
which always exist for $u \geq f(0)$ since $f(k_0)$ in an increasing function of $k_0^2$. We now study the limit $\nu$ fixed and $u \to +\infty$, which implies that $k_0 \to +\infty$ since one has
\be \label{seriesk0}
k_0^2 = \frac{\log (u)}{4 \nu
^2 {T}}+\frac{1}{4}
\left(\frac{\log (4)}{\nu
^2 {T}}-1\right)+\frac{\frac{1}
{4}-w^2}{\log
(u)}+\mathcal{O}\left(\frac{1}{
\log (u)}\right)^2
\ee
In that limit it is convenient
to split the integral in \eqref{startingpoint} in two parts. First, by parity we restrict the integral to $[0,\infty[$ and we split the domain of integration on $[0,k_0]$ and $[k_0,\infty[$.
\begin{itemize}
\item Let us show that the first part, i.e., $k \in [k_0,+\infty[$ is
negligible in the limit. To this aim we set
$k=k_0 + q/(8 k_0 \nu {\sqrt{T}})$ and expand in powers of $1/k_0$
\be
\begin{split}
& 2 \int_{k_0}^{+\infty}  \frac{\rmd k \, }{\pi \nu \left(4
k^2+1\right)  }  \log(1 + \frac{f(k_0)}{f(k)})
\\
&= \frac{1}{4 k_0 \pi \nu^2 {T}} \int_0^{+\infty} dq  (\frac{1}{4 k_0^2}-\frac{\nu
^2+q}{16 k_0^4 \nu
^2 {T}}+ \dots ) \log( 1 + e^{-q}-\frac{e^{-q} q^2}{16
k_0^2 \nu
^2 {T}}+ \dots  ) \nonumber \\
& \simeq  \frac{1}{16 k_0^3 \pi \nu^2 {T}} \frac{\pi^2}{12}
\end{split}
\ee

Hence this part is negligible at large $k_0$.
\item In the second part $k \in [0,k_0]$, we split the logarithm in the integrand as
\be \label{sumlog}
\log(1 + \frac{f(k_0)}{f(k)}) = \log(\frac{f(k_0)}{f(k)}) + \log(1 + \frac{f(k)}{f(k_0)})
\ee
The integral associated to the second term can be bounded from above by
\be
\begin{split}
&2 \int_{0}^{k_0}  \frac{\rmd k \, }{\pi \nu } \frac{1}{4k^2+1} \log(1 + \frac{f(k)}{f(k_0)}) < 2 \int_{0}^{k_0}  \frac{\rmd k \, }{\pi \nu }  \log(1 + \frac{f(k)}{f(k_0)}) \\
&\simeq \frac{2}{8 \pi k_0 \nu^2 {T} } \int_0^{8 k_0^2 \nu {\sqrt{T}}} dq
\log( 1 +
e^{-q}+\frac{e^{-q} q^2}{16
k_0^2 \nu
^2 {T}}+ \dots)  \\
&\simeq \frac{2}{8 \pi k_0 \nu^2 {T}} \frac{\pi^2}{12}
\end{split}
\ee
where we have set $k=k_0 - q/(8 k_0 \nu {\sqrt{T}})$. It is of order $\mathcal{O}(1/k_0)$
and we will neglect it.
\item Hence we only need
to study the first term in \eqref{sumlog}. In summary we
obtain
\be
\begin{split}
& u \Psi'(u)
= 2\int_{ 0}^{k_0} \frac{\rmd k \, }{\pi \nu \left(4
k^2+1\right)  }  (\log(f(k_0))-\log(f(k))) + \mathcal{O}(1/k_0) \\
&= \int_{ 0}^{k_0} \frac{\rmd k \, }{\pi \nu  } \arctan(2k)\frac{f'(k)}{f(k)}  + \mathcal{O}(1/k_0) \\
\end{split}
\ee
It is easy to obtain the expansion of this integral at large $k_0$.
\end{itemize}

 This leads to
\begin{equation}
u \Psi'(u)
= 2 k_0^2 \nu  T-\frac{4 k_0 \nu  T}{\pi }+
\frac{\nu T}{2} + \frac{1}{\nu} B_w  +\mathcal{O}\left(\frac{1}{k_0}\right)
\end{equation}
where
\be B_w = \int_0^{+\infty} \rmd k \,
\frac{2 k \left(1-4 w^2\right)
\arctan(2 k)}{\pi
\left(4 k^2+1\right)
\left(k^2+w^2\right)}  \quad , \quad B_{1/2}=0 \quad , \quad B_0= \log 2
\ee
Inserting \eqref{seriesk0} one finds
\bea
\label{eq:asymptot-u-psip-plus-infty}
u \Psi'(u) = \frac{\log u}{2 \nu} - \frac{2 {\sqrt{T}}}{\pi} \sqrt{\log u}  + b +
\mathcal{O}(1/\sqrt{\log u})\quad , \quad b= \frac{1}{\nu} ( B_w  + \log 2)
\eea
Let us now use \eqref{legendreasymptotic}, up to terms of order $1/u$
\bea
&& J  \simeq  -  u \Psi'(u) + \frac{\log u}{2\nu}     \\
&& \Phi'(J) = 2\nu u \Psi'(u)
\eea
Hence we see that there is a cancellation of the leading term
in $J$ resulting in
\bea
&& J  =  \frac{2 {\sqrt{T}}}{\pi} \sqrt{\log u}  - b + \mathcal{O}(1/\sqrt{\log u})    \\
&& \Phi'(J) = \log u  - \frac{4 \nu {\sqrt{T}}}{\pi} \sqrt{\log u}  + 2 \nu b  + \mathcal{O}(1/\sqrt{\log u})
\eea
Hence we find for $J \to +\infty$
\be \label{tailwasepfinal}
\Phi(J)= \frac{\pi ^2 J^3}{12 {T}}+J^2
\left(\frac{\pi ^2
b}{4 {T}}-\nu \right) + \mathcal{O}(J)
\ee

\subsection{$J \to +\infty$ tail in the SSEP limit}
\label{subsec:lower-tail-ssep}

To obtain the tail of the current distribution in the SSEP limit, we first proceed to the limit $\nu \to 0$ and then take $u \to + \infty$. We show in \eqref{smallnu} that the rate function $\Psi(u)$ admits the following expansion
\begin{equation}
u\Psi'(u)=\frac{\log (1 + u \omega_u)}{2\nu}+u\Psi_0'(u)+\mathcal{O}(\nu)
\end{equation}
where $\Psi_0(u)$ is given explicitly in \eqref{Psi0Apps}. Consider now the equations \eqref{ParamPhi}
and perform the expansion at small $\nu$ and fixed $u$. One first obtains the expansion of $\zeta(u)$
\begin{equation}
\zeta(u)=u + 2\nu \sqrt{\left(\alpha +u -1\right)^2+4 u }  \, u\Psi_0'(u)+\mathcal{O}(\nu^2)
\end{equation}
where we used \eqref{omegauz}.
leading to the expansion of the parametric representation of the current rate function
\begin{equation}
\begin{split}
J&= - \sqrt{\left(\alpha +u -1\right)^2+4 u } \, \Psi_0'(u)+\mathcal{O}(\nu)\\
\Phi'(J)&= \log(1 + u \omega_u) + \mathcal{O}(\nu) =
\log \left(\frac{1+u+\alpha + \sqrt{\left(\alpha +u -1\right)^2+4 u }  }{2}\right)+\mathcal{O}(\nu)
\end{split}
\end{equation}

In the limit of large Laplace parameter $u \to + \infty$, we use the asymptotics of the polylogarithm appearing in \eqref{Psi0Apps}, namely, for $s$ a non-negative integers, one has to leading order
\be
\label{eq:asymptotic-polylog0}
\mathrm{Li}_s(-e^{\mu})\underset{\mu \to +\infty}{\simeq} - \frac{\mu^s}{\Gamma(s+1)}
\ee
From it,
we obtain the leading asymptotics
\begin{equation}
u\Psi_0'(u) = - \frac{2\sqrt{T}}{\pi} \sqrt{\log u}+\mathcal{O}(1)
\end{equation}
independently of $\xi=X/\sqrt{4 T}$. This leads to
to the parametric representation for large $J \to + \infty$
\begin{equation}
\begin{split}
J&\simeq  \frac{2\sqrt{T}}{\pi} \sqrt{\log u}\\
\Phi'(J)&\simeq\log u \simeq  \left(\frac{\pi J}{2\sqrt{T}}\right)^2
\end{split}
\end{equation}
We thus obtain the right tail of the current distribution as
\begin{equation}
\Phi(J) \underset{J \gg 1}{\simeq} \frac{\pi^2 J^3}{12 T}
\end{equation}
which is consistent with the result of \cite{derrida2009currentExact,derrida2009currentMFT}.
Note that it holds for any $X,T$, i.e., the
leading term is independent of $\xi=X/\sqrt{4 T}$
which only appears in subdominant contributions.
\begin{remark}
For the SSEP the result \cite[Eq.~(5)]{derrida2009currentExact}
\be
\Phi(J)= \frac{\pi^2}{12} J^3 - J \log(\rho_1(1-\rho_2)) + \mathcal{O}(1)
\ee

Comparing with our result for the WASEP \eqref{tailwasepfinal}, this suggests that there will
be a crossover from WASEP to SSEP in the subdominant terms $\mathcal{O}(J,J^2)$
of the tail.
\end{remark}

\subsection{$J \to +\infty$ tail in the KPZ limit}
\label{subsec:lower-tail-kpz}
To obtain the tail of the current distribution in the KPZ limit, we first proceed to the limit $\nu \to \infty$ and only later take the rescaled Laplace parameter $\tilde{u}\to + \infty$. We use the results and notations of Section
\ref{sec:kpzlimitlargedev}.
In the limit $\nu \to \infty$ at fixed $\tilde u$ the large deviation function and the auxiliary variables admit the following expansion
\bea  \label{limitKPZparameters}
&& \Psi(u) = \frac{1}{\nu^2} \tilde \Psi(\tilde u)+\mathcal{O}(\frac{1}{\nu^3}) \quad , \quad  u=\frac{4 \tilde{u} e^{\nu ^2 T-\nu  X}}{\nu ^2}
\quad , \quad  \alpha = 1-4\frac{\tilde w}{\nu}+\mathcal{O}\left(\frac{1}{\nu^2}\right), \quad \tilde{w}=\frac{\tilde{\varrho}_1-\tilde{\varrho}_2}{2}
\eea
where $\tilde{\Psi}(\tilde{u})$ is given in
\eqref{PsiKPZgeneral}, and in \eqref{finalPsiKPZ} for $X=0$ and $T=1$.
Consider again the equations \eqref{ParamPhi}
and perform the expansion \eqref{limitKPZparameters}
at small $\nu$ and fixed $\tilde u$. One first finds that
$u \Psi'(u) = \frac{1}{\nu^2} \tilde u \tilde \Psi'(\tilde u)$
and, being careful that one must also expand $\alpha$ near unity,
one obtains the following expansion of $\zeta(u)$
\be
\zeta(u) = \frac{4}{\nu^2}  \tilde u \tilde \Psi'(\tilde u) (  \tilde u \tilde \Psi'(\tilde u) + 2 \tilde w) +\mathcal{O}(\frac{1}{\nu^3})
\ee
Additionally
we define the reduced variables
\begin{equation}
\zeta(u)=\frac{4}{\nu^2}\tilde{\zeta}(\tilde{u}), \quad     \omega_{\zeta(u)} =   \frac{\nu\tilde{\omega}_{\tilde{\zeta}(\tilde{u})}}{2} +\mathcal{O}(1)
, \quad \tilde{\omega}_{\tilde{\zeta}(\tilde{u})} = \frac{ \sqrt{\tilde{w}^2+   \tilde{\zeta}(\tilde{u})}-\tilde{w}}{\tilde{\zeta}(\tilde{u})}
\end{equation}

This leads to the expansion of the parametric representation of the current rate function as
\begin{equation}
\begin{split}
J&=-\frac{1}{2\nu}\log \left(\tilde{\Psi}'(\tilde{u})(\tilde{u}\tilde{\Psi}'(\tilde{u})+2\tilde{w})\right)+\frac{\nu T}{2}-\frac{X}{2}+\mathcal{O}(\frac{1}{\nu^2})\\
\Phi'(J)&=
\frac{2}{\nu}\tilde u \tilde \Psi'(\tilde u)+\mathcal{O}(\frac{1}{\nu^2})\\
\end{split} \label{KPZrateJ}
\end{equation}
From \eqref{PsiKPZgeneral} in the simplest case $\tilde \rho_2=-\tilde \rho_1$ one finds
\be
\tilde u\tilde{\Psi}'(\tilde u) =   \int_{\mathbb{R}}
\frac{\rmd k}{2 \pi}  \log \left( 1 + \frac{\tilde{u} e^{ - k^2 T + \I k X  }}{k^2 + \tilde w^2}\right)
\ee
The large $\tilde u$ asymptotics was performed in \cite[Eq.~(85)]{krajenbrink2017exact}
carefully and including subdominant terms. Here we
show the simplest way to obtain the leading asymptotics for $\tilde u \to + \infty$.
For this, we rescale the integration variable
\be
k = \frac{ \sqrt{\log \tilde u}}{\sqrt{T}} \, p
\ee
so that to leading order one has $\tilde{u} e^{ - k^2 T + \I k X  } \simeq e^{ \log \tilde u (1-p^2)}$.
Replacing $\log(1+ e^Y) \simeq \max(0,Y)$ at large $Y$, one finds the asymptotics
\be \label{uPsiprimelarge}
\tilde{u}\tilde{\Psi}'(\tilde u) \simeq \frac{1}{2 \pi \sqrt{T}} (\log \tilde u)^{3/2}
\int_{-1}^{+1} dp  (1-p^2) = \frac{2}{3 \pi \sqrt{T}} (\log \tilde u)^{3/2}
\ee
at fixed $X$ and $\tilde w$ (the dependence in these parameters
being subdominant in that limit).
Since the KPZ height is related to the WASEP current as $H_{\rm KPZ}= - 2 \nu J  + \nu^2 T - \nu X$ from \eqref{KPZrateJ} and \eqref{uPsiprimelarge}
we thus obtain the right tail of the distribution of the current
\begin{equation}
\Phi(J)\underset{J\gg 1}{\simeq} \frac{16 \sqrt{2\nu}}{15\pi \sqrt{T}}(J-\frac{\nu T}{2}+\frac{X}{2})_+^{5/2} \sim
\frac{1}{\nu^2 \sqrt{T}} \frac{4}{15 \pi } |H_{\rm KPZ}|^{5/2}
\end{equation}
which is in agreement with the known results \cite{le2016exact,krajenbrink2017exact,JanasDynamical,meerson2016large,kolokolov2007optimal}.

\subsection{Crossover between the $J^3$ tail of the WASEP and the $J^{5/2}$ tail of KPZ}
\label{subsec:lower-tail-crossover}

For a large $\nu$, it seems natural to expect a crossover between the exponents $3$ and $5/2$ in the right tail of the current distribution.
The crossover between the two tails occurs when $J-\nu T/2 \sim \nu T$ and we thus define a reduced variable $\tilde{J}=(J-\nu T/2)/(2\nu T)$ and show that the crossover large deviation tail has the scaling form
\begin{equation}
\label{eq:lower-tail-crossover}
P(J,\nu,T)\underset{\varepsilon\ll 1}{\sim} \exp \left(-\frac{T^2 (2\nu)^3}{\varepsilon}\Phi_+\left(\frac{J-\nu T/2}{2\nu T}\right)\right), \quad  J-\nu T/2, \nu \gg 1, \quad \tilde J = \frac{J-\nu T/2}{2\nu T} = \mathcal{O}(1)
\end{equation}

   where $\Phi_+$ is the crossover function which interpolates between the regimes
\begin{equation}
\Phi_+(\tilde{J})=
\begin{cases}
\frac{\pi^2}{12}\tilde{J}^3, \quad &\tilde{J}\gg 1\\
\frac{16}{15\pi}\tilde{J}^{5/2}, \quad &\tilde{J}\ll 1\\
\end{cases}
\end{equation}
The purpose of this section is to obtain the parametric representation of the crossover function $\Phi_+$. Let us start again from the formula \eqref{startingpoint} for $u \Psi'(u)$.
\bea \label{startingpoint3}
u \Psi'(u) = \int_{ \mathbb{R}} \frac{\rmd k \, }{\pi \nu \left(4
k^2+1\right)  }
\log \left(1 + u \frac{\left(4 k^2+1\right)  (2 w+1)^2
e^{- (4 k^2+1) \nu^2 T }}{16
\left(k^2+w^2\right)}\right)
\eea
We now consider the double limit where both $u \to +\infty$ and $\nu \to +\infty$
with $u \sim e^{\nu^2 T (1+ 4 k_0^2) }$ with $k_0$ fixed, i.e., with
\be \label{scale2}
\frac{\log u}{\nu^2T  } = 1 + 4 k_0^2 = \mathcal{O}(1)
\ee
A crossover will occur as $k_0$ increases from $k_0 \to 0$ (KPZ) and $k_0 \to +\infty$ (WASEP). Neglecting the pre-exponential factors in the argument of the logarithm in \eqref{startingpoint3},
it is easy to see that the leading estimate at large $\nu$ is
\bea \label{upsi}
&& u \Psi'(u) \simeq
\nu T\int_{-k_0}^{k_0} \frac{\rmd k \, }{\pi}  \frac{4 \left(k_0^2-k^2\right)
}{4   k^2+1 }  =
\frac{\nu T}{\pi}   \left( (4
k_0^2+1) \arctan(2 k_0)-2
k_0\right)
\eea
We can now use again the large $u$ asymptotics of the parametric representation in \eqref{legendreasymptotic}
where the terms $e^{- 2\nu u \Psi'(u)}$ can be safely neglected, leading to
\bea \label{legendreasymptotic2}
&& J  \simeq  -  u \Psi'(u) + \frac{\log u}{2\nu}
\\
&& \Phi'(J) =  2\nu u \Psi'(u)
\eea
From \eqref{upsi} and \eqref{scale2} we see that the first two terms in $J$ will be proportional to
$\nu$ at large $\nu$. Hence we introduce the scaling variable
\be
\tilde J = \frac{J}{2 \nu T} - \frac{1}{4}
\ee
which will remain of order $\mathcal{O}(1)$ in the crossover region (the factor $-1/4$ was
introduced for later convenience).
One obtains from \eqref{legendreasymptotic2} together with \eqref{upsi} and \eqref{scale2}, that in the double limit
\be
\tilde J = \frac{k_0}{\pi } + k_0^2 - (k_0^2 + \frac{1}{4}) \frac{2}{\pi} \arctan(2 k_0)
\ee
which is a strictly increasing function of $k_0$ with $\tilde J \sim k_0^2$ for $k_0 \ll 1$
and $\tilde J \sim 2 k_0/\pi$ for $k_0 \gg 1$.
The second equation in \eqref{legendreasymptotic2} then
shows that in the crossover region $\Phi(J)$ will
take the scaling form
\be
\Phi(J) \simeq T^2(2 \nu)^3  \Phi_+(\tilde J)
\ee
and one has
\be
\Phi'_+(\tilde J) = \frac{u \Psi'(u)}{2\nu T}  =
\frac{\left(4 k_0^2+1\right)
\arctan(2k_0)-2 k_0}{2 \pi }
\ee

In summary the tail of the large deviation form of the PDF of the integrated current $J$ takes for
large $\nu$ and large $J$ the crossover scaling form of Eq.~\eqref{eq:lower-tail-crossover}
where the function $\Phi_+(\tilde J)$ is determined by the
parametric system (where $k_0 \in [0,+\infty[$ should be eliminated)
\bea
\label{eq:lower-tail-crossover-function}
&& \Phi'_+(\tilde J)  =
\frac{\left(4 k_0^2+1\right)
\arctan(2k_0)-2 k_0}{2 \pi } \\
&& \tilde J = \frac{k_0}{\pi } + k_0^2 - (k_0^2 + \frac{1}{4}) \frac{2}{\pi} \arctan(2 k_0)
\eea

From this representation it is easy to obtain the expansion of $\Phi_+(\tilde J)$ for small $\tilde J$
(corresponding to the KPZ limit, and small $k_0$) and for large $\tilde J$
(corresponding to the WASSEP limit, and large $k_0$). One find
\begin{itemize}
\item For $\tilde J \ll 1$, we have
\begin{equation}
\Phi_+(\tilde{J}) = \frac{16 \tilde{J}^{5/2}}{15 \pi }+\frac{32 \tilde{J}^3}{9 \pi ^2}-\frac{64 \left(3 \pi ^2-70\right) \tilde{J}^{7/2}}{315 \pi ^3}-\frac{256 \left(9 \pi
^2-100\right) \tilde{J}^4}{405 \pi ^4}+\mathcal{O}\left(\tilde{J}^{9/2}\right)
\end{equation}
    \item For $\tilde J \gg 1$ we have
\begin{equation}
\Phi_+(\tilde{J}) =\frac{\pi ^2}{12}  \tilde{J}^3+\frac{1}{16} \left(\pi ^2-8\right) \tilde{J}^2+\left(\frac{\pi ^2}{64}-\frac{1}{6}\right) \tilde{J}+\frac{1}{768} \left(\pi
^2-8\right)+\mathcal{O}\left(\frac{1}{\tilde{J}}\right)
\end{equation}
\end{itemize}

For completeness we also give the parametric representation of $\Phi_+(\tilde J)$
\bea
&& \Phi_+(\tilde J) = \int_0^{k_0} \rmd k_0 \tilde J'(k_0) \Phi'_+(\tilde{J}) \\
&&
=\frac{3 \left(4 k_0^2+1\right) \arctan(2 k_0) \left(-2 \left(4 k_0^2+1\right)
\arctan(2 k_0)+4 k_0 (\pi  k_0+2)+\pi \right)-2 k_0 (4 k_0 (5 \pi  k_0+3)+3 \pi)}{48 \pi ^2} \nn
\eea

\section{Domain of definition of $\Psi(u)$, analytic continuation, solitons and branches}
\label{sec:domain}

For simplicity we restrict in this section to the case $X=0$ and $\rho_1+\rho_2=1$.

\subsection{Extension of the domain of definition of $\Psi(u)$}

The range of definition of $\Psi(u)$ covers naturally $u \in [0,\infty[$. We show in this Section that this range can be extended to $u \in [u_c,\infty[$, with $u_c <0$, and additionally that $\Psi(u)$
possesses several branches. The main branch is given by \eqref{eq:rate-function-wasep-general}.
The secondary branches can be determined by an analytical continuation procedure,
and are required to obtain the tail $J \to -\infty$ of the current distribution.
Such a procedure is quite standard in the computation of the large deviations of weak noise theories, see Refs.~\cite{UsWNTCrossover,UsWNTDroplet2021,tsai2023integrability,krajenbrink2019beyond,NaftaliKMP2}. In order to simplify the analysis, we restrict to $X=0$, $\rho_1= \frac{1}{2} + w$, $\rho_2= \frac{1}{2} - w$, where $ 0 \leq w \leq 1/2$, and $\delta=1/2$ and start from the expression valid for the main branch, from
\eqref{eq:rate-function-wasep-general}
\be
\begin{split} \label{startingagain}
u \Psi'(u) &= \int_{ \mathbb{R}} \frac{dk}{\pi \nu \left(4
k^2+1\right)  }
\log \left(1 + u \frac{\left(4 k^2+1\right)  (2 w+1)^2
e^{- (4 k^2+1) \nu^2 T }}{16
\left(k^2+w^2\right)}\right)\\
\end{split}
\ee
Since $w \leq 1/2$, the function appearing inside the logarithm
\begin{equation}
\label{eq:analytical-continuation-h-k}
h(k): \quad  k \mapsto \frac{\left(4 k^2+1\right)  (2 w+1)^2
e^{- (4 k^2+1) \nu^2 T }}{16
\left(k^2+w^2\right)}
\end{equation}
is decreasing for $k\geq 0$, increasing for $k \leq 0$ and reaches its maximum at $k=0$ for the value
\begin{equation}
h(k=0)=\frac{ (2 w+1)^2 e^{-\nu ^2 T}}{16 w^2}
\end{equation}
Hence, for $u\Psi'(u)$ and by extension $\Psi(u)$ to be properly defined, we need notably $\log(1+u h(0))$ to be well defined and thus $u\geq -1/h(0)$ which provides a lower bound as a necessary condition for the definition of $u\Psi'(u)$
\begin{equation}
u\geq u_c = - \frac{16 w^2}{ (2 w+1)^2 e^{-\nu ^2 T}} = - (\sqrt{\alpha}-1)^2 e^{\nu ^2 T}
\end{equation}
where $\alpha=\frac{\rho_2(1-\rho_1)}{\rho_1(1-\rho_2)}=(1-2w)^2/(1+2 w)^2$.
Upon reaching $u=u_c$, one can obtain the corresponding $P=P_c$ and $J=J_c$ from
\eqref{ParamPhi} as
\begin{equation}
\begin{split}
J_c  &= - \frac{1}{2\nu}  \log \left(\frac{(1- e^{-  2\nu u_c \Psi'(u_c)}) (e^{  2\nu u_c \Psi'(u_c)}-\alpha)}{u_c}\right) \\
P_c  &=2 \nu u_c \Psi'(u_c)
\end{split}
\end{equation}
where $u_c \Psi'(u_c)$ is given by the above integral \eqref{startingagain} which is well defined at $u=u_c$.
The parametric equations \eqref{ParamPhi} with $u \in [u_c,\infty[$ and
$u \Psi'(u)$ given by \eqref{startingagain}
allow to compute the large deviation function $\Phi(J)$ for $J\in [J_c,\infty[$.
We now address how to compute $\Phi(J)$ for $J<J_c$. Note that (since $u_c<0$) one
has $J_c < \bar{J}$
where $\bar{J}$ is the mean integrated current computed in Section~\ref{sec:cum} (which corresponds
to $u=0$).\\

As in Refs.~\cite{UsWNTCrossover,UsWNTDroplet2021,NaftaliKMP2,krajenbrink2019beyond,tsai2023integrability}, one needs to determine the other branches of $\Psi(u)$. This allows to extend the range of the solution for $\Phi(J)$ to $J\in [J_c,\infty[$. The interpretation within the scattering approach to the equations of the MFT,
is that solitons are spontaneously generated in the regime $J\leq J_c$, which then provide an additional contribution to the large deviation function. The solitonic structure is unveiled by finding the locations of the zeros of the argument of the logarithm in $u\Psi'(u)$ in the complex plane, and their positions will define the rapidities of the solitons. We present here a simple derivation
leaving a more refined one using inverse scattering methods (which are in principle
possible from the integrable Lax structure that we have provided in the main text)
to a future work.

\subsection{Location of solitons}

To study potential solitons, we need to solve for $k \in \I \R$ the equation $1+uh(k)=0$, equivalent to
\begin{equation}
\label{eq:soliton-rapidity-equation}
\frac{u}{u_c}    = \frac{\left(w^2-\kappa ^2\right) e^{-4 \kappa ^2 \nu ^2 T}}{\left(1-4 \kappa ^2\right) w^2}:=g(\kappa)
\end{equation}
Note that since we restrict here to $X=0$, we only need to consider $k \in \I \R$, for $X\neq 0$, the solitons might have a more complicated structure, akin to the one observed in \cite{UsWNTCrossover,NaftaliKMP2}. Its Jacobian reads
\begin{equation}
\label{eq:jacob-soliton-kappa-u}
\frac{\rmd u}{u} = \frac{g'(\kappa)}{g(\kappa)} \rmd \kappa = (-8T \nu^2\kappa + \frac{\kappa  \left(2-8 w^2\right)}{\left(4 \kappa ^2-1\right) \left(w^2-\kappa ^2\right)})  \rmd \kappa
\end{equation}

\begin{figure}[t!]
\centering
\includegraphics[width=0.4\linewidth]{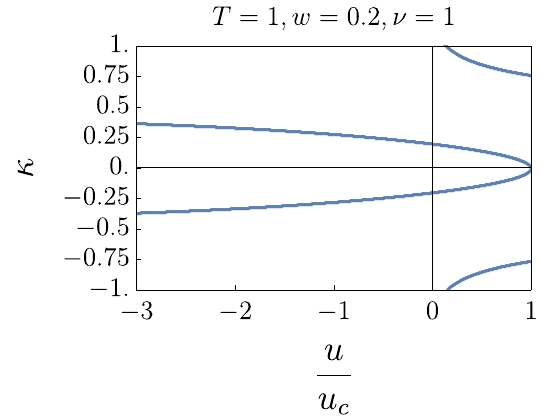}\\
\includegraphics[width=0.4\linewidth]{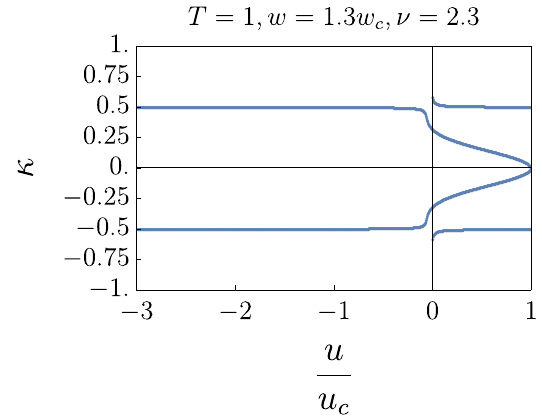}
\includegraphics[width=0.4\linewidth]{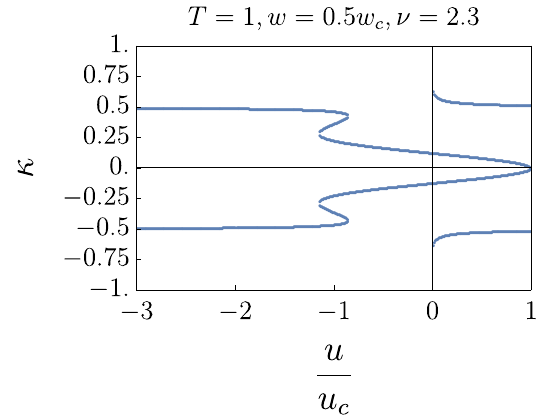}
\caption{Numerical solution of the soliton rapidity equation \eqref{eq:soliton-rapidity-equation} in different regimes. \textbf{(a) top} $T=1$, $w=0.2$, $\nu=1$ where the physical branch of $\kappa(u)$ (i.e., the one departing to the right from $\kappa(u_c)=0$) is single-valued, \textbf{(b) bottom-left} $T=1$, $w=1.3w_c$, $\nu=2.3$ where the physical branch is single-valued, \textbf{(c) bottom-right} $T=1$, $w=0.5w_c$, $\nu=2.3$ above the transition where the physical branch is tri-valued between $u_{c1}$ and $u_{c2}$.
We will not study this case in detail in this paper.}
\label{fig:soliton-rapidity-structure}
\end{figure}

Let us study the structure of the solutions, see Fig.~\ref{fig:soliton-rapidity-structure},
\begin{itemize}
\item For $u/u_c>0$ there are two pairs solutions $k=\pm \I \kappa_0$ and $k=\pm \I\kappa_{-1}$ with $\kappa_0<w$ and $\kappa_{-1}>1/2$. Only $\kappa_0$ will be of use here at we need the solution $\kappa(u)$ to vanish for $u=u_c$.
\item For $u/u_c<0$ there are one or three solutions $k=\pm \I\kappa_{0,1,2}$ (with $w<\kappa_{0,1,2}<1/2$). The number of solutions is controlled by the existence of a critical point defined as a real root of
\begin{equation}
g'(\kappa)=0 \quad \Leftrightarrow \quad 16 T\nu^2 \kappa^4 - 4 T\nu^2 (1 + 4 w^2) \kappa^2 + 1
+ 4 w^2 (T\nu^2-1) = 0 \quad \text{or} \quad \kappa=0
\end{equation}
Note that by definition $w^2 \leq 1/4$ in our model.
For $T \nu^2 \geq  4/(1- 4 w^2)$, equivalently if one fixes
$\nu^2 T >4$, for $w\in [0, w_c=\sqrt{\frac{T\nu^2-4}{4T\nu^2}}]$,
there are two pairs of critical points (besides the trivial root $\kappa=0$)
\begin{equation}
\begin{split}
\kappa_{c1}^2 &= \frac{1}{8} \left(4 w^2+1-\frac{\sqrt{\nu ^2 T \left(4 w^2-1\right) \left(\nu ^2 T \left(4 w^2-1\right)+4\right)}}{\nu ^2 T}\right)\\
\kappa_{c2}^2 &= \frac{1}{8} \left(4 w^2+1+\frac{\sqrt{\nu ^2 T \left(4 w^2-1\right) \left(\nu ^2 T \left(4 w^2-1\right)+4\right)}}{\nu ^2 T}\right)\\
\end{split}
\end{equation}
while for $T \nu^2 <  4/(1- 4 w^2)$ there are no such critical points.
If there are no critical points then there is only one pair of solutions $k=\pm \I \kappa_0$ for $u/u_c<0$,
see Figure~\ref{fig:soliton-rapidity-structure} panels a) and b).
If there are critical points then there exists an interval such that for
$u\in [u_{c2},u_{c1}]$ there are three pairs of solutions (with $u_{c1}/u_c<0$, $u_{c2}/u_c<0$),
see Figure~\ref{fig:soliton-rapidity-structure} panel c). We further associate $u_{c1} \leftrightarrow \kappa_{c1}$ and $u_{c2} \leftrightarrow \kappa_{c2}$ so that the function $\kappa(u)$ is tri-valued in the interval $[u_{c2},u_{c1}]$, see Fig.~\ref{fig:soliton-rapidity-structure}.

\end{itemize}
\subsection{Analytical continuation of $u\Psi'(u)$}
Knowing the solitonic structure, we can now obtain the analytical continuation of $u\Psi'(u)$ using the same derivation as in Ref.~\cite{krajenbrink2017exact} (see also \cite{UsWNTFlat2021} for subsequent work). We start by proceeding to an integration by part on the integral definition of  $u\Psi'(u)$ as
\be
\begin{split}
u \Psi'(u)
&= 2\int_{ 0}^{+\infty} \frac{dk}{\pi \nu \left(4   k^2+1\right)  }  \log \left(1 + u h(k)\right)\\
&=-\int_{ 0}^{+\infty} \frac{\rmd k}{\pi \nu} \arctan(2k) \frac{u h'(k)}{u h(k)+1}
\end{split}
\ee

where we have defined $h(k)$ in \eqref{eq:analytical-continuation-h-k}. We apply a further change of variable $b=\frac{1}{h(k)}$ so that
\begin{equation}
\begin{split}
u \Psi'(u) &=  \int_{ 1/h(0)=-u_c}^{1/h(+\infty)=+\infty} \frac{\rmd b}{\nu \pi }  \arctan(2k(b)) \frac{u }{b}\frac{1}{u +b}
\end{split}
\end{equation}

Note that $u_c$ is a branching point for $u\Psi'(u)$, around which one can turn around in the complex plane at the expense of changing the Riemann sheet on which the function is defined \cite{krajenbrink2017exact}. This amounts to add a jump contribution $u\Psi'(u) \to u\Psi'(u)+u\Delta'(u)$ where the jump reads
\begin{equation}
\label{eq:jump-u-delta-prime-u}
u\Delta'(u) = -\frac{2}{\nu} \mathrm{arctanh}(2\kappa(u))
\end{equation}
where $\kappa(u)=g^{-1}(u/u_c)$ where $g(\kappa)$ is defined in  \eqref{eq:soliton-rapidity-equation}. Note that this requires $\kappa<1/2$. A number of subtleties arise at this stage
\begin{itemize}
\item If there are multiple solutions $\kappa_{0,1,2}$ to the equation $\frac{u}{u_c}=g(\kappa)$, see the previous subsection, then there are multiple possible choices for the value of $\kappa(u)$ in the jump function $\Delta(u)$ or its derivative $u\Delta'(u)$ in \eqref{eq:jump-u-delta-prime-u}. In the analysis of the weak noise theory of the KPZ equation with Brownian initial condition in \cite{krajenbrink2017exact}, similar multiple choices have led to the existence of a phase transition, which in this context is a non-analyticity of the large deviation function $\Phi(J)$, see \cite{krajenbrink2017exact,UsWNTFlat2021,UsWNTCrossover} and \cite{JanasDynamical,NaftaliKMP2}. This subtlety is more easily understood in the context of the inverse scattering methods and thus we leave the analysis of the case of multiple solutions to a future work.
\item  In the present paper we focus on the case where $\kappa(u)$ is single-valued, i.e., on the case
(see previous subsection)
\be \label{domain}
T \nu^2 <4, \quad \mathrm{or}, \quad  T\nu^2 \geq 4, 1/2\geq w\geq w_c=\sqrt{\frac{T\nu^2-4}{4T\nu^2}}
\ee

    The function $\kappa(u)$ is given by the root $\kappa_0$ (which belongs to the physical branch, see Fig.~\ref{fig:soliton-rapidity-structure}) where $0\leq \kappa_0 \leq w$ for $u/u_c\geq 0$ and $w\leq \kappa_0 \leq 1/2$ for $u/u_c \leq 0$ (with $\kappa(u_c)=0$ and $\kappa(0)=w$).  In that case one can further integrate this relation to obtain $\Delta(u)$, which is the continuation of $\Psi(u)$ as
\begin{equation}
\label{eq:jump-delta-u}
\begin{split}
\Delta(u)&=-\frac{2}{\nu}\int_{u_c}^u \frac{\rmd u'}{u'}   \mathrm{arctanh}(2\kappa(u'))\\
&= \frac{2}{\nu}\int_{0}^{\kappa(u)}\rmd \kappa (8T \nu^2\kappa +\frac{2 \kappa }{(w-\kappa ) (\kappa +w)}+\frac{8 \kappa }{(2 \kappa -1) (2 \kappa +1)})  \mathrm{arctanh}(2\kappa)
\end{split}
\end{equation}
    where we have used from the first line to the second line the expression of the Jacobian \eqref{eq:jacob-soliton-kappa-u}. For general $w<1/2$ Eq.~\eqref{eq:jump-delta-u} provides a parametric representation of $\Delta(u)$, where $u \in [u_c,\infty[$ (equivalently $\kappa \in [0,1/2[$).
    The integral in \eqref{eq:jump-delta-u} can be computed explicitly but leads to a complicated expression
    for general $w$.

    For the step initial condition, i.e., $w=1/2$, the expression of the jump simplifies and reads
\begin{equation}
\label{eq:jump-delta-u-step-initial}
\Delta(u)=2 \nu  T \left(\left(4 \kappa ^2-1\right) \mathrm{arctanh}(2 \kappa )+2 \kappa \right), \quad \kappa=\kappa(u)=\sqrt{\frac{\log(u_c/u)}{4\nu^2 T}}, \quad u_c=-e^{\nu^2 T}
\end{equation}
    where $u$ now varies in $[u_c,-1]$.

\end{itemize}

\subsection{Continuation of the parametric representation of $\Phi(J)$ for $J<J_c$}
\label{subsec:paramsmallJ}

Let us recall that the main branch \eqref{ParamPhi} corresponds to $u\in [u_c,+\infty[$ and $J>J_c$.
Let us now give our result for the parametric representation of $\Phi(J)$ for $J<J_c$,
where again $u\in [u_c,+\infty[$
\begin{equation}
\begin{split} \label{PhiContinued}
\Phi'(J)&=2\nu u (\Psi'(u)+\Delta'(u))\\
J  &= - \frac{1}{2\nu}  \log \frac{\zeta(u)}{u} \\  \zeta(u) &: = e^{  2\nu u (\Psi'(u)+\Delta'(u))} - (1 + \alpha) + \alpha e^{  - 2\nu u (\Psi'(u)+\Delta'(u))}\\
&= \left(\frac{1-2\kappa(u)}{1+2\kappa(u)} \right)^2 e^{  2\nu u \Psi'(u)} - (1 + \alpha) + \alpha \left(\frac{1+2\kappa(u)}{1-2\kappa(u)} \right)^2 e^{  - 2\nu u \Psi'(u)}\\
\end{split}
\end{equation}
where $\Delta(u)$ and $\kappa(u)$ where defined in the previous subsection.
In the last line we have used \eqref{eq:jump-u-delta-prime-u}.
We recall that \eqref{PhiContinued} is valid for $X=0$, $\rho_1+\rho_2=1$ and
in the parameter range defined in \eqref{domain}. In addition we
also assumed $\alpha>0$, i.e., we excluded the step initial condition
which we now address.
\begin{remark} \label{remarkstep} Case of step initial condition: in that case
the integrated current at $X=0$, $J(0,T)$, is
the total number of particles which are on the right of zero at time $T$ and
must thus be positive. Hence the large deviation rate function $\Phi(J)$ is
only defined for $J>0$. It is obtained parametrically from
\begin{equation}
\begin{split} \label{PhiContinuedstep}
\Phi'(J)&=2\nu  (u \Psi'(u)+ u \Delta'(u))\\
J  &= - \frac{1}{2\nu}  \log \frac{(  e^{  2\nu ( u \Psi'(u)+ u \Delta'(u))} -
1) }{u}  = - \frac{1}{2\nu}  \log \left(\frac{ \left(\frac{1-2\kappa(u)}{1+2\kappa(u)} \right)^2 e^{  2\nu u \Psi'(u)} -   1 }{u} \right)
\end{split}
\end{equation}

where for $J>J_c$, $u \Delta'(u)=0$ (and one can set $\kappa(u)\equiv 0$) and $u$
varies from $u=+\infty$ (i.e., $J=+\infty$)
to $u=u_c=- e^{\nu^2 T}$ (i.e., $J=J_c$),
while for $0<J<J_c$ one has
$u \Delta'(u)$ given by \eqref{eq:jump-u-delta-prime-u}, and $u\in [u_c,-1]$
(where $u=-1$ corresponds to $J=0$). In addition one has
\be
\begin{split} \label{startingagainstep}
u \Psi'(u) &= \int_{ \mathbb{R}} \frac{\rmd k \, }{\pi \nu \left(4
k^2+1\right)  }
\log \left(1 + u e^{- (4 k^2+1) \nu^2 T }\right)\\
\end{split}
\ee
Hence
\be \label{Jcformula}
J_c = \frac{\nu T}{2} - \frac{1}{2\nu}  \log (1-  e^{  2 \nu u_c \Psi'(u_c)})
\quad , \quad 2 \nu u_c \Psi'(u_c) = \frac{2}{\pi}  \int_{ \mathbb{R}} \frac{\rmd k \, }{  \left(4
k^2+1\right)  } \log \left(1 - e^{- 4 k^2 \nu^2 T }\right)
\ee
Since the second term is always positive we see that $J_c>\frac{\nu T}{2}$ (which is
positive since we assume everywhere $\nu>0$).

\end{remark}

These subtleties cleared, we will show in the next appendix how to obtain the upper tail of the current distribution from the jump $u\Delta'(u)$ in the case \eqref{domain}.

\begin{remark}[Analytical continuation of the observable]
Let us assume $\alpha>0$.
Although \eqref{PhiContinued} always holds, there is a hidden branching point at $u=u_{\zeta_c}>u_c$
and $\zeta(u)=\zeta_c$, where $\zeta_c$ is defined below. This is
associated to the fact that the observable itself contains polylogarithm functions and also
needs to be continued. It
has no consequence for \eqref{PhiContinued}.
We show that this continuation amounts to change the branch of $\omega$ obtained in \eqref{omegauz}.
Starting from the relation (obtained in Section~\ref{subsec:leading-order-ssep-identity} for any $\zeta(u)$) for the derivative of the observable with respect to $u$
\begin{equation}
\log(1+\zeta(u) \omega_{\zeta(u)})=   \int_{\I \mathbb{R} + \delta}
\frac{\rmd y}{2\I \pi y(1-y)}  \log\left( 1 + \zeta(u)\frac{\rho _1 \left(1-\rho _2\right) (y-1) y}{\left(y-\rho_1\right)
\left(y-\rho _2\right)}  \right)
\end{equation}

we see that this integral is the same one as the one which defines $2\nu u\Psi'(u)$, upon choosing $\nu \to 0$ and $u \to \zeta(u)$. We can thus use the results on the continuation of $u\Psi'(u)$ and obtain that $\omega_{\zeta(u)}$ and this expression are well defined as long as $\zeta(u)  \geq \zeta_c=-(\sqrt{\alpha}-1)^2=-16w^2/(1+2w)^2$. This corresponds to $P_{\zeta_c}=\log \sqrt{\alpha}$,  $\omega(\zeta_c)=1/(1-\sqrt{\alpha})$. The jump of the observable is now obtained in a similar way as in \eqref{eq:jump-u-delta-prime-u} and reads
\begin{equation}
\begin{split}
\log(1+\zeta(u) \omega_{\zeta(u)}) &\to \log(1+\zeta(u) \omega_{\zeta(u)})    -4 \mathrm{arctanh}\, \left(\frac{2w\sqrt{\zeta(u)/\zeta_c-1}}{\sqrt{4\zeta(u)w^2/\zeta_c-1}}\right)\\
&= -\log (1+\zeta(u)\omega_{\zeta(u)})+\log \alpha
\end{split}
\end{equation}

We note that $-\log (1+\zeta(u)\omega_{\zeta(u)})+\log \alpha=\log (1+\zeta(u)\hat{\omega}_{\zeta(u)})$ where $\hat{\omega}$ is the second branch of the solution of \eqref{omegauz}. For the continuations to be compatible, we assume that $u_{\zeta_c}\geq u_c$ and $\zeta(u_c)\geq \zeta_c$. This implies that
\begin{equation}
-(\sqrt{\alpha}-1)^2 e^{2\nu J_{\zeta_c}}  \geq -(\sqrt{\alpha}-1)^2 e^{\nu^2 T} \Longleftrightarrow  J_{\zeta_c} < \nu T/2
\end{equation}

     and
\begin{equation}
-(\sqrt{\alpha}-1)^2 e^{-2\nu J_c+\nu^2 T}\geq -(\sqrt{\alpha}-1)^2 \Longleftrightarrow J_c > \nu T/2
\end{equation}
 Denoting by $J_{\zeta_c}$ the current at the position $u_{\zeta_c}$, this further implies that $J_c \geq  J_{\zeta_c}$. Finally, we need to complement the parametric representation \eqref{PhiContinued} with the final equations
\begin{itemize}
\item For $u\in [u_c,u_{\zeta_c}]$ and $J\in [J_{\zeta_c},J_c]$, we have
\begin{equation}
\begin{split}
\log (1+ \zeta(u) \omega_{\zeta(u)}  ) &= 2\nu u(\Psi'(u)+\Delta'(u)) \\
\end{split}
\end{equation}

\item For $u\in [u_{\zeta_c},+\infty[$ and $J\in ]-\infty,J_{\zeta_c}]$ we have
\begin{equation}
\begin{split}
-\log (1+\zeta(u)\omega_{\zeta(u)})+\log \alpha & =  2\nu u(\Psi'(u)+\Delta'(u)) \\
\end{split}
\end{equation}
\end{itemize}

\end{remark}

\subsection{Plots of $\Phi(J)$ and $\Phi'(J)$}

\begin{figure}[t!]
\centering
\includegraphics[width=0.45\linewidth]{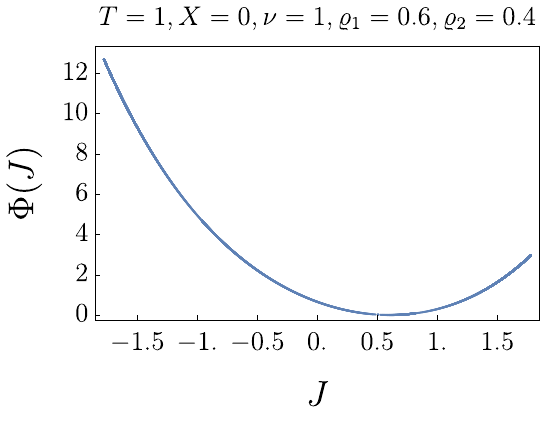}
\includegraphics[width=0.47\linewidth]{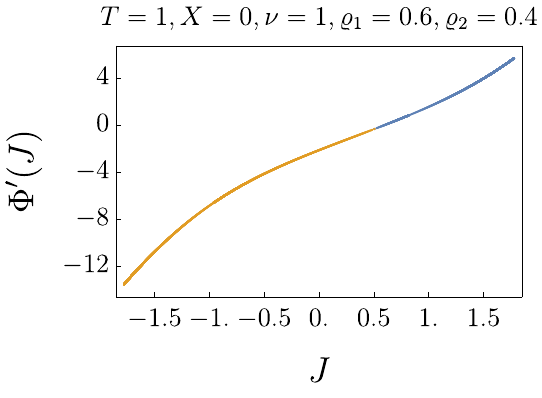}\\
\includegraphics[width=0.45\linewidth]{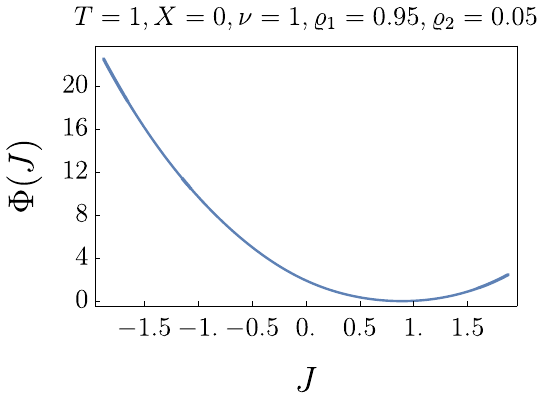}
\includegraphics[width=0.47\linewidth]{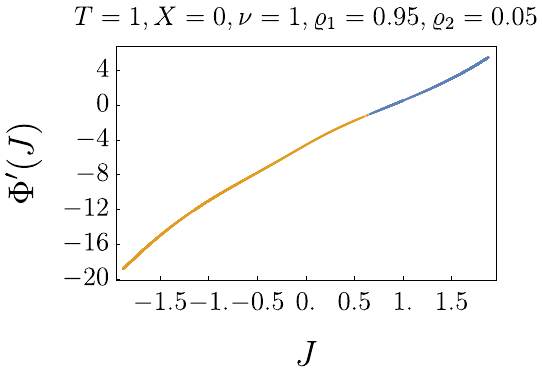}\\
\includegraphics[width=0.45\linewidth]{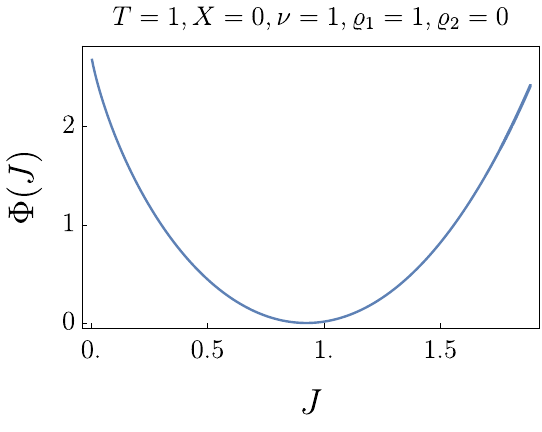}
\includegraphics[width=0.47\linewidth]{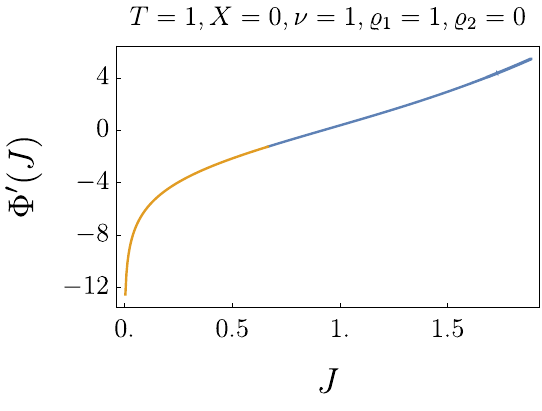}\\
\caption{Plot of $\Phi(J)$ obtained numerically parametrically for the values $X=0$, $T=1$, $\nu=1$ and various densities $\varrho_1=1-\varrho_2=\{0.6,0.95,1\}$. In the right plots representing $\Phi'(J)$ the blue curve represents the main branch and the orange curve represents the second branch where the continuation is introduced.}
\label{fig:phiJ}
\end{figure}

We plot in this Section the functions $\Phi(J)$ and $\Phi'(J)$ using the parametric representations in the main branch \eqref{ParamPhi} and in the second branch \eqref{PhiContinued}. We report the plots in Fig.~\ref{fig:phiJ}.
\section{Tail for $J\to -\infty$ (or $J \to 0$ for step initial condition)}
This tail corresponds to the {\it upper tail} for the height field $h=-J$.
For simplicity, as in the previous Section, we restrict
to the case $X=0$ and $\rho_1+\rho_2=1$, and to the parameter range defined in \eqref{domain}.
\subsection{Tail for $J\to -\infty$ for the WASEP (double-sided Bernoulli initial condition)}
The upper tail of the WASEP for $\alpha>0$ (which excludes the step initial condition)
is obtained by taking $u \to +\infty$ adding the contribution of the jump to the large deviation function. We will use the results from Section~\ref{subsec:lower-tail-wasep} for the asymptotics of $u\Psi'(u)$, which we will complement with the one of $u\Delta'(u)$ obtained above.\\

We first provide the asymptotics of the derivative of the jump for $u \to +\infty$
\begin{equation}
\begin{split}
u\Delta'(u)=- \frac{2}{\nu} \mathrm{arctanh}(2\kappa(u))
&=-\frac{\left(2\log (u)-\log \alpha \right)}{2\nu} -\frac{-4 \nu ^2 T+16 \nu ^2 T w^2+8 w^2+2}{\nu u (2   w+1)^2}+\mathcal{O}\left(\frac{1}{u}\right)^2\\
\end{split}
\end{equation}

where we recall that $\alpha=\left(\frac{1-2w}{1+2w}\right)^2$. This series is obtained by expanding \eqref{eq:soliton-rapidity-equation} around $\kappa=1/2$ (equivalent to $u\to +\infty$), inverting the series and injecting it inside the hyperbolic arctangent. Adding this contribution to the
asymptotics in \eqref{eq:asymptot-u-psip-plus-infty} we obtain for $u \to +\infty$
\begin{equation}
\label{eq:asymptot-u-psip-deltap-plus-infty}
u \Psi'(u)+u\Delta'(u) =  -\frac{\log u}{2 \nu} - \frac{2\sqrt{T}}{\pi} \sqrt{\log u}  + \tilde{b} +
\mathcal{O}(1/\sqrt{\log u})\quad , \quad \tilde{b}= \frac{1}{\nu} ( B_w  + \log 2 + \frac{\log \alpha}{2})
\end{equation}

Let us now examine the parametric equations \eqref{PhiContinued}. From
\eqref{eq:asymptot-u-psip-deltap-plus-infty} we see that in the equation
of $\zeta(u)$
it now the last term which is dominant at large $u$
leading to
\bea
&& J  \simeq    u (\Psi'(u)+\Delta'(u)) + \frac{\log u-\log \alpha}{2\nu}     \\
&& \Phi'(J) = 2\nu u (\Psi'(u)+\Delta'(u))
\eea
up to much smaller terms of order $\mathcal{O}(1/u)$ which we can neglect in the present
expansion which is in $1/\log u$.

Let us now use \eqref{eq:asymptot-u-psip-deltap-plus-infty}, up to leading order
\bea
&& J  = - \frac{2\sqrt{T}}{\pi} \sqrt{\log u}  + b + \mathcal{O}(1/\sqrt{\log u})    \\
&& \Phi'(J) =  -\log u - \frac{4 \nu \sqrt{T}}{\pi} \sqrt{\log u}  + 2 \nu \tilde{b} +
\mathcal{O}(1/\sqrt{\log u})
\eea
Hence we find for $J \to -\infty$ the following tail
\be \label{upper-tailwasepfinal}
\Phi(J)= \frac{\pi ^2 |J|^3}{12T}+J^2
\left(\frac{\pi ^2
b}{4T}+\nu \right) + \mathcal{O}(J)
\ee
Note that the first difference with the $J\to +\infty$ tail \eqref{tailwasepfinal} appears in the subdominant $\mathcal{O}(J^2)$ term.

\subsection{Tail near the wall $J \to 0$ for the WASEP (step initial condition)}

In the case of the step initial condition there is a "wall" at $J=0$, since
$\Phi(J)$ is defined only for $J>0$. One can ask about the asymptotic
behavior near the wall. It corresponds to $u \to -1^-$
and $\kappa(u) \to 1/2^-$. Setting $u= - (1+ \delta)$
and using Eqs.~\eqref{eq:jump-u-delta-prime-u}, \eqref{eq:jump-delta-u-step-initial} and
\eqref{PhiContinuedstep} we obtain
\begin{equation}
u\Delta'(u)=\frac{1}{\nu}\log \left(\frac{\delta}{4\nu ^2 T}\right)+\frac{\delta  \left(1-\nu ^2 T\right)}{2 \nu ^3 T}+\mathcal{O}\left(\delta
^2\right)
\end{equation}
\begin{equation}
J = \frac{\delta }{2 \nu }+\frac{\delta ^2 }{32 \nu ^5}\left(\frac{e^{-2\nu  \Psi'(-1)}}{T^2}-8
\nu ^4\right)+\mathcal{O}\left(\delta ^3\right)
\end{equation}
So that for small $\delta$
\be
\Phi'(J)= 2 \log \left(\frac{\delta}{4\nu ^2 T}\right) - 2 \nu \Psi'(-1) + \mathcal{O}(\delta)
\ee
This leads to
\begin{equation}
\Phi'(J)=2 \log (\frac{J}{2\nu T})-2\nu \Psi'(-1)+\mathcal{O}(J)
\end{equation}
Hence near the wall for $J \to 0$ we find that
\be
\Phi(J) = \Phi(0^+) + 2 J \log (\frac{J}{2\nu T}) - (2\nu \Psi'(-1) + 2) J + \mathcal{O}(J^2)
\ee
where $\Phi(0^+)$ is determined by the fact that one must have simultaneously
$\Phi(J_{\rm typ})=\Phi'(J_{\rm typ})=0$ for the typical value (also
equal to the mean) $J_{\rm typ}= \langle J \rangle$, which was computed in \eqref{Jav}.

\subsection{Large $\nu$ limit and crossover from the wall to the typical value (step initial condition)}

Here we study the step initial condition in the regime where $\nu \to +\infty$.
In that regime we will find that $J_c \simeq \langle J \rangle \simeq \frac{\nu T}{2}$.
We study here the region $J<J_c$.
Interestingly, it describes a crossover between the "wall" region described
in the previous section and a regime where KPZ behavior emerges. Surprisingly,
it can be matched completely to a recent result on ASEP, obtained in an a priori
different limit (see below).\\

Let us start with the estimation of $J_c$ and of the typical value, $\langle J \rangle$, in the limit of large $\nu$ (which corresponds to $\kappa=0$).
One has from \eqref{Jcformula} for large $\nu$
\bea
&& 2 \nu u_c \Psi'(u_c)
= - \frac{\zeta(3/2)}{\sqrt{\pi T}\nu } +\frac{\zeta \left(5/2\right)}{2 \sqrt{\pi } \nu ^3 T^{3/2}} + \mathcal{O}(\frac{1}{\nu^5})\\
&& J_c
\simeq \frac{\nu T}{2} + \frac{1}{2\nu} \log(  \frac{\nu \sqrt{\pi T}}{\zeta(3/2)} ) + \mathcal{O}(1/\nu^2)
\label{Jcestimate}
\eea
while we recall that
\be
\langle J \rangle =-\frac{\log (\text{Erfc}(\nu \sqrt{T})
))}{2 \nu } =
\frac{\nu T}{2}   + \frac{1}{2 \nu} \log(\nu \sqrt{\pi}) + \mathcal{O}(1/\nu^3) \quad \nu \to + \infty
\ee
Hence we see that $J_c$ and $\langle J \rangle$ are both large, but their
difference vanish at large $\nu$ as $\langle J \rangle - J_c \simeq \frac{1}{2\nu} \log(\zeta(3/2))$.
These results can be compared with the KPZ equation, as provided in the following remark.
\begin{remark}
In \cite{le2016exact} the KPZ equation was studied for short time $T_{\rm KPZ} \ll 1$ with droplet initial condition.
Let us denote here ${\cal H}$ the variable $H$ defined in that paper.
It is related to $H_{\rm KPZ}$ defined here in \eqref{eq:def-hkpz-compendium} as follows
\be
{\cal H} = H_{\rm KPZ}  + \log(\sqrt{4 \pi T_{\rm KPZ}}) -\log (2 \nu \varepsilon)
\ee

The correspondence is read from \eqref{eq:generating-func-convergence-kpz}
and from \cite{le2016exact}.
It was shown there that the position of the branching point is
${\cal H}_c=
\log(\zeta(3/2)) $. This is consistent with the present
result, using $T_{\rm KPZ} = \nu^4 \varepsilon^2 T$ (for general $T$) since here we obtain
\be
H^c_{\rm KPZ} = -2 \nu (J_c- \frac{\nu T}{2}) = {\cal H}_c - \log(\nu \sqrt{\pi T}) + o(1)
\ee
\end{remark}

We now study the crossover regime corresponding to $u_c = - e^{\nu^2 T} < u \ll -1$, with $\kappa=\kappa(u) = \mathcal{O}(1)$ and $u \to -\infty$ with (from \eqref{eq:jump-delta-u-step-initial})
\be \label{scalingu}
\frac{ \log(-u)}{\nu^2 T} = 1 - 4 \kappa^2    \quad , \quad 0< \kappa < \frac{1}{2}
\ee
so that $\kappa=0$ corresponds to $u=u_c$ and $\kappa=1/2$ to $u/u_c \to 0$. In the regime $\kappa=\mathcal{O}(1)$ it is easy to see that $u \Psi'(u)+ u\Delta'(u) \simeq
u\Delta'(u)$ up to exponentially small correction terms, indeed one has, from \eqref{startingagainstep}
\be
u \Psi'(u) = \frac{1}{\nu} \int_{ \mathbb{R}} \frac{\rmd k \, }{\pi \left(4
k^2+1\right)  }
\log \left(1 - e^{- (4 k^2+4 \kappa^2) \nu^2 T }\right)
\simeq - \frac{1}{\nu} \int_{ \mathbb{R}} \frac{\rmd k \, }{\pi \left(4
k^2+1\right)  }  e^{- (4 k^2+4 \kappa^2) \nu^2 T } \sim \frac{1}{\nu^2}
e^{- 4 \kappa^2 \nu^2 T }
\ee
which is negligible for $\kappa \gg 1/\nu^2$.\\

Using Eqs.~\eqref{PhiContinuedstep}, \eqref{eq:jump-u-delta-prime-u} and \eqref{scalingu}, we then obtain the parametric representation for $0 < J \leq J_c$
\bea
&& \Phi'(J) \simeq 2 \nu u \Delta'(u) + \mathcal{O}(1/\nu) = - 4 \, \mathrm{arctanh}(2 \kappa) + \mathcal{O}(1/\nu) \\
&& J \simeq \frac{1}{2 \nu} \log |u| - \frac{1}{2 \nu} \log\left( 1 - (\frac{1- 2 \kappa}{1 + 2 \kappa})^2\right)
\simeq (1- 4 \kappa^2) \frac{\nu T}{2} + \mathcal{O}(1/\nu) \label{Ju}
\eea
To this leading order we also find
$2 \kappa = \sqrt{1 - \frac{J}{J_c}} + \mathcal{O}(1/\nu^2)$ and
\be
\Phi'(J) =  - 4 \, \mathrm{arctanh}(\sqrt{1 - \frac{J}{J_c}}) + \mathcal{O}(1/\nu)
\ee
leading to
\bea  \label{eq:jump-phi-u-step-initial}
\Phi(J) &&\simeq 2 \nu  T \left(\left(4 \kappa ^2-1\right) \mathrm{arctanh}(2 \kappa )+2 \kappa \right) \\
&& \simeq
4 J_c \left(  \sqrt{1 - \frac{J}{J_c}} - \frac{J}{J_c} \mathrm{arctanh}(\sqrt{1 - \frac{J}{J_c}}) \right)
\eea
which vanishes at $J=J_c \simeq \langle J \rangle$ as it should, and
where corrections are of order $1/\nu$.
Note that $\Phi(J)$ coincides exactly with $\Delta(u)$ obtained in \eqref{eq:jump-delta-u-step-initial}.
The reason is that for the step initial condition one has, from the Legendre
inversion formula at the saddle point
see Section~\ref{sec:derivation}
\be
\begin{split}
\Phi(J)= \Psi(u)+\Delta(u) + \frac{1}{2 \nu} {\rm Li}_2(- u e^{- 2\nu J})
\end{split}
\ee
and since $u e^{- 2\nu J}=\mathcal{O}(1)$ from \eqref{Ju}, and
that $\Psi(u) = \mathcal{O}( \nu T)$ we see that
the third term is subdominant in the limit $\nu \to +\infty$. One can perform the expansions around $J\to 0$ and $J \to J_c$ and
one finds
\begin{equation}
\Phi(J) =
\begin{cases}
4 J_c\left(1- \frac{J}{2J_c} (1- \log
\frac{J}{4J_c})\right)+\mathcal{O}(J^2)\\
\frac{8}{3} J_c
\left(1-\frac{J}{J_c}\right){}^{3/2}+ \frac{8}{15} J_c
\left(1-\frac{J}{J_c}\right){}^{5/2}+\mathcal{O}\left( (1-\frac{J}{J_c})^{7/2}\right)
\end{cases}
\end{equation}
\begin{remark}
Remarkably, the expression \eqref{eq:jump-phi-u-step-initial} coincides with the large deviation
function for the upper tail of the ASEP for the step initial condition
obtained recently in \cite[Eq.~(1.4)]{Das_2022}. Let us recall that result. With the same initial
condition as in the present paper, they find that
$H_0(t)={\sf J}_{\sf t}(0)$, i.e., the number of particles to the right of zero,
satisfies the large deviation principle
\be
{\rm Prob}\left(\frac{\gamma {\sf t}/4- H_0({\sf t})}{\gamma {\sf t}/4} > y\right) \sim e^{- \gamma {\sf t} \Phi_+(y)} \quad , \quad
\Phi_+(y) = \sqrt{y}- (1-y) \mathrm{arctanh}(\sqrt{y}) \quad , \quad 0<y<1
\ee
where $\gamma=R-L$. At small $y$ one has $\Phi_+(y) \simeq \frac{2}{3} y^{3/2}$, i.e.
it matches both the tail of the GUE Tracy-Widom distribution  as it should, see \cite[Remark 1.5]{Das_2022}, and
the upper tail of the KPZ equation. This result is a priori in a
completely different regime (fixed asymmetry $R-L = \mathcal{O}(1)$, and large ASEP time)
from the one studied here. The coincidence between the two results,
which can be checked using that $\gamma {\sf t} \to 2 \nu T/\varepsilon=4 J_c/
\varepsilon$, indicates
that no intermediate regime exists in the large deviations between these two limits.
\end{remark}
\end{document}